\documentclass[pdflatex,sn-mathphys-ay,twocolumn, a4paper]{sn-jnl}

\usepackage{geometry}
\usepackage{tcolorbox}
\usepackage{graphicx}%
\usepackage{multirow}%
\usepackage{amsmath,amssymb,amsfonts}%
\usepackage{amsthm}%
\usepackage{mathrsfs}%
\usepackage[title]{appendix}%
\usepackage{xcolor}%
\usepackage{textcomp}%
\usepackage{manyfoot}%
\usepackage{booktabs}%
\usepackage{algorithm}%
\usepackage{algorithmicx}%
\usepackage{algpseudocode}%
\usepackage{listings}%
\usepackage{supertabular}
\usepackage{caption}
\usepackage{tabularx}
\usepackage{array}
\usepackage{enumitem}

\theoremstyle{thmstyleone}%
\theoremstyle{thmstyletwo}%

\theoremstyle{thmstylethree}%

\begin{document}

\title[The New Compiler Stack: A Survey on the Synergy of LLMs and Compilers]{The New Compiler Stack: A Survey on the Synergy of LLMs and Compilers}
\author[1,2]{\fnm{Shuoming} \sur{Zhang}}\email{zhangshuoming21s@ict.ac.cn}
\author*[1,2]{\fnm{Jiacheng} \sur{Zhao}}\email{zhaojiacheng@ict.ac.cn}
\author[1,2]{\fnm{Qiuchu} \sur{Yu}}\email{yuqiuchu19@mails.ucas.ac.cn}
\author[3]{\fnm{Chunwei} \sur{Xia}}\email{C.Xia@leeds.ac.uk}
\author[3]{\fnm{Zheng} \sur{Wang}}\email{Z.Wang5@leeds.ac.uk}
\author[1,2]{\fnm{Xiaobing} \sur{Feng}}\email{fxb@ict.ac.cn}
\author[1,2]{\fnm{Huimin} \sur{Cui}}\email{cuihm@ict.ac.cn}

\affil*[1]{\orgdiv{SKLP}, \orgname{Institute of Computing Technology, CAS}, \orgaddress{\street{6th Kexueyuan South Rd}, \city{Beijing}, 
\country{China}}}
\affil[2]{\orgname{University of Chinese Academy of Sciences}, \orgaddress{\city{Beijing}, \country{China}}}
\affil[3]{\orgname{University of Leeds}, \country{UK}}

\abstract{
  This survey has provided a systematic overview of the emerging field of LLM-enabled compilation by addressing several key research questions. We first answered how LLMs are being integrated by proposing a comprehensive, multi-dimensional taxonomy that categorizes works based on their Design Philosophy (Selector, Translator, Generator), LLM Methodology, their operational Level of Code Abstraction, and the specific Task Type they address. In answering what advancements these approaches offer, we identified three primary benefits: the democratization of compiler development, the discovery of novel optimization strategies, and the broadening of the compiler's traditional scope. Finally, in addressing the field's challenges and opportunities, we highlighted the critical hurdles of ensuring correctness and achieving scalability, while identifying the development of hybrid systems as the most promising path forward. By providing these answers, this survey serves as a foundational roadmap for researchers and practitioners, charting the course for a new generation of LLM-powered, intelligent, adaptive and synergistic compilation tools.
}

\keywords{ survey, LLM, compiler, code translation, code optimization}

\maketitle

\section{Introduction}\label{sec:intro}

For decades, the compiler has stood as a cornerstone of the computing stack, undertaking the critical and complex task of translating human-readable source code into efficient machine-executable programs. A primary challenge in this process is optimization, a domain traditionally governed by intricate, handcrafted heuristics designed by human experts to navigate a vast and complex decision space. The advent of machine learning~\citep{alexnet2012,resnet2016} introduced a new paradigm, employing data-driven models for tasks like phase-ordering and flag selection~\citep{mlincompilersurvey2018}. 
However, these traditional machine learning approaches often rely on an intensive process of feature engineering, where experts must meticulously design and extract program features to train a model, leaving the core components of compiler design largely unchanged.

The recent emergence of Large Language Models (LLMs) represents a fundamental shift in this landscape. Pre-trained on vast corpora of text and code, LLMs have demonstrated a remarkable capacity to understand, generate, and transform programming languages as raw text, largely eliminating the need for explicit feature engineering. By training on codebases orders of magnitude larger than any human could study, LLMs internalize a deep understanding of programming patterns, syntax, and semantics across numerous languages. These capabilities have been rapidly integrated into the software development workflow through LLM-driven chatbots~\citep{gpt4,gemini2023,claude3}, code assistants~\citep{copilot, tabnine}, and semi-automated agents~\citep{cursor_ide,gemini_cli,claude_code}, boosting developer efficacy at every stage.
This wave of innovation has, in turn, catalyzed new research within the broader compiler domain. The scope of tasks has expanded from narrow problems like pass selection to ambitious, end-to-end objectives like code transpilation and automated program repair. Consequently, the compiler's role is being reimagined from a static tool to a dynamic, interactive partner in software development.

This rapid evolution has led to a surge of new research across a wide spectrum of compiler-related tasks. Researchers are now applying LLMs to ambitious goals, including source-to-source Code Transpilation to migrate code across different languages~\citep{177_transcoder, 38_jana2024cotran, 60_zhang2025_pldi, 91_alphatrans2025_fse} and architectures~\citep{153_babeltower2022_icml, 174_coderosseta2024_neurips, 74_palkowski2024automatic, 10_qimengxpiler2025_osdi}, high-level Code Optimization~\citep{17_gao2025search, 52_peng2025perfcodegen,58_purschke2025speedgen,63_xu2025efficient, 12_lin2025eco} to surpass traditional heuristics, low-level IR Optimization~\citep{11_deng2024compilerdream_arxiv,metallmcompiler-cc25,51_decos2025_ics} to fine-tune performance and code size, and even the notoriously difficult problem of using LLMs to act as compilers~\citep{71_gao-etal-2024-virtual, 70_c-x86-emnlp24, 44_vega2025_cgo} or decompilers~\citep{130_xu2023lmpa,72_armengol2024slade,73_tan-etal-2024-llm4decompile,75_hu2024degpt,173_wong2025decllm}. While promising, the sheer volume and diversity of these studies can hinder a clear understanding of the current landscape. 
To address this gap, this paper provides a systematic review of recent advancements in LLM-enabled compiler research. We offer a comprehensive multi-dimensional taxonomy of the state-of-the-art and a clear analysis of its core advancements and challenges, establishing a roadmap for this exciting direction in compiler research.

This paper makes the following primary contributions:

\begin{itemize}
\item We conduct a systematic literature review, identifying and curating a corpus of 159 primary studies that represent the state-of-the-art in LLM-enabled compilation.
\item We propose a novel, multi-dimensional taxonomy to structure the field. This taxonomy classifies existing work based on the LLM's \textbf{Design Philosophy} (Selector, Translator, Generator), the \textbf{LLM Methodology}, its operational \textbf{Level of Code Abstraction}, and the specific \textbf{Task Type} it addresses.

\item We provide a comprehensive analysis of the primary advancements offered by LLM-based approaches, highlighting their role in democratizing compiler development, discovering novel optimizations, and broadening the scope of compilation.

\item We synthesize the common challenges facing the field—including correctness, scalability, and interpretability—and discuss promising future research directions, such as the development of hybrid systems and self-improving compilers.
\end{itemize}

The remainder of this paper is organized as follows.~\autoref{sec:method} details the research questions we target to answer and our systematic methodology for literature selection.~\autoref{dimension1} and~\autoref{dimension2} present our multi-dimensional taxonomy in detail, categorizing the existing work.~\autoref{benchmark} systematically presents representative datasets and benchmarks in this field, and associated state-of-the-art advancements.
~\autoref{sec:discussion} provides an in-depth discussion of the field's advancements, challenges, and future opportunities. Finally,~\autoref{sec:conclusion} concludes the paper.

\section{Methodology}\label{sec:method}

This section details the systematic methodology we employed to survey the landscape of LLM-enabled compiler researches. A rigorous and transparent protocol is essential for ensuring that our review is both comprehensive and reproducible. To that end, we first present the research questions that steered our investigation in~\autoref{sec:method:RQ}. Following that, we describe the structured, three-phase literature search and selection protocol used to assemble the final corpus of primary studies for our analysis in~\autoref{sec:method:protocol}.

In this section, we introduce the detailed steps of conducting a literature review. To ensure a comprehensive and unbiased review of the field, we adopted a systematic literature review (SLR) protocol following standard practice~\citep{literaturereview2007}. This protocol defines the research questions that guide our survey, the search process for identifying relevant studies, and the criteria for including or excluding papers.

\subsection{Research Questions (RQs)}\label{sec:method:RQ}

To guide our literature review and provide a clear structure for our analysis, we formulated the following four key research questions regarding the application of LLMs in the compiler domain:
\begin{itemize}
    \item \textbf{RQ1: How are LLMs being integrated into the compilation process?} 
    We explore this question from two complementary perspectives: first, the \textbf{Design Philosophy}, which defines the LLM's architectural role within the system (task side); and second, the \textbf{LLM Methodology}, which details how the model is technically developed and applied to that task (LLM side).
    \item \textbf{RQ2: What are the primary compiler-related tasks addressed by LLMs?} This question aims to identify and categorize the specific compiler-related tasks, such as code optimization, transpilation, and code generation, that are being targeted by recent research.
    \item \textbf{RQ3: What are the primary advancements offered by LLM-based approaches?} This question focuses on summarizing the novel contributions and breakthroughs that these techniques bring to both the compiler and machine learning communities.
    \item \textbf{RQ4: What are the common challenges and future opportunities in this emerging field?} This question seeks to synthesize the key obstacles reported in the literature and identify promising directions for future work.

\end{itemize}

\subsection{Literature Search and Selection Protocol}\label{sec:method:protocol}

\begin{figure*}
    \centering
    \includegraphics[width=0.8\textwidth]{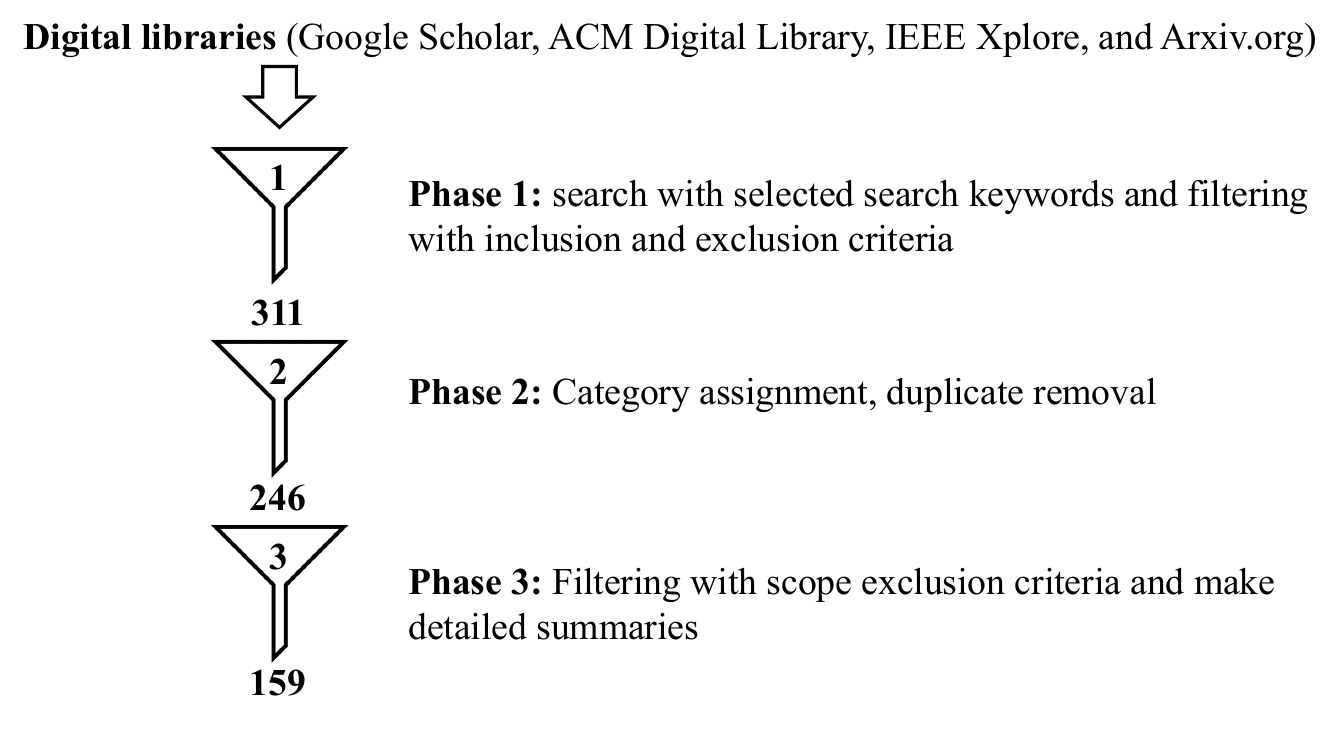}
    \caption{Literature Search and Selection Protocol Overview}
\end{figure*}

To systematically answer the research questions defined above, we designed and executed a comprehensive literature search and selection protocol. This process was organized into three distinct phases, which progressively filtered a large pool of initial candidates down to a focused and relevant corpus of 159 papers.
\begin{enumerate}
    \item \textbf{Phase 1: Initial Search and Candidate Collection.} The first phase involved an extensive search for candidate papers across four major digital libraries: arXiv, Google Scholar, the ACM Digital Library, and IEEE Xplore. We utilized a broad set of search queries, including keywords such as ``LLM compiler'', ``large language model for code optimization'', ``AI in compilers'', ``transformer for compilation'', and ``LLM for code generation''. To complement this automated search, we also manually included several seminal papers identified by domain experts. We then applied a snowballing technique, examining the forward citation chain of these core papers to ensure comprehensive coverage and include other closely related studies who cited them.
    \item \textbf{Phase 2: Deduplication and Initial Screening.} In the second phase, the initial pool of papers was processed to remove duplicates. Subsequently, we performed a preliminary screening of the remaining articles. This step involved reviewing titles and abstracts to filter out studies that were clearly outside the scope of our survey, such as those not related to compilers or not utilizing LLMs.
    \item \textbf{Phase 3: Full-Text Review and Final Selection.} The final phase consisted of a full-text review of each paper that passed the initial screening. This in-depth analysis allowed us to determine a study's final eligibility based on a strict set of inclusion and exclusion criteria. 

    \textbf{Inclusion Criteria}:\begin{itemize}
        \item The study must primarily focus on using a Large Language Model (specifically a Transformer-based model) for a task related to compilation or code optimization.
        \item The study must be a peer-reviewed conference paper, journal article, or a significant, relavant pre-print manuscript (to include the latest advancements often published on arXiv).
        \item The full text of the study must be publicly available through digital libraries or websites for open access.
    \end{itemize}
    \textbf{Exclusion Criteria}:\begin{itemize}
        \item Studies that use traditional machine learning (e.g., SVM, Decision Trees) without language models (LMs).
        \item Articles that are not technical research papers, such as editorials, keynotes, tutorials, posters or extended abstracts with insufficient detail.
        \item Studies where the core application of the model was not related to the interest of broader compilation domain.
    \end{itemize}
\end{enumerate}

Through this rigorous three-phase process, we curated a final collection of 159 primary studies. This curated corpus forms the foundation for the systematic review, categorization, and analysis presented in the subsequent sections of this paper.

\subsection{Scope of the Taxonomy}

Before presenting our detailed taxonomy, it is essential to define the scope of this survey. The term ``LLM-enabled compiler'' encompasses a wide range of emerging research. For the purpose of this review, we define our scope to include any study that utilizes a large, pre-trained, Transformer-based language model to perform or augment a task traditionally associated with the compilation workflow or the broader code development and optimization lifecycle.

This definition deliberately draws a distinction between the new paradigm of LLM-based approaches and prior work. Specifically, \textbf{our survey excludes}:
\begin{itemize}
    \item \textbf{Traditional ML Techniques}: Studies that rely on classic machine learning models (e.g., Support Vector Machines, Decision Trees, Random Forests) that require extensive, handcrafted feature engineering from source code (such as code quality metrics) and have been well-studied and surveyed. 
    \item \textbf{Purely NLP-based SE Tasks}: Studies that apply language models to software engineering artifacts without directly analyzing, transforming, or generating code are not included. For example, the classification of bug reports or the summarization of developer comments, while related, fall outside our compiler-centric focus.
\end{itemize}

By establishing this scope, we aim to provide a focused and coherent review of the state-of-the-art in the specific, transformative paradigm of applying large language models to the science and engineering of compilation. 
The core function of any compiler is fundamentally rooted in two processes: \texttt{translation} and \texttt{optimization}. Consequently, these two areas form the primary focus of our survey. We include a significant body of work on \textbf{Code Optimization}, where LLMs are used to rewrite programs to improve performance or reduce code size, operating at either high-level source code and low-level Intermediate Representation (IR). 

Equally important is \textbf{Code Transpilation}, which we define broadly as the translation between different program representations. This includes narrowly-defined compilation (e.g., programming language to assembly), decompilation, source-to-source transpilation, and binary translation. This category also covers unique tasks like translating programming languages into hardware description languages (HDLs) or even into neural network weights.

It is important to note that these task categories are not always mutually exclusive; in fact, they often overlap. A prominent example is the translation of C code to CUDA, which can be viewed simultaneously as a Code Transpilation task for language migration and a Code Optimization task aimed at unlocking parallel performance on specific hardware.

Beyond these core functions, a significant portion of our survey is dedicated to tasks related to correctness and verification. This focus is directly motivated by a fundamental characteristic of LLMs: they are powerful, but not perfectly reliable, generative models. Unlike traditional deterministic compilers, any workflow that uses an LLM as a direct operator to modify code necessitates a corresponding validation process to ensure the correctness of its output. 

Consequently, we include a comprehensive review of \textbf{Automated Program Repair} and \textbf{Bug Fixes}, where LLMs are employed to correct defects. Furthermore, we cover \textbf{Program Synthesis and Code Generation} with a special emphasis on tasks that bolster this verification ecosystem related to compilers, such as generating effective test cases for compiler fuzzing or aiding in the development of the compiler's own source code.

The 159 primary studies within our scope are highly diverse. To bring structure to this landscape, we propose a multi-dimensional taxonomy that classifies research along four key axes:

\begin{itemize}
\item \textbf{Design Philosophy}, which describes how the LLM is architecturally integrated into the compilation workflow.
\item \textbf{LLM Methodology}, which details how the model is technically developed and applied to that task.
\item \textbf{Level of Code Abstraction}, which characterizes the representational level of the code being processed (e.g., source code, IR).
\item \textbf{Task Type}, which defines the specific compiler-related goal being accomplished (e.g., optimization, transpilation).
\end{itemize}

To provide a clear analysis, we group these four dimensions based on the research questions they answer.

\autoref{dimension1} will analyze the first two dimensions, Design Philosophy (the ``task side'') and LLM Methodology (the ``LLM side''), to comprehensively answer \textbf{RQ1: How are LLMs integrated?}.
Subsequently, \autoref{dimension2} will analyze the final two dimensions, Level of Code Abstraction and Task Type. 
We observe that these two are highly correlated (e.g., source-to-source transpilation operates at a different abstraction level than IR optimization). We therefore analyze them together to provide a cohesive overview and answer \textbf{RQ2: What tasks are addressed?}.

\section{Dimension 1\&2: Design Philosophy \& LLM Methodology}\label{dimension1}

\begin{figure*}
    \centering
    \includegraphics[width=0.8\textwidth]{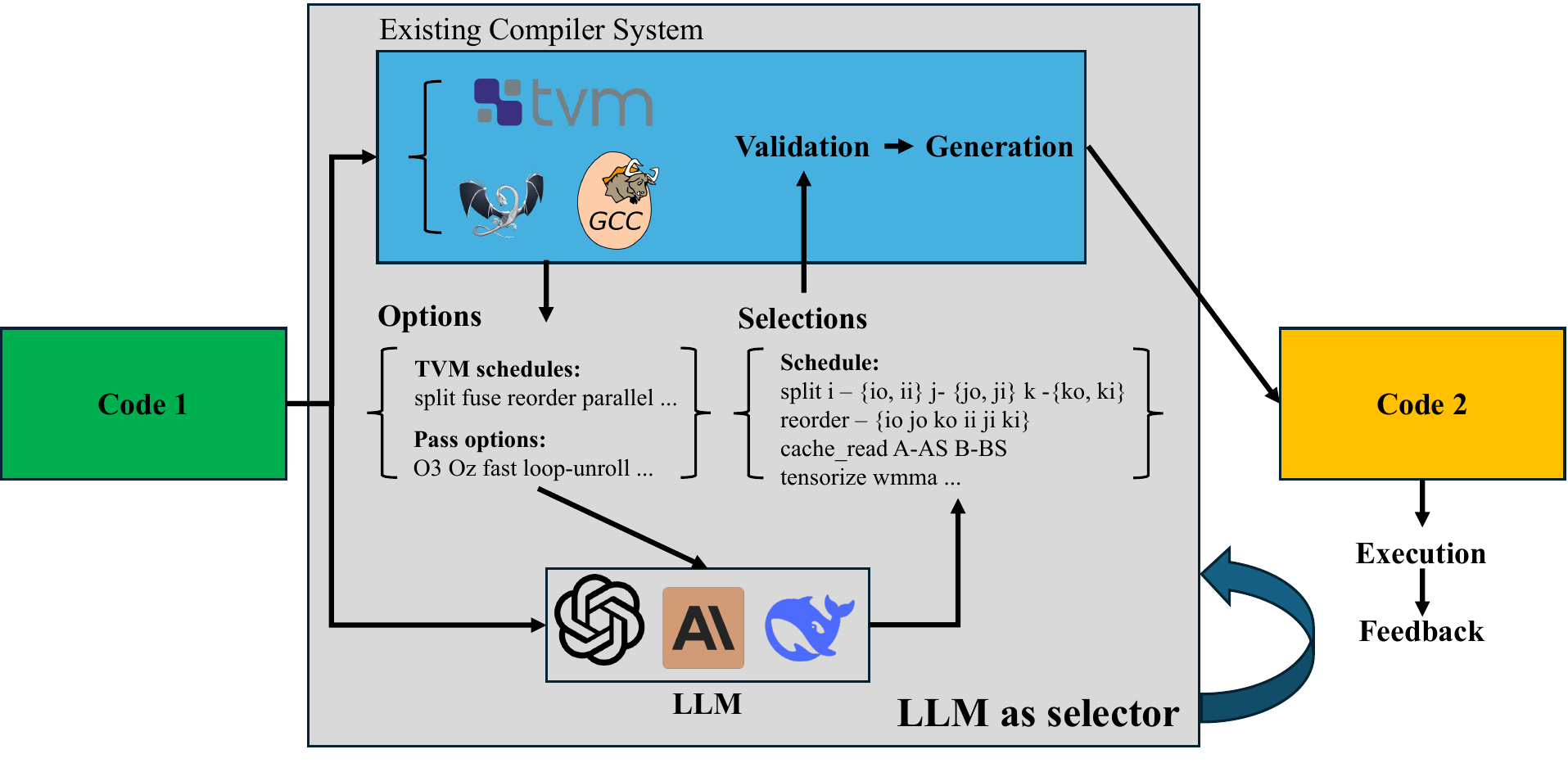}
    \caption{Overview of LLM-as-Selector methodology}
    \label{fig:llm_as_selector}
\end{figure*}

This section addresses our first research question—\textit{How are LLMs being integrated into the compilation process?}—by analyzing two complementary dimensions: \textbf{Design Philosophy} and \textbf{LLM Methodology}. The former describes the LLM's architectural role, while the latter details the technical methods used to apply the model to that role.

The first dimension, \textbf{Design Philosophy}, classifies approaches based on the conceptual role the LLM plays within the broader compilation system. This choice is a critical architectural decision, as it fundamentally determines how the LLM's capabilities are leveraged and constrained. It significantly influences the system's degree of autonomy, its trustworthiness, and the complexity of verifying its outputs. As we will detail, we identify three dominant philosophies: LLM as selector (\autoref{llm_as_selector}), LLM as translator (\autoref{llm_as_translator}), and LLM as generator (\autoref{llm_as_generator}).

Complementing this ``task-side'' view, the second dimension, \textbf{LLM Methodology}, examines the ``LLM-side'' of the integration. This dimension addresses the technical methods used to make an LLM capable of its designated task. These methods exist on a spectrum, from ``training-free'' approaches (e.g., in-context learning, prompt engineering, Retrieval Augmented Generation) to ``training-required'' adaptations (e.g., fine-tuning, domain-specific pre-training, reinforcement learning). We will analyze these methodologies and their trade-offs in detail in \autoref{llm_methodology}.

\subsection{LLM as selector}\label{llm_as_selector}

In this design, the LLM functions as a sophisticated policy engine or a ``hyper-optimizer''. As depicted in~\autoref{fig:llm_as_selector}, its primary role is not to generate new code, but to select the best course of action from a predefined and finite set of options or operations provided by the compilation system. The LLM is prompted with the source code context and a set of valid choices, and it uses its deep contextual understanding to make an informed decision. The compilation system itself then performs the corresponding transformation based on the LLM's selection.

This approach is the most direct evolution from traditional ML-in-compiler techniques, with the key difference being the replacement of models trained on handcrafted features with a powerful LLM that can reason directly over raw source code (can be finetuned as well). Common applications include selecting optimal compiler flags or determining the most effective sequence of optimization passes for a given program. The main advantage of this model is its inherent safety and control; since the LLM only chooses from a list of valid, human-defined actions, the resulting transformation is guaranteed to be valid if a valid selection is generated by LLM. However, its creative potential is limited by this predefined search space, and it cannot discover entirely novel optimizations that are not already encoded in the available options.

The works in this category can be broadly grouped based on their primary goal and methodology:

\textbf{Replacing or Augmenting Compiler Heuristics:} This is the most common application, where the goal is to find better compiler flag or pass sequences than the default heuristics (e.g., -O2, -O3). This area has a long history, with earlier works using techniques like NeuroEvolution~\citep{127_neuroevolutionary2023_gecco} and~\citet{40_mammadli2020static_llvmhpc} to tune compiler heuristics,~\citet{162_tavarageri2021poly_arxiv} using DNNs to optimize polyhedral loop generation strategy or graph-based algorithms~\citep{125_graphbased2024_hpec} to optimize GCC flag settings. Modern approaches now use LLMs to apply this same principle to more complex, domain-specific areas like selecting optimization strategies for efficient model serving~\citep{4_tang2025compiler_arxiv}, optimizing quantum compilation~\citep{89_quantum2024_dac}, or guiding the generation of high-performance tensor programs~\citep{167_zhai2024_osdi}.

\textbf{Improving the Efficiency of the Search Process:} Rather than just selecting the final configuration, several works use the LLM to make the search for good configurations more intelligent and efficient. For example, CompilerDream~\citep{11_deng2024compilerdream_arxiv} uses an LLM to help build a ``world model'' of the optimization space, which then guides a more effective search. Others use an LLM for priority sampling~\citep{27_grubisic2024priority_arxiv} to decide which compiler options are most promising to test, or to enhance black-box fuzzing~\citep{107_wang2024enhancing_issre} by intelligently mutating flags based on feedback from previous compilation attempts.

\textbf{Agentic and Reinforcement Learning (RL) Frameworks:} The most advanced selector systems employ LLMs as the core of an autonomous agent that can interact with the compilation environment. These agents can perform a series of selections to achieve a goal. Several works use RL for compiler auto-tuning, such as Compiler-R1~\citep{45_compilerr1-2025_arxiv}, an agentic framework for exploring the option space, and DeCOS~\citep{51_decos2025_ics}, which uses an LLM to ``ignite'' the RL process for more data-efficient learning. Taking this a step further, CompileAgent~\citep{8_hu2025compileagent_acl} demonstrates a high-level agent that selects and orchestrates a series of command-line tools (git, make, etc.) to automate the complex task of compiling entire real-world software repositories.

\subsection{LLM as translator}\label{llm_as_translator}

\begin{figure*}
    \centering
    \includegraphics[width=0.8\textwidth]{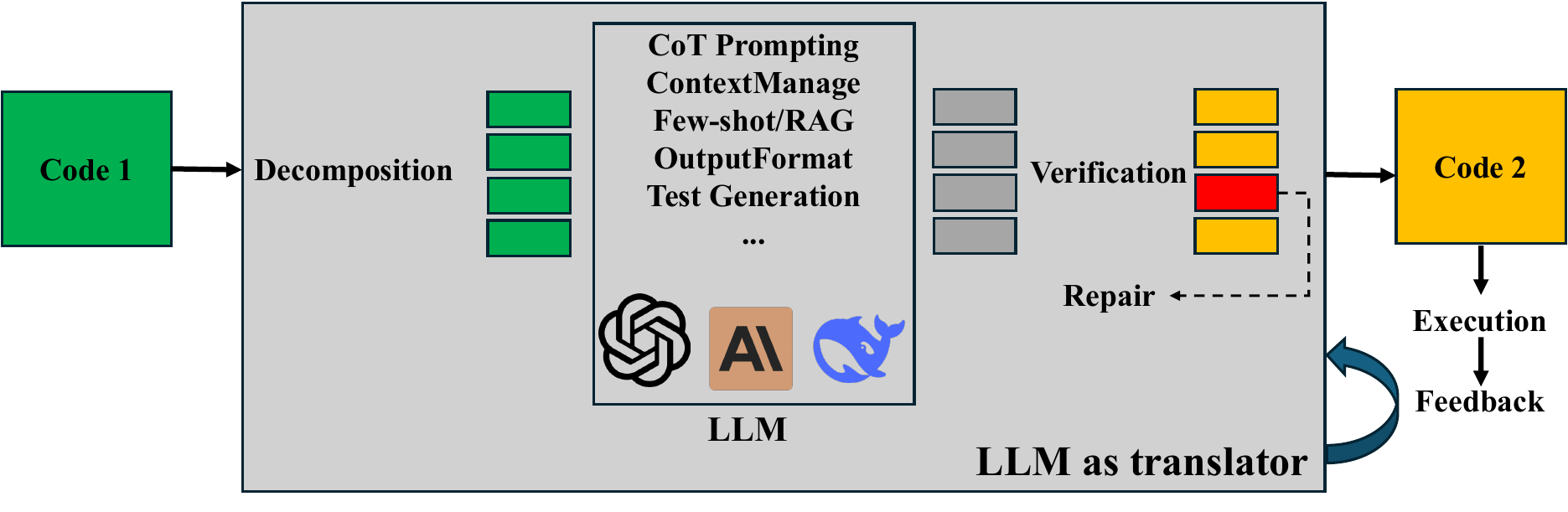}
    \caption{Overview of LLM-as-Translator methodology}
    \label{fig:llm_as_translator}
\end{figure*}

Using LLM as translator is the most direct and ambitious application, where the LLM itself acts as a partial or complete translator. It directly performs one or more  transformations, such as translation, optimization, or refactoring, on a given scope of a program or code snippet. This philosophy treats code transformation as a generative, sequence-to-sequence task, leveraging the full power of the LLM to rewrite the input program into a new version.

~\autoref{fig:llm_as_translator} describes an ideal LLM translator system. First of all, the translator must scale to enough code input to be used in real scenarios. Thanks to programming language's composibility nature, a large project-level can be decomposed into multiple function-level code snippets, or even some finer-granularity like basicblocks or statements, with proper context management to make sure the decomposed translation results can be combined again. Later, each code fragment is prompted to LLM to perform code translation, during this stage, multiple techniques could be applied to improve translation quality, for example, Chain-of-Thought prompting~\citep{cot_2022_neurips}, which decomposes a large translation task into several sequential subtasks to reduce translation complexity. Few-shot examples or retrieval augmented generation (RAG)~\citep{rag_2020_neurips} can improve LLM's hallucination by providing necessary information needed for the translation, while output format of LLM~\citep{95_macedo2024exploring} can also impact LLM's generation quality. During the translation, LLM can also generates other useful code, such as test code (if no golden test is provided), and assertions/preconditions needed in verification.

After LLM translation, a post verification is needed to verify some quality-important attributes, for example, check if the code can compile is the most trivial attribute, as for more specific attributes, such as memory safety check, overflow check and for-loop index check, requiring sophisticated verification methods like SMT solver~\citep{z3smt_2008} or transform validator like alive2~\citep{alive2_2021_pldi}. Translation fails to pass verification will need to be repaired through an automated program repair process using compiler,runtime or behavioral feedbacks.

This is the dominant approach for a variety of source-to-source tasks, which can be categorized as follows:

\begin{itemize}
\item \textbf{Language Transpilation:} This involves translating code from one high-level language to another. Examples in the literature are diverse, including:
\begin{itemize}
\item Translating from a high-resource language like Java to a low-resource one such as OCaml~\citep{106_knowledgetransfer2024_oopsla}.
\item Converting sequential C into parallel CUDA to leverage GPU architectures~\citep{153_babeltower2022_icml, 174_coderosseta2024_neurips}.
\item Migrating between different parallel paradigms, such as from SIMT-parallel CUDA to SIMD-parallel BANG C~\citep{10_qimengxpiler2025_osdi}.
\end{itemize}
\item \textbf{Code Optimization:} This focuses on rewriting specific code segments to improve performance or other non-functional properties. A common example is automatically rewriting \texttt{for}-loop pragmas to achieve better performance on a given target hardware~\citep{47_llmvectorizer2025_cgo}.

\item \textbf{Automated Program Repair:} In this context, bug fixing is treated as a translation problem. The goal is to leverage the LLM to translate a flawed program into a semantically correct version~\citep{131_wei2023_fse,133_xia2023_icse}.
\end{itemize}

The key advantage of this methodology is its immense potential to learn and generate complex, non-trivial transformations that may surpass human-designed rule-based translators. However, the primary disadvantage is the significant challenge of ensuring correctness. The probabilistic nature of LLMs means they can ``hallucinate'' and produce code that is syntactically correct but semantically flawed, making rigorous verification a critical and difficult component of this approach.

Notably, we only list some remarkable studies in this category and will detail the rest in \autoref{dimension2}, as the level of code abstraction can help categorize the major studies of this survey better.

\subsection{LLM as generator}\label{llm_as_generator}

\begin{figure*}
    \centering
    \includegraphics[width=0.8\textwidth]{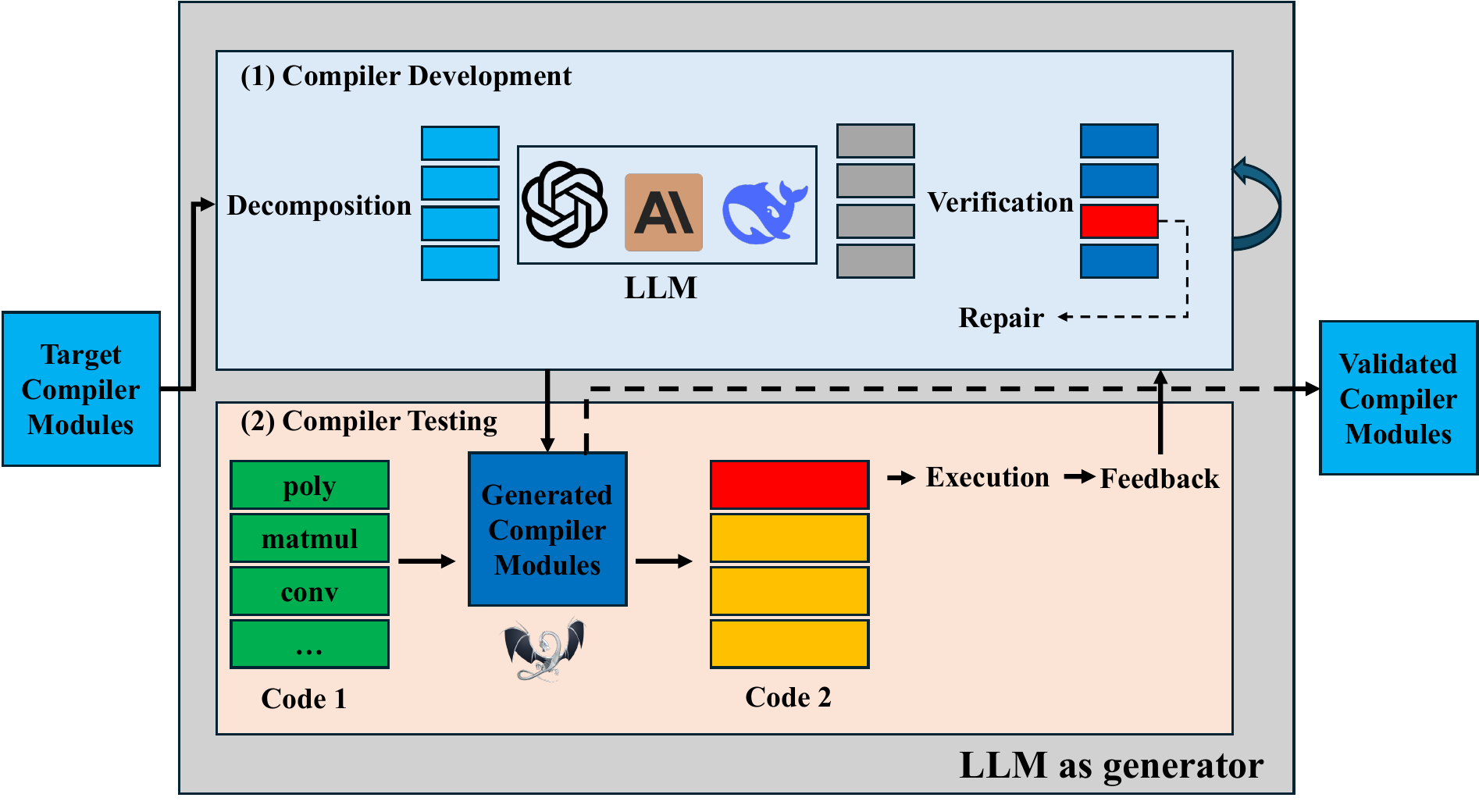}
    \caption{Overview of LLM-as-Generator methodology}
    \label{fig:llm_as_generator}
\end{figure*}

This is a more indirect, meta-level approach where the LLM's role is to generate the source code of a program that, in turn, performs the desired code transformation. As depicted in~\autoref{fig:llm_as_generator}, the workflow is typically a two-step process: first, the LLM writes a script or program that implements the transformation logic, which can be seen as a compiler development process if applied to a compiler system development; second, this generated program is compiled and executed to apply the changes to the target code, which can also be seen as a compiler testing process.

Currently, this design philosophy is most popularly realized in the form of AI-powered code assistants and agents, which generate functional code snippets, scripts, or entire applications based on natural language prompts. Prominent examples include tools like GitHub Copilot~\citep{copilot}, Cursor~\citep{cursor_ide}, and command-line interfaces applications like gemini-cli~\citep{gemini_cli} and claude-code~\citep{claude_code}.

The same principle can be extended to the compiler domain, where an LLM could theoretically generate new optimization passes or other compiler components. However, due to the immense complexity and the need for deep, specialized knowledge of compiler internals, data structures, and APIs, applying this approach effectively is a significant challenge. As a result, well-designed studies that successfully employ the LLM-as-generator paradigm for core compiler tasks are still scarce. Nonetheless, some pioneering works have begun to explore this frontier. For example, VEGA~\citep{44_vega2025_cgo} demonstrates a method for automatically generating compiler backends by using a pre-trained transformer model, while ComBack~\citep{88_comback2024_neurips} provides a versatile dataset specifically designed to facilitate research on generating and enhancing compiler backend code. On the source code side, CodeTransform~\citep{178_cummins2024donttransformcodecode} also preliminarily explores the code rewriting potential, in a word: \textit{Don't Transform the Code, Code the Transforms}.

This hybrid approach combines the pattern-recognition strength of LLMs with the rigor of a traditional, deterministic program. The main advantage is that the generated artifact—the transformation script—can be inspected, verified, and reused, offering a higher degree of trust than a direct generative model. The primary challenges are the increased complexity of the workflow and the requirement for the LLM to possess a sophisticated understanding of the specific compiler APIs and frameworks it needs to use in the code it generates.

In summary, the design philosophy dictates the fundamental architecture of an LLM-enabled compiler. The three paradigms discussed—\textbf{Selector}, \textbf{Translator}, and \textbf{Generator}—represent a spectrum of integration strategies, each presenting a distinct trade-off between control and creative potential. The Selector model offers the highest degree of safety by operating within a predefined action space, while the Translator model unleashes the full generative power of LLMs to discover novel transformations, albeit with significant verification challenges. The Generator provides a hybrid approach, balancing generative flexibility with the determinism of traditional code system.

\subsection{LLM Methodology}
\label{llm_methodology}

\begin{table*}[t]
    \centering
    \begin{tabularx}{\textwidth}{|l|X|X|}
    \hline
    
    \textbf{Aspect} & \textbf{Training-Required} & \textbf{Training-Free} \\
    \hline
    
    \textbf{Core Principle} & Adapt Parameters & Guide Inference \\
    \hline
    
    \textbf{Primary Cost} & Data Curation \newline Training Computation & Prompting Effort \newline Inference w/ Extra Context \\
    \hline
    
    \textbf{Performance} & High (Specialized) & Varies (Generalized) \\
    \hline
    
    \textbf{Flexibility} & Low (Task-Locked) & High (Adaptable) \\
    \hline
    
    \textbf{Knowledge Source} & Internal (Weights) & External (Context) \\
    \hline
    
    \textbf{Typical Examples} & 
        \begin{itemize}[leftmargin=*, nosep]
             \item Domain Pre-training \newline LLMCompiler~\citep{metallmcompiler-cc25}
             \item Supervised Fine-Tuning \newline VirtualCompiler~\citep{71_gao-etal-2024-virtual}
             \item Reinforcement Learning \newline CUDA-L1~\citep{cudal1_2025_arxiv}
        \end{itemize} & 
        \begin{itemize}[leftmargin=*, nosep]
             \item Prompt Engineering \newline LEGOCompiler~\citep{5_legocompiler-arxiv25}
             \item Retrieval Augmented Generation \newline CoCoGen~\citep{110_bi-etal-2024-iterative}
             \item Agentic Workflows \newline Compiler-R1~\citep{45_compilerr1-2025_arxiv}
        \end{itemize} \\
    \hline
    \end{tabularx}
    \caption{Qualitative Comparison of LLM Methodologies}
    \label{tab:methodology_comparison}
\end{table*}

Complementing the architectural \textit{Design Philosophy}, \textit{LLM Methodology} details the technical methods used to make a model capable of its designated task. This dimension provides a crucial comparative analysis of \textit{how} LLMs are technically employed, distinguishing the ``LLM-side'' development from the ``task-side'' role. These methods can be broadly categorized into two main methodologies: \textbf{Training-Required} approach that adapt the model's weights, and \textbf{Training-Free} approach that guide a pre-trained model's inference process. The choice between them involves a fundamental trade-off between the cost of data curation and training versus the complexity of prompt engineering and inference-time systems.

\subsubsection{Training-Required: Adapting Model Weights}
These methods modify the LLM's parameters to instill domain-specific knowledge and specialize its behavior. This is often necessary when the task is highly specialized or when the target representation (like compiler IR) is scarce in the model's original pre-training data.
\begin{itemize}
    \item \textbf{Domain-Specific Pre-training:} This is the most resource-intensive approach, where a model is trained from scratch or undergoes continued pre-training on a massive, domain-specific corpus. This is essential for tasks operating on representations unseen by general models. For example, \textbf{Meta LLMCompiler}~\citep{metallmcompiler-cc25} created foundation models pre-trained specifically on LLVM IR, while \citet{167_zhai2024_osdi} trained their \textbf{TLM} model on a large corpus of TVM's search space to act as a selector. The \textbf{TransCoder} series~\citep{177_transcoder, 176_dobf, 175_transcoderst,  154_transcoderir2023_iclr} also rely on this to learn cross-lingual representations before tackling translation.

    \item \textbf{Supervised Fine-Tuning (SFT):} 
    This is the most common adaptation methodology. It takes a general-purpose, pre-trained foundation model (e.g., CodeLlama, GPT) and fine-tunes it on a curated dataset of ``input-output'' examples for a specific task. This adaptation can be performed using full finetuning or, more commonly, through parameter-efficient finetuning (PEFT) techniques like LoRA. This approach has proven highly effective across various tasks, such as fine-tuning models for C-to-x86 neural compilation~\citep{70_c-x86-emnlp24}, Verilog generation~\citep{104_verigen2024}, and source-level optimization~\citep{63_xu2025efficient}. \textbf{VirtualCompiler}~\citep{71_gao-etal-2024-virtual} also leverages SFT to translate source code to assembly by matching semantics. The \textbf{CUDA-L1} study~\citep{cudal1_2025_arxiv} also begins with SFT to teach the model the basics of CUDA optimization.
    
    \item \textbf{Reinforcement Learning (RL) \& Feedback-Based Tuning:} When a task has a clear, non-differentiable metric for success (e.g., performance speedup, code size reduction, or passing a test suite), RL or other feedback mechanisms can be used to optimize the model directly for that objective. This is often applied after SFT. For instance, \textbf{PerfRL}~\citep{134_duan2025perfrlsmalllanguagemodel} uses RL with metric feedback for code optimization. Similarly, \textbf{VerilogLLM}~\citep{64_wang2025insights} uses feedback from testbenches, \textbf{CoTran}~\citep{38_jana2024cotran} uses compiler and symbolic execution feedback, and \textbf{CUDA-L1}~\citep{cudal1_2025_arxiv} employs GRPO~\citep{grpo_2024_arxiv} to surpass its initial SFT performance. \textbf{DeCOS}~\citep{51_decos2025_ics} also uses an LLM to ``ignite'' an RL process for data-efficient learning.
\end{itemize}

\subsubsection{Training-Free: Guiding Model Inference}
These methods leverage a powerful, general-purpose foundation model's existing capabilities without altering its weights. The focus shifts from data curation and training to designing sophisticated inference-time systems that provide the model with the necessary context and guidance.
\begin{itemize}
    \item \textbf{Prompt Engineering:} 
    This involves crafting a detailed prompt that instructs the model (zero-shot) or provides it with a few in-context examples (few-shot learning) of the task. For more complex reasoning tasks, this is often extended to multi-step prompting strategies like Chain-of-Thought (CoT). For example, \textbf{LEGOCompiler}~\citep{5_legocompiler-arxiv25} employs both few-shot learning and CoT-prompting to perform neural compilation, \textbf{CodeOptCoT}~\citep{116_xu2024code} and \textbf{SBLLM}~\citep{17_gao2025search} both explicitly leverage CoT to improve code optimization by forcing the model to ``think step-by-step.'' Similarly, \textbf{RTLLM}~\citep{114_lu2024rtllm} uses a ``self-planning'' prompting strategy, and \textbf{BuiltRome}~\citep{120_nakkab2024rome} also finds that hierarchical prompt structuring is critical for LLM-based hardware design.

    \item \textbf{Retrieval-Augmented Generation (RAG):} 
    To combat hallucinations and provide the model with specialized knowledge it may not have been trained on (e.g., obscure APIs, project-specific context, or optimization rules), a RAG system coule be used. This system first retrieves relevant documents from an external knowledge base and adds them to the model's context. This is used by \textbf{Autoiot}~\citep{59_autoiot2025_mobicom} for background knowledge, \textbf{CoCoGen}~\citep{110_bi-etal-2024-iterative} for project-level context retrieval, and \textbf{SBLLM}~\citep{17_gao2025search} for retrieving optimization knowledge.

    \item \textbf{Agentic \& Iterative Workflows:} 
    This advanced methodology treats the LLM as a reasoning engine within a larger loop. The ``agent'' can plan, use external tools (like compilers, verifiers, or SMT solvers), and iteratively refine its output based on feedback from these tools. This approach is central to systems like \textbf{Compiler-R1}~\citep{45_compilerr1-2025_arxiv} and \textbf{CompileAgent}~\citep{8_hu2025compileagent_acl}. It is also the core principle behind iterative repair loops, such as those in \textbf{DecLLM}~\citep{173_wong2025decllm} and \textbf{ProblemOriented}~\citep{123_ye2024problem}, and multi-agent frameworks like \textbf{WhiteFox}~\citep{31_yang2024whitefox}. \textbf{Qimeng-Xpiler}~\citep{10_qimengxpiler2025_osdi} also uses an MCTS-based search, which can be seen as a form of guided, iterative generation.
\end{itemize}

\subsubsection{Methodology Comparison}

The choice between Training-Required and Training-Free methodologies presents a fundamental design trade-off, which we summarize qualitatively in \autoref{tab:methodology_comparison}.

Training-Required methods focus on \textbf{adapting model parameters (weights)}, which incurs significant upfront \textbf{cost} in data curation and training computation. However, this internalizes domain knowledge, leading to high performance on specialized or narrow tasks (e.g., Meta LLMCompiler~\citep{metallmcompiler-cc25} is specialized on LLVM IR code size optimization). The trade-off is low flexibility, as the resulting model is ``locked-in'' to its specific trained task.

Conversely, Training-Free methods \textbf{guide a fixed model's inference} process. This shifts the \textbf{cost} to prompt engineering and inference-time resources. These approaches offer high flexibility, as prompts and tools can be quickly adapted. Their performance is more generalized and highly dependent on the base model's capabilities, with knowledge externalized and injected at runtime via context (e.g., \citet{116_xu2024code}, \citet{17_gao2025search}, \citet{8_hu2025compileagent_acl}).

In practice, these methodologies are not mutually exclusive. A common pattern is to use a \textit{Training-Required} method (like SFT) to create a specialized model, which is then deployed within a \textit{Training-Free} methodology (like an agentic workflow with RAG). This combination leverages the model's specialized knowledge while simultaneously grounding its output with real-time context and verification.

Finally, combining the analyzed two dimensions, we can now provide a comprehensive answer to our first research question (\textbf{RQ1}), as summarized below.

\begin{tcolorbox}[colback=white,colframe=black]
    \textbf{RQ1}: How are LLMs being integrated into the compilation process?

    \textbf{Answer}:
    
    On the task side, LLMs have been integrated into compilation systems through three primary design philosophies that define their role: (1) as a \textbf{Selector} to choose from predefined compiler actions (\autoref{llm_as_selector}), (2) as a direct \textbf{Translator} to rewrite code (\autoref{llm_as_translator}), or (3) as a \textbf{Generator} to create new compiler tools and components (\autoref{llm_as_generator}).
    
    On the LLM side, these roles are realized through two primary \textbf{LLM Methodologies}: (1) \textbf{Training-Required} methods (e.g., fine-tuning, RL) that adapt model weights for specialized tasks, and (2) \textbf{Training-Free} methods (e.g., prompt engineering, RAG, agentic workflows) that guide a general model's inference.

\end{tcolorbox}

\section{Dimension 3\&4: Level of Code Abstraction \& Task Type}\label{dimension2}

\begin{figure*}
    \centering
    \includegraphics[width=0.9\textwidth]{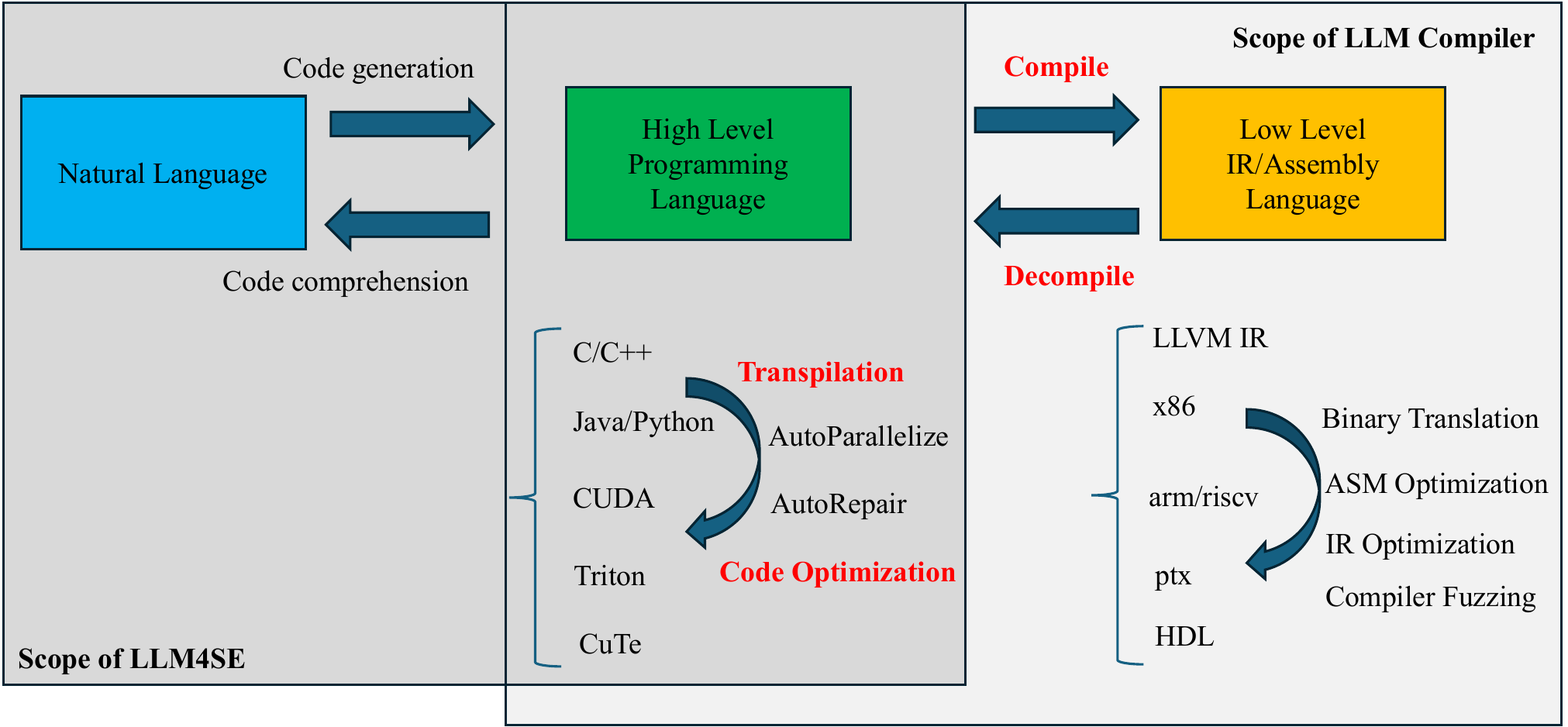}
    \caption{Code level view of LLM Compiler}
    \label{fig:code_level}
\end{figure*}

This section details the third and fourth dimensions of our taxonomy: the Level of Code Abstraction and the specific Task Type. We analyze these two axes together as they are deeply intertwined; the level of program representation fundamentally dictates the nature of the tasks that can be performed. As illustrated in~\autoref{fig:code_level}, we categorize these representations into three primary strata: 
\begin{itemize}
    \item Natural Language (NL), the highest level of human intent; 
    \item High-Level Programming Language (PL), the source code developers write;
    \item Low-Level Intermediate Representation and Assembly (ASM), the machine-centric representations used by the compiler backend and executed by computers.
\end{itemize}

The choice of abstraction level fundamentally dictates the nature of the tasks that can be performed, which in turn forms the third dimension of our taxonomy. In this section, we first detail the tasks that operate within a single level of abstraction (intra-level transformations) and then discuss the more complex tasks that bridge these different levels (cross-level transformations).

\begin{table*}
    \centering
    \resizebox{\textwidth}{!}{%
    \begin{tabular}{|c|c|c|c|}
    \hline
    \textbf{Acronym} & \textbf{Citation} & \textbf{Tasks} & \textbf{Code level} \\
    \hline
    Qimeng-Xpiler & \citet{10_qimengxpiler2025_osdi} & Transpile & VNNI,CUDA,HIP,BANG \\
    G-TransEval & \citet{34_jiao2023_ase} & Transpile & C++,Java \\
    CoTran & \citet{38_jana2024cotran} & Transpile & Java,Python\\
    Oxidizer & \citet{60_zhang2025_pldi} & Transpile & Go-Rust\\
    OpenCLGen & \citet{74_palkowski2024automatic} & Transpile & PolyC-OpenCL \\
    LLMLift & \citet{81_bhatia2024_neurips} & Transpile & C,C++,Java-DSLs \\
    Rectifier & \citet{84_yin2024rectifier} & Transpile & C++,Java,Python \\
    AlphaTrans & \citet{91_alphatrans2025_fse} & Transpile & Java-Python \\
    OutputFormat & \citet{95_macedo2024exploring} & Transpile & C,C++,Go,Java,Python \\
    LostInTranslation & \citet{99_pan2024_ICSE} & Transpile & C,C++,Go,Java,Python \\
    KnowTransfer & \citet{106_knowledgetransfer2024_oopsla} & Transpile & Java,Python,JS-Ocaml,Racket \\
    CanLLMParallel & \citet{108_nichols2024can} & Transpile & C++-MPI+OpenMP\\
    SALLM & \citet{146_sallm_asew} & Transpile & python+test \\
    TransCoder & \citet{177_transcoder} & Transpile & C++,Java,Python \\
    DOBF & \citet{176_dobf} & Transpile & C++,Java,Python \\
    TransCoder-ST & \citet{175_transcoderst} & Transpile & C++,Java,Python \\
    TransCoder-IR & \citet{154_transcoderir2023_iclr} & Transpile & LLVM IR-C++,Java,Rust,Go \\
    BabelTower & \citet{153_babeltower2022_icml} & Transpile & C-CUDA \\
    CodeRosseta & \citet{174_coderosseta2024_neurips} & Transpile & C-CUDA \\
    HPCTrans & \citet{2_HPCTransCompile2025_arxiv} & Dataset Generator & C-CUDA \\
    UnsuperBinTrans & \citet{168_unsupervised2023_emnlp} & Binary Transpile & ARM-x86 \\
    \hline
    TFix & \citet{166_tfix2021_icml} & Code Repair & JavaScript \\
    LLMAPR & \citet{133_xia2023_icse} & Code Repair & Java,Python,C \\
    ChatGPTRepair & \citet{148_zhang2023critical} & Code Repair & Java \\
    EISP & \citet{16_chen2024test} & Code Repair & Python,JavaScript \\
    MacroConfig & \citet{19_albuquerque2024evaluating} & Code Repair & C/C++/Java \\
    CoCoGen & \citet{110_bi-etal-2024-iterative} & Code Repair & Python \\
    ZS4C & \citet{29_kabir2025zs4c} & Code Repair & Python \\
    Repilot & \citet{131_wei2023_fse} & Code Repair & Java \\
    RustAssistant & \citet{113_deligiannis2024rustassistant} & Code Repair & Rust \\

    LIBRO & \citet{150_kang2023large} & CVE test Generation & Defects4J \\
    SecurityTestGen & \citet{144_zhang2023doesllmgeneratesecurity} & CVE test Generation & Java CVE test \\
    GeneticImprove & \citet{147_brownlee2023enhancing} & Compiler Fuzzing & Java \\
    BenchDirect & \citet{139_tsimpourlas2023benchdirectdirectedlanguagemodel} & Compiler Fuzzing & OpenCL \\
    WhiteFox & \citet{31_yang2024whitefox} & Compiler Fuzzing & PT-Inductor/XLA/TF-Lite \\
    MetaMut & \citet{76_ou2024mutators_asplos} & Compiler Fuzzing & C/C++ \\
    ClozeMaster & \citet{53_gao2025clozemaster} & Compiler Fuzzing & Rust \\
    FMCSO & \citet{14_italiano2025finding} & Compiler Fuzzing & C/C++ \\
    
    \hline
    CORL & \citet{40_mammadli2020static_llvmhpc} & Pass Optimization & LLVM pass \\
    GraphFlagOpt & \citet{125_graphbased2024_hpec} & Pass Optimization & GCC pass \\
    NeuroEvolution & \citet{127_neuroevolutionary2023_gecco} & Pass Optimization & LLVM pass \\
    CompilerR1 & \citet{45_compilerr1-2025_arxiv} & Pass Optimization & LLVM pass \\
    TLM & \citet{167_zhai2024_osdi} & Autotuning Optimization & TVM IR \\
    ReasoningCompiler & \citet{4_tang2025compiler_arxiv} & Autotuning Optimization & TVM IR \\
    Effi-Learner & \citet{79_huang2024effilearner} & Code Optimization & Python \\
    OMPar & \citet{93_kadosh2024omparautomaticparallelizationaidriven} & Code Optimization & C/C++ + OpenMP \\
    ProblemOriented & \citet{123_ye2024problem} & Code Optimization & C++ \\
    CodeOptCoT & \citet{116_xu2024code} & Code Optimization & Python \\
    LangProp & \citet{30_ishida2024langprop} & Code Optimization & Python \\
    AutoComp & \citet{6_hong2025autocomp} & Code Optimization & C-C+Intrinsic \\
    SBLLM & \citet{17_gao2025search} & Code Optimization & Python,C++ \\
    PCAOT & \citet{22_romero2025should} & Code Optimization & C+OpenMP \\
    CompilerGPT & \citet{3_compilergpt2025_arxiv} & Code Optimization & C/C++ \\
    Perfcodegen & \citet{52_peng2025perfcodegen} & Code Optimization & Python \\
    SpeedGen & \citet{58_purschke2025speedgen} & Code Optimization & Python \\
    CodeOPT & \citet{63_xu2025efficient} & Code Optimization & C/C++ \\
    PerfRL & \citet{134_duan2025perfrlsmalllanguagemodel} & Code Optimization & C++,Java,Python \\
    ECO & \citet{12_lin2025eco} & Code Optimization & C++ \\
    CUDA-L1 & \citet{cudal1_2025_arxiv} & Code Optimization & CUDA/Pytorch \\
    LLMVectorizer & \citet{47_llmvectorizer2025_cgo} & Code Optimization & C/C++ \\
    RACL & \citet{46_wang2025_pldi} & Code Optimization & C/C++ \\
    CodeTransform & \citet{178_cummins2024donttransformcodecode} & Code Optimization & Python \\
    \hline
    \end{tabular}
    }
    \caption{Summary of LLM for intra-level code transformation}
    \label{tab:intra-level}
\end{table*}

\begin{table*}
    \centering
    \begin{tabular}{|c|c|c|c|}
    \hline
    \textbf{Acronym} & \textbf{Citation} & \textbf{Tasks} & \textbf{Code level} \\
    \hline
    NatGen & \citet{172_chakraborty2022natgen} & Code Optimization & Java,Python \\
    CodeOptEdu & \citet{54_rong2025integrating} & Code Optimization & Python \\
    RTLrewriter & \citet{81_bhatia2024_neurips} & Code Optimization & RTL \\
    SymRTLO & \citet{9_wang2025symrtlo} & Code Optimization & RTL \\
    InstCombiner & \citet{152_mannarswamy2022learningcombineinstructionsllvm} & ASM optimization & arm \\
    peephole & \citet{15_fang2024towards} & ASM optimization & arm \\
    VeriLOCC & \citet{85_jin2025verilocc} & ASM optimization & SASS,RDNA \\
    CompilerDream & \citet{11_deng2024compilerdream_arxiv} & IR Optimization & LLVM IR \\
    MetaLLMCompiler & \citet{metallmcompiler-cc25} & IR Optimization & LLVM IR \\
    PrioritySampling & \citet{27_grubisic2024priority_arxiv} & IR Optimization & LLVM IR \\
    DeCOS & \citet{51_decos2025_ics} & IR Optimization & LLVM IR \\
    \hline
    \end{tabular}
    \caption*{\textbf{Table 2}: (Continued)}
\end{table*}

\subsection{Intra-Level Transformations}
Intra-level transformations are those where the input and output of the process remain at the same level of abstraction. These tasks typically focus on refinement, optimization, or migration within a given representation.
\begin{itemize}
    \item \textbf{High-Level Programming Language (PL) Transformations:} This is currently the most active area of research, focusing on source-to-source transformations, a joint hotspot for both software engineering, programming language and LLM communities. LLMs are applied to a wide array of PLs, from general-purpose languages like C/C++, Java, and Python to specialized, domain-specific languages for parallel computing such as CUDA, Triton~\citep{triton2019mapl}, and CuTe~\citep{cute2025web}. Key tasks at this level include:
    \begin{itemize}
        \item Transpilation: Translating between different PLs, a task crucial for migrating legacy codebases or converting programs between different programming ecosystems.
        \item Code Repair: A major goal of production code design is maintainability, therefore, LLM-generated code should be tested more or less, and automated test-generation, or fuzzing test is an important direction. Besides, the flawed code should be automated repaired with minimal human intervention.
        \item Source Optimization: Rewriting PL code to improve its quality, readability, or performance without changing the language.

    \end{itemize}
    \item \textbf{Low-Level IR / Assembly Transformations:} At this level, LLMs perform transformations on machine-centric representations, often invisible to the original programmer but critical for final performance. These operate on common compiler IRs like LLVM IR and instruction set architectures such as x86, PTX, ARM, and RISC-V. The main tasks include:
    \begin{itemize}
        \item IR Optimization: Applying optimization passes directly on an intermediate representation to improve the efficiency of the generated code.
        \item Assembly Optimization: Applying assembly-level optimizations like register-allocation, instruction scheduling and peephole optimizations to optimize the code performance or code size.
        \item Binary Translation: Translating low-level code from one instruction set architecture (ISA) to another, for example, from x86 to ARM. 
    \end{itemize}
\end{itemize}

As summarized in~\autoref{tab:intra-level}, the application of LLMs to intra-level code transformation has become a highly active area of research. These tasks, which operate within the same level of abstraction (e.g., source-to-source), primarily leverage the generative capabilities of LLMs to rewrite programs. In this section, we categorize and review these studies based on their primary objective: Code Transpilation, which focuses on migrating between languages; Code Repair, which aims to correct defects; and Code Optimization, which seeks to improve program performance.

\subsubsection{Transpile}

Code transpilation, or source-to-source translation, is one of the most prominent tasks in this domain. The core challenge is to accurately translate a program from a source language to a target language while preserving its semantic correctness and functionality. This capability is critical for modernizing legacy codebases, improving interoperability between systems, and migrating applications to new hardware ecosystems, such as converting sequential C code to parallel CUDA for GPUs. The following studies demonstrate the breadth of this task, tackling a wide range of language pairs and programming paradigms.

OpenCLGen~\citep{74_palkowski2024automatic} uses LLM to translate polyhedral C kernels into OpenCL.
Rectifier~\citep{84_yin2024rectifier} introduces an LLM correcter to handle different transpilation errors to improve the code transpilation accuracy.
OutputFormat~\citep{95_macedo2024exploring} studies the impact of LLM output format in code translation tasks.
LostInTranslation~\citep{99_pan2024_ICSE} empirically summarizes bugs introduced by LLM transpilation into detailed categories across multiple programming languages.
SALLM~\citep{146_sallm_asew} benchmarks the capabilities of LLMs to generate secure Python code by examining the Common Weakness Enumeration (CWE)-related test generation.

Besides using LLMs directly, earlier work focuses on a data-centric problem: \textit{how to obtain parallel code corpora to train a model? }
To solve this, TransCoder-series~\citep{177_transcoder, 176_dobf, 175_transcoderst,154_transcoderir2023_iclr} have established how to learn a transpiler language model unsupervisedly: (1) using mask language modeling (MLM) to learn cross-lingual languages first, (2) using back translation (BT) to learn the translation rules later. Other techniques like denoising autoencoding~\citep{177_transcoder}, deobfuscation~\citep{176_dobf}, unittest filtering~\citep{175_transcoderst} and utilizing compiler representations~\citep{154_transcoderir2023_iclr} gradually improve the learned transpiler's capability. These studies are typically between general programming languages like C++, Java and Python.

Besides transpiling between general programming languages, BabelTower~\citep{153_babeltower2022_icml} proposes an improved TransCoder-based approach to unsupervisedly learn a transpiler between C++ and CUDA with new parallel metrics.
CodeRosseta~\citep{174_coderosseta2024_neurips} further improves BabelTower with AST Entity Recognition and customized denoising auto-encoding. Now these learning-based methods are more used as dataset generator for more powerful LLM training. Besides the learning-based approaches, HPCTrans~\citep{2_HPCTransCompile2025_arxiv} modifies the AI compiler TVM~\citep{tvm_2018_osdi} to auto-generate semantically equivalent CUDA and C code without TVM dependency, thereby creating rich CUDA-to-CPU transpilation corpora.

CanLLMParallel~\citep{108_nichols2024can} studies the auto-parallel capability within LLMs by using LLM to translate sequential C++ code into parallel code like MPI, OpenMP and CUDA.

CoTran~\citep{38_jana2024cotran} proposes a RL-based feedback mechanism using compiler feedback and symbolic execution to improve both compilation accuracy and functional accuracy in Python-to-Java translation.

AlphaTrans~\citep{91_alphatrans2025_fse} proposes a LLM-based project-level transpiler within GraalVM to support partial transpilation verification, they also implement a divide-and-conquer strategy to decompose both source code and test to achieve scalable Java-to-Python transpilation.
Oxidizer~\citep{60_zhang2025_pldi} also presents a LLM-based Go-to-Rust transpiler for entire projects, where a project partition module divides-and-conquers the translation complexity and a type-checking feature mapping module handles to migrate semantics between languages. 

KnowTransfer~\citep{106_knowledgetransfer2024_oopsla} proposes that transpilation to low-resource languages (e.g. Ocaml, Racket) can be improved and verified with deterministic test case transpilation from high-resource languages (e.g. Java, Python), thereby synthesizing rich bilingual corpora effectively.

LLMLift~\citep{81_bhatia2024_neurips} proposes that LLM can be used to transpile sequential programs into Tensor-processing framework DSLs like PyTorch and NumPy, with Floyd-Hoare Logic (FHL) to validate generated programs.

Qimeng-Xpiler~\citep{10_qimengxpiler2025_osdi} uses LLMs to transpile parallel code between different architectures, including SIMT GPUs (CUDA and HIP), SIMD CPU (C+VNNI) and SIMD NPU (BANG C), using SMT-solver to check key transpilation results correctness and Monte-Carlo Tree Search (MCTS) to explore higher performance candidates within the target architecture. 

Besides source code transpilation, there are also some interesting work like UnsuperBinTrans~\citep{168_unsupervised2023_emnlp} that performs binary code translation to preliminarily study the vulnerability discovery problem in a low-resource ISA from high-resource ISA (e.g. x86). However, many binary level work cannot be directly neurally transpiled due to code difficulties, instead, many work focuses on other utility tasks, such as binary similarity detection or compiler property identification, which we will detail in~\autoref{sec:utility}.

\subsubsection{Code Repair}

Beyond translating between different languages, another critical application is translating a program from a flawed state to a correct one. Automated Program Repair (APR) using LLMs treats bug fixing as a specialized translation task: from an incorrect source program to a corrected version in the same language. This approach aims to automate the often tedious and error-prone process of debugging by leveraging the model's learned knowledge of common programming mistakes and their corresponding fixes. The works reviewed here showcase various strategies for applying LLMs to automatically identify and repair bugs. Additionally, generating specialized tests to detect bugs (fuzzing) is also highly-related to this category, which can be seen as either an intra-level code transformation (sythesizing tests) task or a utility-based task.

TFix~\citep{166_tfix2021_icml} learns a transformer model to perform specialized sequence-to-sequence code repair task.
LLMAPR~\citep{133_xia2023_icse} perform the first extensive study on directly applying LLMs for APR, suggesting that both sample size increase and incorporation with fix template information can help improve LLM-based APR.
ChatGPTRepair~\citep{148_zhang2023critical} reveals an important overlooked data leakage issue of automated program repair (APR), and finds ChatGPT is more powerful in APR on its proposed decontaminated benchmark EvalGPTFix than PLBart and CodeT5.
EISP~\citep{16_chen2024test} introduces a test-free semantic mistakes localization framework using LLM-based static-analysis.
MacroConfig~\citep{19_albuquerque2024evaluating} studies LLM's capabilities in resolving configurable macros errors with compilation feedback.
CoCoGen~\citep{110_bi-etal-2024-iterative} uses an iterative generation and verification process to do project-level code repair with careful context retrieval.
ZS4C~\citep{29_kabir2025zs4c} proposes a zero-shot LLM-based synthesizer to autocomplete incomplete code to a compilable one.

Except using LLMs for automated code repair, LLMs can also be used in a repair assistant way.
Repilot~\citep{131_wei2023_fse} uses LLM as completion engine for program repair by aggresively synthesizing valid patches during the repair process.
RustAssistant~\citep{113_deligiannis2024rustassistant} is another assistant tool to suggest potential Rust code fixes based on compiler feedbacks.

On the fuzzing test side, 
LIBRO~\citep{150_kang2023large} showcases LLMs can be used to reproduce bugs by synthesizing test programs from bug reports. 
SecurityTestGen~\citep{144_zhang2023doesllmgeneratesecurity} also studies the capabilities of LLMs to generate security tests from CVEs.

GeneticImprove~\citep{147_brownlee2023enhancing} finds that LLMs can be used as mutation operators combined with genetic improvment to generate diverse valid programs. 
BenchDirect~\citep{139_tsimpourlas2023benchdirectdirectedlanguagemodel} trains a language model to generate compiler testing benchmarks with high readability. 
WhiteFox~\citep{31_yang2024whitefox} adopts a multi-agent framework to automatically synthesize white-box compiler fuzzing tests.
MetaMut~\citep{76_ou2024mutators_asplos} proposes to guide an LLM as mutator, it uses LLM to fill-in a carefully crafted templates to generate non-trivial mutator designs, then randomly applied to test programs to synthesize fuzzing programs, which successfully harvested 131 GCC/Clang compiler bugs.
ClozeMaster~\citep{53_gao2025clozemaster} similarly uses LLMs to do fuzzing tests for Rust compiler, where it first generates cloze-masked code snippets then fill-in-the-blank to generate diverse compiler test programs.
FMCSO~\citep{14_italiano2025finding} presents a mutation testing methodology by using LLMs to iteratively modify starting code seed then develop differential testing strategies to find missing code size optimizations in LLVM compilers.

\subsubsection{Code Optimization}

The third major category of intra-level transformation is Code Optimization. This can be viewed as translating a program from a semantically correct but sub-optimal version to an improved version that is more efficient in terms of performance, memory usage, or code size. While this is a classic compiler goal, the application of LLMs is particularly broad in this domain. Unlike transpilation and repair which primarily operate on source code, optimization tasks are being explored at multiple levels of abstraction. This includes not only high-level, source-to-source rewriting but also critical low-level tasks such as IR Optimization and assembly-level tuning. Across all these levels, LLMs offer a novel, data-driven approach to discovering complex optimization strategies that may be difficult for traditional heuristic-based compilers to identify.

First, like previous studies in ML-powered compiler studies, using language model as selector is also feasible and can be learned through proper feature engineering. CoRL~\citep{40_mammadli2020static_llvmhpc}, NeuroEvolution~\citep{127_neuroevolutionary2023_gecco}, GraphFlagOpt~\citep{125_graphbased2024_hpec} and CompilerR1~\citep{45_compilerr1-2025_arxiv} are used as selector to optimize compiler pass ordering. TLM~\citep{167_zhai2024_osdi} uses an LLM trained from scratch to generate predicted schedule sequences from the large schedule space to accelerate TVM autotuning searching. ReasoningCompiler~\citep{4_tang2025compiler_arxiv} similarly uses MCTS to explore the the search space through LLM reasoning and also integrated to TVM to accelerate the sampling efficiency.

Besides using LLM as selector, there are more work exploring to use LLM to perform generative optimizations, acting as a direct translator, where the vast pretrained code knowledge in LLM is believed to guide the LLM to perform reasonable optimizations. The majority of these optimizations happen at the source code level. 

Effi-Learner~\citep{79_huang2024effilearner} proposes a self-optimization framework utilizing execution overhead profiles to improve the efficiency of LLM-generated code.
OMPar~\citep{93_kadosh2024omparautomaticparallelizationaidriven} studies the capability of OpenMP pragma generation to auto-parallelize C/C++ code.
ProblemOriented~\citep{123_ye2024problem} proposes an anchor verification mechanism, first synthesizing test inputs based on slow code, constructing verified test case through executing with slow code, and iteratively refining the optimized code with execution feedback.
CodeOptCoT~\citep{116_xu2024code} applies the Chain-of-Thought techniques to augment the structure understanding and self-checking to improve code optimization.
LangProp~\citep{30_ishida2024langprop} has a specific focus on autodriving code optimization in CARLA.

In 2025, studies within this task have boomed. AutoComp~\citep{6_hong2025autocomp} studies using LLM to optimize C code into different tensor accelerators with distinct intrinsics. 
SBLLM~\citep{17_gao2025search} proposes a search-based LLM framework using optimization knwoledge retrieval and genetic operator-inspired CoT prompting.
PCAOT~\citep{22_romero2025should} compares LLM-based optimization with traditional optimization compilers and finds LLMs struggle to optimize large programs (measured in LOCs) directly.
CompilerGPT~\citep{3_compilergpt2025_arxiv} proposes using LLMs to analyze and act on compiler optimization reports, making tailored code rewriting optimizations to fit the optimization reports.

Perfcodegen~\citep{52_peng2025perfcodegen} proves that small LLMs with proper execution feedback can generate performant code compared to naively prompted flagship LLMs.
SpeedGen~\citep{58_purschke2025speedgen} uses LLM to optimize Python code performance by rewriting to use high performance libraries like PyTorch and NumPy.
CodeOPT~\citep{63_xu2025efficient} finetuned a LoRA adapter to perform optimization strategies like loop unroll, inline expansion, constant folding and dead code elimination in the source code level in C/C++.
PerfRL~\citep{134_duan2025perfrlsmalllanguagemodel} trains a small language model within a reinforcement learning environment with direct metric feedback to obtain better code optimization results.

ECO~\citep{12_lin2025eco} uses historical commits to record anti-patterns addressed, and using a finetuned LLM to automatically refactor code, auto-verify, and submit to code review, constructing an automated code optimization pipeline that scales to warehouse-scale computers in Google production.
CUDA-L1~\citep{cudal1_2025_arxiv} reveals a remarkable capability of RL in autonomous learning for CUDA optimization, with a SFT+GRPO~\citep{grpo_2024_arxiv} finetuned model, capable of generating 249 out of 250 KernelBench~\citep{kernelbench_2025_icml} cases and optimizing 240 of them better than native PyTorch.
LLMVectorizer~\citep{47_llmvectorizer2025_cgo} studies to auto-vectorize C code with SIMD intrinsics in source code level and formally verified through symbolic verification with Alive2~\citep{alive2_2021_pldi}. RACL~\citep{46_wang2025_pldi} uses reductive analysis to divide-and-conquer program complexity and perform input-centric code optimizations, significantly reduces profiling efforts.

CodeTransform~\citep{178_cummins2024donttransformcodecode} explores to preliminarily use LLMs to code the transform instead of directly transforming code, using the LLM-as-generator methodology.

Besides pure performance optimization, there are work like NatGen~\citep{172_chakraborty2022natgen} which introduces a novel task of ``code naturalization'' to transform unnatural code into natural one that fits programming paradigms, where it pretrained a LM to naturalize code well in Java and Python. Other work like CodeOptEdu~\citep{54_rong2025integrating} is used to refine code for educational purposes.

\subsection{Cross-Level Transformations}

\begin{table*}
    \centering
    \resizebox{\textwidth}{!}{%
    \begin{tabular}{|c|c|c|c|}
    \hline
    \textbf{Acronym} & \textbf{Citation} & \textbf{Tasks} & \textbf{Code level} \\
    \hline
    ANPL & \citet{138_ANPL23_neurips} & Code Generation & NL-Python \\
    AIOSCompiler & \citet{23_xu2024aios} & Code Generation & NL-CoRE DSL \\
    Autoiot & \citet{59_autoiot2025_mobicom} & Code Generation & NL-AIoT code \\
    ArabicLLMCompiler & \citet{25_sibaee2024llms} & Code Generation & Arabic-Python \\
    Unicoder & \citet{92_sun-etal-2024-unicoder} & Code Generation & Pseudo Code-Python,JS \\
    \hline
    transformer-x86 & \citet{39_armengol-estape2021learning} & Compilation & C-x86 \\
    transformer-llvm & \citet{171_guo2022enabling} & Compilation & C-LLVM IR \\
    llm-x86 & \citet{70_c-x86-emnlp24} & Compilation & C-x86 \\
    LEGO-Compiler & \citet{5_legocompiler-arxiv25} & Compilation & C-x86,arm,riscv \\
    VirtualCompiler & \citet{71_gao-etal-2024-virtual} & Compilation & C-x86 \\
    VeriLOCC & \citet{85_jin2025verilocc} & Compilation & MIR-SASS,RDNA \\
    InstCombiner & \citet{152_mannarswamy2022learningcombineinstructionsllvm} & Compilation & arm \\
    VEGA & \citet{44_vega2025_cgo} & Compilation & LLVM C++,TableGen \\
    ComBack & \citet{88_comback2024_neurips} & Compilation & LLVM C++, TableGen \\
    \hline
    AutoChip & \citet{135_thakur2023autochip} & RTL Generation & Verilog \\
    Origen & \citet{124_cui2024origen} & RTL Generation & Verilog \\
    VeriGen & \citet{104_verigen2024} & RTL Generation & Verilog \\
    VerilogLLM & \citet{64_wang2025insights} & RTL Generation & Verilog \\
    
    MakeMoveCount & \citet{111_delorenzo2024make} & RTL Generation & Verilog \\
    RTLLM & \citet{114_lu2024rtllm} & RTL Generation & Verilog \\
    BuiltRome & \citet{120_nakkab2024rome} & RTL Generation & Verilog \\
    
    HLSPilot & \citet{98_xiong2024hlspilot} & HLS & C++-HLS C++ \\
    \hline
    Slade & \citet{72_armengol2024slade} & Decompilation & x86,arm-C \\
    LLM4Decompile & \citet{73_tan-etal-2024-llm4decompile} & Decompilation & (obfuscated) x86-C \\
    Degpt & \citet{75_hu2024degpt} & Decompilation & x86-C \\
    Lmpa & \citet{130_xu2023lmpa} & Decompilation & x86-C \\
    BoostDecompile & \citet{155_cao2022boosting} & Decompilation & x86-C \\
    BinSum & \citet{129_binsum2023_arxiv} & Decompilation & x86-NL \\
    DecLLM & \citet{173_wong2025decllm} & Decompilation & x86-C \\
    IR-LLM & \citet{13_jiang2025can} & IR Decompilation & LLVM IR-C \\
    Forklift & \citet{169_armengol-estape2024forklift} & Decompilation to IR & x86,arm,riscv-LLVM IR \\
    \hline
    NeuralShapeCompiler & \citet{145_luo2023neural} & Multimodal & Text-PointCloud-Code \\

    RASP & \citet{165_weiss2021thinking} & NeuralCompilation & RASP DSL  \\
    Tracr & \citet{132_lindner2023tracr} & NeuralCompilation & RASP-transformer weight \\
    ALTA & \citet{83_shaw2025alta} & NeuralCompilation & ALTA-transformer weight \\
    TransformerProgram & \citet{143_friedman2023learning} & NeuralDecompilation & transformer weight-RASP \\
    AlgorithmicLM & \citet{20_saldyt2025algorithmiclanguagemodelsneurally} & NeuralDecompilation & transformer weight-python \\
    \hline
    \end{tabular}
    }
    \caption{Summary of LLM for inter-level code transformation}
    \label{tab:inter-level}
\end{table*}

Cross-level transformations are those that bridge the different strata of abstraction. These tasks are fundamental to the very definition of a compiler and its related tools, representing some of the most complex and impactful applications of LLMs in this domain. As outlined in~\autoref{tab:inter-level}, we will list these cross-level transformations in the following section.

\subsubsection{Code Generation: NL-PL}

This is the process of translating a Natural Language specification into a structured Programming Language program. In this role, the LLM acts as the ultimate compiler front-end, directly converting a developer's description of a problem into a working solution.

ANPL~\citep{138_ANPL23_neurips} introduces an interactive way to decompose program generation problem to sketch and holes, and fill later with flexible user interruption.

AIOSCompiler~\citep{23_xu2024aios} uses LLM as interpreter for natural language programming and flow programming of AI agents, where they introduce a domain-specific language CoRE to unify natural language programming, pseudo-code programming and flow programming.
Unicoder~\citep{92_sun-etal-2024-unicoder} also proposes a code representation to unify different programming languages and is LLM-friendly for code generation.

Autoiot~\citep{59_autoiot2025_mobicom} focuses on AIoT applications code generation using natural language programming, supported by a background knowledge retrieval module, a CoT program synthesis module and an automated code improvement module.

ArabicLLMCompiler~\citep{25_sibaee2024llms} uses LLM as interpreter to translate Arabic-based programming languages into executable Python code.

The reverse of code generation: code comprehension is also an important task for software engineering, however, as our study has a compiler-centric focus, studies with code comprehension goal are therefore treated as out of scope in this survey. 

\subsubsection{Compilation: PL-ASM}

This is the core classic compilation task, where an LLM is used to translate a human-readable PL into a low-level representation like IR or assembly language. This is a core challenge where LLMs are being explored to augment or even replace components of traditional compiler backends. Some similar subtasks, like high-level synthesis, HDL generation, also fall into this category.

Transformer-x86~\citep{39_armengol-estape2021learning} preliminarily studies to learn a transformer from compiler-generated C-x86 corpora.
Transformer-llvm~\citep{171_guo2022enabling} similarly studies C-LLVM and C-x86 with optimization level setting. Neither of them achieve fair translation accuracy, typically for optimized setting, the translation accuracy is 0\%.

VirtualCompiler~\citep{71_gao-etal-2024-virtual} models the LLM compilation as similar task of assembly code search and proposes using an LLM model to compile any source code of any language to assembly code by matching assembly results.

Llm-x86~\citep{70_c-x86-emnlp24} uses a data-centric augmentation pipeline to improve compiler-generated corpora quality, with specific focus on numerical representation and label generation, using the improved corpora to finetune an LLM, achieve substantial C-x86 compilation improvement, over 91\%.

LEGO-Compiler~\citep{5_legocompiler-arxiv25} studies the scalability problem of neural compilation, decompositing function-level code into finer granularities like basicblock-level or statement-level with managed context, and compiles each fragment accordingly with necessary symbol information. Together with a stepwise workflow that can verify intermediate results and an auto repair loop, it achieves over 99\% neural compilation accuracy across three architectures on O0 setting and achieve an order of magnitude scalability improvement without model advancement.
However, LLM-based compilation is still preliminary, because it fails to outperform existing compiler systems.

Besides full compilation process performed by LLM, there are also works on specific step within the compilation pipeline. VeriLOCC~\citep{85_jin2025verilocc} studies the register allocation capabilities using LLMs on both CUDA SASS and AMD RDNA assembly generation. While InstCombiner~\citep{152_mannarswamy2022learningcombineinstructionsllvm} studies the instruction combination capabilities on ARM64 assembly using LLMs.

Except LLM-as-translator methodology, there are also interesting works investigating to generate specific compiler code. VEGA~\citep{44_vega2025_cgo} proposes a LLVM backend generation method using pretrained LM with carefully crafted compiler backend dataset ComBack~\citep{88_comback2024_neurips}. By treating the LLVM code as features to learn and different backends as training data, the model can generate sketches of over 60\% of the LLVM backend functions without human intervention. However, fully automated compiler code generation is still far from reality, as compilers are one of the most complicated softwares to maintain.

Except compilation to assembly code, there are also works focusing on RTL code generation, which can also be seen as a broader compiler task. Among them, 
AutoChip~\citep{135_thakur2023autochip} proposes a self-reflection loop to fix trivial generation errors. 
Origen~\citep{124_cui2024origen} proposes a data augmentation pipeline using claude3 distilled data and a similar self-reflection mechanism.
VeriGen~\citep{104_verigen2024} finetunes a series of verilog generation LLMs using supervised finetuning.
VerilogLLM~\citep{64_wang2025insights} further trains a verilog generation LLM with reinforcement learning (RL) with testbench feedback. 

MakeMoveCount~\citep{111_delorenzo2024make} studies to perform MCTS on RTL code generation.
RTLLM~\citep{114_lu2024rtllm} proposes both benchmark for LLM-based RTL generation and a simple-yet-effective self-planning prompting strategy to boost performance of RTL generation with GPT3.5.
BuiltRome~\citep{120_nakkab2024rome} also found that hierarchical prompt structuring can dramatically improve LLM performance on hardware design tasks, enabling successful generation of complex modules that would otherwise be impossible.

HLSPilot~\citep{98_xiong2024hlspilot} instead of focusing on verilog RTL generation, it generates High-Level Synthesis (HLS) code translated from software C++ code, it outperforms manually written FPGA kernels with the integration of profiling tools and DSE tools to enable automatic hardware/software partition and pragma tuning.

\subsubsection{Decompilation: ASM-PL}

This involves the reverse process of compilation, where an LLM attempts to reconstruct a human-readable and semantically meaningful PL from a low-level representation like a binary or assembly file. This is a notoriously difficult task for which the pattern recognition capabilities of LLMs are a promising research direction.

Slade~\citep{72_armengol2024slade} and LLM4Decompile~\citep{73_tan-etal-2024-llm4decompile} are both learned language models used for decompilation tasks, each has a centric for either optimized code recovery and obfuscated code recovery.

Degpt~\citep{75_hu2024degpt} focuses on using LLMs to interpret and refine decompiler output to improve readability and simplicity, which can assist the reverse engineering process.

Lmpa~\citep{130_xu2023lmpa} develops a program analysis assisted method for symbol name recovery in decompilation, by querying ChatGPT with managed context, the model can generate 75\% of the recovered names considered good by users.

BoostDecompile~\citep{155_cao2022boosting} uses a multi-layer decompilation pipeline, where the recovery of program is jointly performed by rules and neural networks.

BinSum~\citep{129_binsum2023_arxiv} focues on assembly code understanding capabilities of LLMs, and finds that stripping debug symbols has a significant loss to the decompilation accuracy.

DecLLM~\citep{173_wong2025decllm} proposes a recompilable centric decompilation system, with an iterative LLM-based repair loop to improve decompiler outputs, which combines static recompiling and dynamic runtime feedback.

Besides full decompilation process, there are also works focusing on some steps. 
IR-LLM~\citep{13_jiang2025can} focuses on the LLVM IR-to-C decompilation process, while Forklift~\citep{169_armengol-estape2024forklift} focuses on the assembly-to-LLVM IR decompilation process.

\subsubsection{Special cases of cross-level transformations}

Beyond transformations within the traditional compilation stacks, an emerging line of research explores more special cases of compilation which is not limited to typical text-based or symbolic domains. 

For example, NeuralShapeCompiler~\citep{145_luo2023neural} proposes a compiler-inspired, unified framework for translating data between multiple modules. This framework first converts all data into a unified, discrete intermediate code and then uses a transformer model to perform the translation across text, code and pointcloud.

A distinct and emerging research direction, also termed with the name of ``neural compilation'', explores the direct translation of \textit{programs into the parameters of a neural network}. This approach is not aimed at traditional execution but at understanding the algorithmic capabilities of models like transformers, thereby contributing to the study of explainability in LLMs. A foundational line of work in this area began with RASP~\citep{165_weiss2021thinking}, an abstract programming language designed to express algorithms using primitives that mirror the core operations of a transformer. Building on this theoretical framework, Tracr~\citep{132_lindner2023tracr} was developed as a compiler that translates RASP programs directly into the weights of a standard transformer model. This capability was further extended by ALTA~\citep{83_shaw2025alta}, which adds support for dynamic control flow operations like loops.

Complementing this compilation process, some research explores the reverse direction—a form of decompilation. For example, TransformerProgram~\citep{143_friedman2023learning} introduces a method to train a modified transformer that can be automatically converted back into a discrete, human-readable Python program. Furthermore, AlgorithmicLM~\citep{20_saldyt2025algorithmiclanguagemodelsneurally} investigates methods for composing these ``neurally compiled'' weights to execute combined algorithms through an augmented interpreter.

Despite its theoretical importance, this area of research currently also faces significant limitations in both the scale of the programs and the variety of operations that can be compiled, marking it as a key area for further investigation.

As we can see in the number of studies about intra-level (\autoref{tab:intra-level}) and inter-level (\autoref{tab:inter-level}) code transformations, there are more studies handling with high-level code (59+34) than low-level IR/assembly (14+16). 
The quantitative imbalance can be viewed in two ways. First, Low-level compiler tasks do receive less attention from the community as there are much more LLM researchers than compiler researchers (who are interested in tasks about low-level representations). Second, low-level code representations are less amenable to LLM approaches, because: 1. there are less low-level code corpora than high-level corpora in most LLMs' pretraining stage; 2. the quality of low-level corpora is relatively low, as many low-level code representations are automatically generated through compilers, which lack readability and is hard for LLMs to learn with.

\subsection{Non-Transformed Utilities}\label{sec:utility}

\autoref{tab:utility} outlines related utility-based studies. Except for generative tasks, LLMs also enable utility-based tasks in broader compiler domain, such as the vulnerability test generation and compiler fuzzing test, which we have outlined earlier.

As for other utility studies, CompileAgent~\citep{8_hu2025compileagent_acl} proposes to use LLMs to auto-configure project setup. DCC~\citep{80_dcc2024} develops a tool that integrates a Large Language Model into the Debugging C Compiler to generate context-aware, novice-focused explanations for compile- and run-time errors to help introductory programming students. Quantum~\citep{89_quantum2024_dac} employs a Seq2Seq model to solve the qubit routing problem in quantum compilation, reducing the number of added gates and compilation runtime compared to heuristic algorithms. ComPAT~\citep{94_compat2024} introduces a LLM-based teaching assistant for compiler principles courses. HPC-GPT~\citep{149_hpcgpt2023_scw} finetunes a LLM with automatically generated, domain-specific data to improve performance on High-Performance Computing tasks like data race detection and AI model management. Fair~\citep{122_fair2024_icse} proposes a pre-trained model for IR that uses a novel flow-type-aware graph input, a Graph Transformer architecture, and five specific pre-training tasks to improve LLM's ability to understand IR.

There is a growing interest in integrating deep learning models into binary analysis, a domain critical for reverse engineering, vulnerability detection, and software supply chain security. This research primarily focuses on two key tasks: binary similarity detection and compiler provenance identification (CPI).

In binary similarity detection, the goal is to determine if two binary functions are semantically equivalent despite variations introduced by different compilers or optimization settings. Foundational work like OSCAR~\citep{170_how2021_icml} established a pre-training paradigm to learn code representations from LLVM IR, using contrastive learning to handle diverse optimizations. Subsequent approaches have refined this concept. JTrans~\citep{160_jtrans2022_issta}, for example, introduces a jump-aware Transformer model to better capture control flow, while DiEmph~\citep{137_2023_issta} improves robustness by de-emphasizing compiler-induced noise in the binary code. Similarly, OPTango~\citep{141_optango2023_issre} utilizes multi-central representation learning to create a binary diffing tool resilient to the complex effects of compiler optimizations.

A specialized application of this is Compiler Provenance Identification (CPI), which aims to identify the compiler and optimization level used to generate a given binary. Here, researchers have explored various neural architectures. For instance, MuCPI~\citep{117_2024} enhances a Gated Graph Neural Network (GGNN) with attention mechanisms to learn features from a binary's control flow graph. In a more unconventional approach, ObfuscateCPI~\citep{118_2024} demonstrates a resilient method by converting binaries into images and applying pre-trained computer vision models to identify the compiler's visual fingerprint, even through obfuscation.

\begin{table*}
    \centering
    \begin{tabular}{|c|c|c|c|}
    \hline
    \textbf{Acronym} & \textbf{Citation} & \textbf{Tasks} & \textbf{Code level} \\
    \hline
    LIBRO & \citet{150_kang2023large} & CVE test Generation & Defects4J \\
    SecurityTestGen & \citet{144_zhang2023doesllmgeneratesecurity} & CVE test Generation & Java CVE test \\
    GeneticImprove & \citet{147_brownlee2023enhancing} & Compiler Fuzzing & Java \\
    BenchDirect & \citet{139_tsimpourlas2023benchdirectdirectedlanguagemodel} & Compiler Fuzzing & OpenCL \\
    WhiteFox & \citet{31_yang2024whitefox} & Compiler Fuzzing & PT-Inductor/XLA/TF-Lite \\
    MetaMut & \citet{76_ou2024mutators_asplos} & Compiler Fuzzing & C/C++ \\
    ClozeMaster & \citet{53_gao2025clozemaster} & Compiler Fuzzing & Rust \\
    FMCSO & \citet{14_italiano2025finding} & Compiler Fuzzing & C/C++ \\

    CompileAgent & \citet{8_hu2025compileagent_acl} & Auto Configuration & Cmake/Make \\
    DCC & \citet{80_dcc2024} & Error Explanation & Compiler Log \\
    Quantum & \citet{89_quantum2024_dac} & Qubit Routing & Quantum Program \\
    ComPAT & \citet{94_compat2024} & Course Assistant & Compiler Textbook \\
    OSCAR & \citet{170_how2021_icml} & Binary Detection & LLVM IR \\
    JTrans & \citet{160_jtrans2022_issta} & Binary Detection & Assembly \\
    DiEmph & \citet{137_2023_issta} & Binary Detection & Assembly \\
    OPTango & \citet{141_optango2023_issre} & Binary Detection & Assembly \\
    MulCPI & \citet{117_2024} & Binary Detection & Assembly \\
    ObfuscateCPI & \citet{118_2024} & Binary Detection & Assembly \\
    
    HPC-GPT & \citet{149_hpcgpt2023_scw} & HPCknowledge Pretrain & HPC knowledge \\
    Fair & \citet{122_fair2024_icse} & IR Understanding & LLVM IR \\
    \hline
    \end{tabular}
    \caption{Summary of LLM for compiler utilities}
    \label{tab:utility}
\end{table*}

In the end of this section, we could finally answer \textbf{RQ2} based on recent LLM compiler studies.

\begin{tcolorbox}[colback=white,colframe=black]
    \textbf{RQ2}: What are the primary compiler-related tasks addressed by LLMs?

    \textbf{Answer}: \textbf{Generative tasks}, such as code generation, code transpilation, code repair, code optimization and decompilation are now the primary tasks addressed by LLM-enabled compilers, which have made significant advancements with more powerful LLMs and more complete system design, and are now surpassing traditional methods in one or more dimensions. \textbf{Utility tasks} such as system autoconfiguration, similarity detection and compiler fuzzing are also significantly improved with LLMs.

\end{tcolorbox}

However, on the narrowly defined compilation tasks, although there are studies on the end-to-end compilation process~\citep{39_armengol-estape2021learning, 70_c-x86-emnlp24,71_gao-etal-2024-virtual,5_legocompiler-arxiv25}, part of the compilation process~\citep{85_jin2025verilocc, 152_mannarswamy2022learningcombineinstructionsllvm}, and the IR optimization process~\citep{11_deng2024compilerdream_arxiv, 27_grubisic2024priority_arxiv,51_decos2025_ics,metallmcompiler-cc25}, they are more or less preliminary and mostly cannot surpass traditional compilers in either performance, cost or scalability. Some works~\citep{88_comback2024_neurips,44_vega2025_cgo} try to address the compiler construction problem in an ``agile-development'' view, they still require significant compiler experts intervention and cannot generalize to arbitrary compiler development tasks due to the lack of dataset. 

Nevertheless, with increasing number of studies on this field year after year, and the advancements made. We believe LLM-powered compiler system is a promising research direction. 

\section{Benchmarks \& State-of-the-Art Evolution}
\label{benchmark}

\subsection{Benchmark, Metrics \& Scale}
\label{benchmark_metrics_scale}

\begin{table*}[t]
  \centering
  \begin{tabularx}{\textwidth}{|l|l|X|r|}
    \hline
    \textbf{Benchmark/Dataset} & \textbf{Languages} & \textbf{Primary Task} & \textbf{Size (Units)} \\
    \hline
    CodeNet~\citep{codenet} & Multi-lingual (55) & Code Understanding, Transpilation & 13.9M \\
    \hline
    AVATAR~\citep{avatar} & Java, C++ & Code Transpilation & \(\sim\)9.5k \\
    \hline
    TransCoder-Test~\citep{177_transcoder} & C++, Java, Py & Code Transpilation & 1.4k \\
    \hline
    PIE~\citep{pie} & Python, Java & Code Transpilation, Edit & \(\sim\)77k \\
    \hline
    MiBench~\citep{mibench} & C & Embedded Compilation, Optimization & 35 \\
    \hline
    KernelBench~\citep{kernelbench_2025_icml} & CUDA, PyTorch & GPU Kernel Generation / Optimization & 250 \\
    \hline
    TritonBench~\citep{tritonbench_2025_acl} & Triton DSL & GPU Kernel Generation / Optimization & 184 \\
    \hline
    AnghaBench~\citep{anghabench} & C, ASM & Neural Compilation & \(\sim\)1M \\
    \hline
    ExeBench~\citep{exebench_2022_maps} & C, ASM & Neural Compilation & 0.7M \\
    \hline
    HumanEval~\citep{humaneval} & Python & Code Generation & 164 \\
    \hline
    MBPP~\citep{mbpp} & Python & Code Generation & 974 \\
    \hline
    Defects4J~\citep{defects4j} & Java & Code Repair & 854 \\
    \hline
  \end{tabularx}
  \caption{
    Summary of commonly used benchmarks/datasets for LLM-Compiler tasks. 
    The ``Size (Units)'' column is simplified to show the approximate number of core entities to refelect the evaluation scale, 
    e.g., (M)illions or (k)thousands of the major problems, functions, or code samples.}
  \label{tab:benchmarks-summary}
\end{table*}

A critical aspect of evaluating this field is to systematically analyze the benchmarks themselves. \autoref{tab:benchmarks-summary} provides a broad summary of the benchmarks and datasets used across various LLM-Compiler tasks. As shown, we can analyze these benchmarks along three recurring dimensions:

\textbf{Benchmark construction} generally follows three patterns: (1) \textit{adapting large public benchmarks/datasets} (e.g., CodeNet, AVATAR), often with new filters or metadata. We summarize these commonly used benchmarks/datasets in \autoref{tab:benchmarks-summary}.

For general understanding and transpilation, \textbf{CodeNet}~\citep{codenet} provides a large-scale dataset of nearly 14 million code samples spanning more than 55 languages and accompanied by rich metadata. \textbf{AVATAR}~\citep{avatar} builds on CodeNet by curating 9,515 problems and deriving 3,391 parallel function pairs for fine-grained alignment. \textbf{TransCoder-Test} offers a standardized parallel set used since~\citet{177_transcoder} (948 test instances) and remains a staple for translation accuracy. \textbf{PIE (Performance-Improving Edits)}~\citep{pie} is a performance-optimization benchmark built from roughly 77k submission pairs (mostly sourced from CodeNet) and evaluates runtime improvements under the gem5 simulator, making it a widely adopted testbed for optimization-oriented LLMs.

For compilation and generation tasks, \textbf{MiBench}~\citep{mibench} contributes 35 domain-realistic embedded C programs used to stress end-to-end compilation. \textbf{KernelBench}~\citep{kernelbench_2025_icml} focuses on GPU kernel generation, packaging 250 tasks with correctness and speedup metrics. \textbf{TritonBench}~\citep{tritonbench_2025_acl} similarly focuses on Triton-based GPU kernel evaluation. For neural compilation, \textbf{AnghaBench}~\citep{anghabench} contains 1M compilable C functions, while \textbf{ExeBench}~\citep{exebench_2022_maps} scales this to 680K executable C functions with a 40K unittest-based verifiable split. For code generation, \textbf{HumanEval}~\citep{humaneval} (164 Python problems) is the de facto standard for \textit{pass@k} evaluation, and \textbf{MBPP}~\citep{mbpp} (974 Python tasks) targets everyday programming competence.

The second pattern, (2) \textit{curating real open-source repositories}, aims to reflect end-to-end realism. For example, \textbf{AlphaTrans}~\citep{91_alphatrans2025_fse} decomposes ten real-world repositories to assess runtime behavior at repository scale. Similarly, \textbf{Oxidizer}~\citep{60_zhang2025_pldi} curates several real-world Go projects to evaluate Go-to-Rust transpilation. The well-known \textbf{Defects4J}~\citep{defects4j} benchmark, a curated collection of reproducible bugs from real Java projects, also follows this pattern and serves as a foundational benchmark for many program repair studies~\citep{133_xia2023_icse, 131_wei2023_fse}.

The third pattern, (3) \textit{synthesizing test cases}, probes capabilities on tailored downstream tasks. For example, \textbf{QiMeng-Xpiler}~\citep{10_qimengxpiler2025_osdi} assembles a kernel-level tensor-program suite to test cross-DSL transcompilation.
This pattern is also the foundation of \textbf{compiler fuzzing}, where works like \textbf{WhiteFox}~\citep{31_yang2024whitefox} and \textbf{MetaMut}~\citep{76_ou2024mutators_asplos} synthesize diverse and non-trivial test programs specifically designed to find compiler bugs.

At the level of \textbf{granularity}, most benchmarks remain function-centric, which makes side-by-side comparison tractable. For example, among the 21 transpile-related tasks listed in \autoref{tab:intra-level}, only a small but important slice moves to whole-program/repo contexts, e.g., in \textbf{Oxidizer}~\citep{60_zhang2025_pldi} and \textbf{AlphaTrans}~\citep{91_alphatrans2025_fse}, success is defined end-to-end: the code must compile, execute, and exhibit correct behavior, reflecting real deployment conditions rather than snippet-level fidelity.

On \textbf{metrics}, evaluation coalesces around two families of metrics: \textit{text-based} similarity and \textit{semantic-based} correctness across these benchmarks. Text-based metrics such as BLEU~\citep{bleu} measure n-gram overlap with references and are inexpensive and reproducible, but they neither account for code syntax/grammar nor data- and control-flow, which limits their faithfulness for programs; CodeBLEU~\citep{codebleu} explicitly amends this by combining standard n-gram overlap with keyword-weighted n-grams, AST (syntax) matching, and data-flow (semantics) matching to better correlate with expert judgments on code tasks. 

Semantics-centric metrics execute or analyze the artifact: pass@k estimates the probability that at least one of k generated candidates passes all unit tests (now standard for HumanEval-style synthesis); Computational Accuracy (a.k.a. functional-equivalence accuracy) runs reference tests on translated code and reports the fraction that compile and produce correct outputs; and optimization-focused works report SpeedUp and Percent Optimized (share of problems where the new code exceeds a fixed improvement threshold, e.g., \>10\% faster). Beyond these shared metrics, task-specific measures capture domain goals: \textbf{SALLM}~\citep{146_sallm_asew} augments pass@k with secure@k and vulnerable@k to quantify the security of generated samples. \textbf{CompilerDream}~\citep{11_deng2024compilerdream_arxiv} evaluates end-to-end code-optimization by reporting code size reduction (e.g., IR instruction count) relative to compiler baselines such as LLVM -Oz, aligning the metric with embedded deployment objectives. \textbf{CoTran}~\citep{38_jana2024cotran} augments correctness with error-position statistics.

\subsection{Evolution}
\label{benchmark_evolution}

\begin{figure*}[t]
    \centering
    \includegraphics[width=\textwidth]{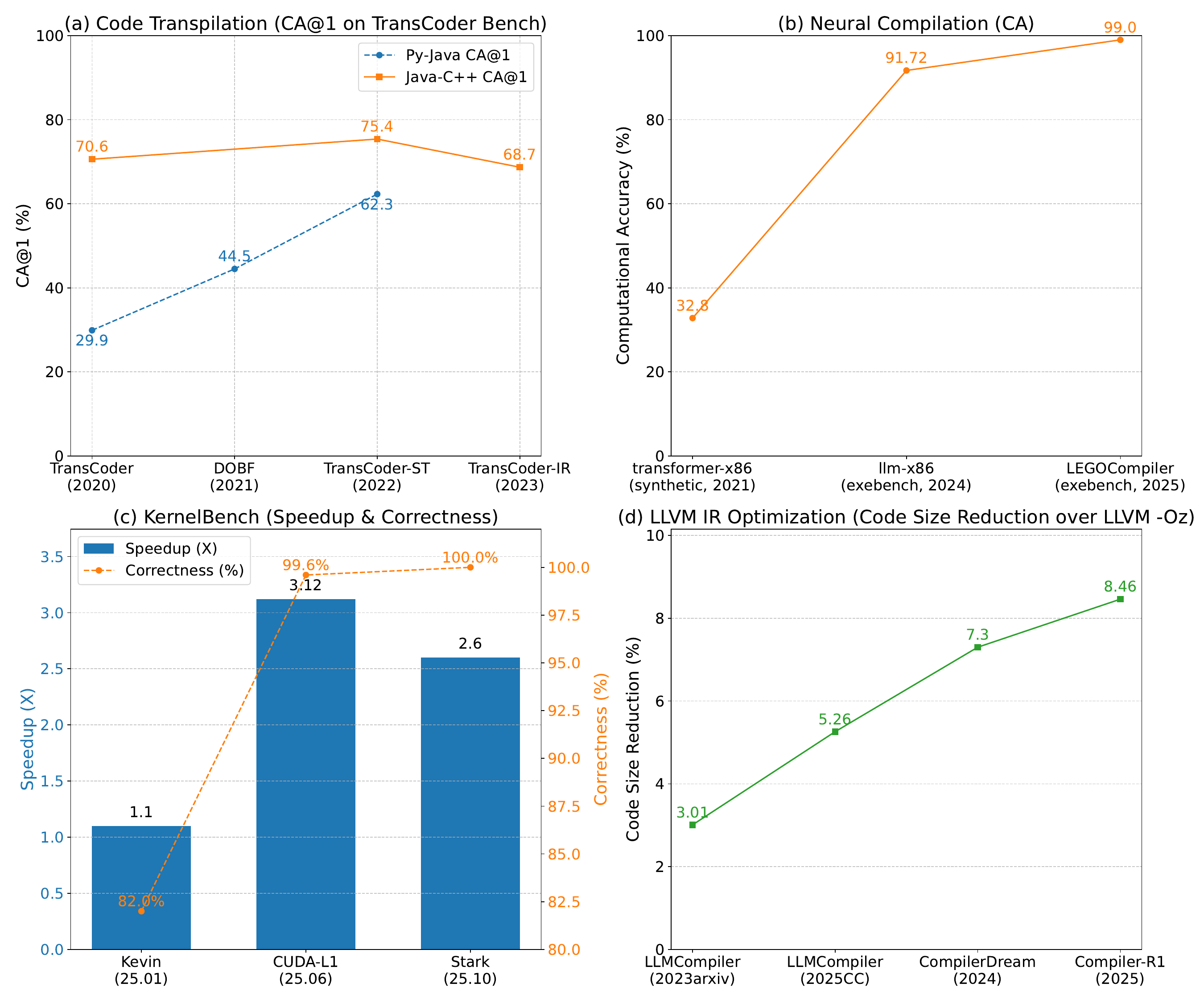}
    \caption{
        Visualization of State-of-the-Art (SOTA) progress across four key LLM-Compiler tasks: 
        \textbf{(a)} Code Transpilation (CA@1 on TransCoder), 
        \textbf{(b)} Neural Compilation (CA on ExeBench), 
        \textbf{(c)} GPU Kernel Generation (Speedup \& Correctness on KernelBench), and 
        \textbf{(d)} LLVM IR Optimization (Code Size Reduction).
        }
    \label{fig:sota_progress}
\end{figure*}

While \autoref{benchmark_metrics_scale} provided a broad overview of the evaluation landscape, this section provides a deeper, chronological analysis of SOTA evolution for four representative tasks. This allows us to track concrete technical progress made in recent years.

We have selected four tasks to illustrate this evolution: \textbf{Code Transpilation}, \textbf{Neural Compilation}, \textbf{GPU Kernel Generation}, and \textbf{LLVM IR Optimization}.

We will analyze the key benchmarks and metrics for each, tracking the chronological progress of state-of-the-art results, which are visually summarized in~\autoref{fig:sota_progress}. It is crucial to add a caveat: \textbf{direct comparisons are not always perfectly fair}. Different studies often use non-uniform experimental setups, with different dataset splits, baseline compiler versions, or testing conditions. Nonetheless, this analysis clearly demonstrates the rapid and significant technical progress in the field.

    \textbf{Code Transpilation}:
    The TransCoder-series of works established a foundational benchmark for source-to-source translation, focusing on programming language transpilations, e.g, Python, Java, and C++. While early metrics included BLEU and Exact Match (EM), Computational Accuracy (CA@1)—whether the translated code passes a set of unit tests—emerged as the most meaningful and enduring metric. As illustrated in~\autoref{fig:sota_progress}(a), the original \textbf{TransCoder}~\citep{177_transcoder} achieved CA@1 scores such as 29.9\% (Py-Java) and 70.6\% (Java-C++). Subsequent works, like \textbf{DOBF}~\citep{176_dobf} which focused on deobfuscation, improved on these (e.g., 44.5\% on Py-Java). A significant leap came with \textbf{TransCoder-ST}~\citep{175_transcoderst}, which introduced test-based filtering and improved data curation to boost CA@1 to 62.3\% (Py-Java) and 75.4\% (Java-C++). \textbf{TransCoder-IR}~\citep{154_transcoderir2023_iclr} explored a novel compiler-IR-based approach, achieving a comparable 68.7\% (Java-C++), demonstrating robustness across more compiler backends.

    \textbf{Neural Compilation}
    The goal of end-to-end neural compilation (e.g., C-to-x86) is extremely challenging. Progress here is clearly marked by a shift from superficial metrics to rigorous, execution-based benchmarks, as depicted in~\autoref{fig:sota_progress}(b). Initial work by~\citet{39_armengol-estape2021learning} trains a standard transformer on AnghaBench for testing the neural compilation capability, reporting high BLEU (90.2) and Syntax (98.5) scores. However, these metrics failed to capture semantic correctness; on a small 64-item synthetic benchmark, which has similar benchmark difficulty compared to ExeBench (not published yet), they achieved only 32.8\% CA. 
    A major advancement came from~\citet{70_c-x86-emnlp24}, who fine-tuned a large language model CodeLlama-13B with extensive data augmentation. Evaluated on the rigorous ExeBench (developed from AnghaBench, with over 17k verifiable testcases), they achieved a remarkable 91.72\% CA. Most recently, LEGO-Compiler~\citep{5_legocompiler-arxiv25} employed a training-free, divide-and-conquer methodology, with simplified step-by-step translation and basicblock-level code complexity reduction, it further pushes the CA on ExeBench to over 99\% across multiple models.

    \textbf{GPU Kernel Generation}
    For GPU kernel generation, the evaluation standard shifts from mere semantic correctness to achieving \textbf{performance speedup} over strong baselines like \texttt{torch eager}. The primary benchmark in this area is \textbf{KernelBench}~\citep{kernelbench_2025_icml}, which is used by several key studies, as visualized in~\autoref{fig:sota_progress}(c).
    However, direct comparison remains challenging because \textbf{different works may evaluate on different GPU hardware} (e.g., A100 vs. L40), which can significantly affect speedup results. Therefore, the following progression should be viewed as a demonstration of the trend rather than a direct, fair comparison.
    Early work (\textbf{Kevin}~\citep{kevin2025arxiv}) highlighted the correctness challenge, achieving only \textbf{82\%} correctness with a minimal \textbf{1.10x} average speedup over \texttt{torch eager}. A significant advancement came with \textbf{Stark}~\citep{stark2025arxiv}, which first achieved \textbf{100\% correctness} across all tasks, and delivered substantial speedups ranging from \textbf{1.58x} to \textbf{3.03x} on different benchmark levels, with an average \textbf{2.60x}. Another well-known work \textbf{CUDA-L1}~\citep{cudal1_2025_arxiv} employed both SFT and reinforcement learning (GRPO), achieving near 100\% correctness (249/250) while pushing the SOTA average speedup to \textbf{3.12x} over \texttt{torch eager}, a result that is further supported by its open-sourced evaluation results across multiple GPUs.

    \textbf{LLVM IR Optimization}
    For LLVM IR optimization, progress is often tracked by code size reduction relative to the compiler's most aggressive size-optimization flag (e.g., -Oz). As shown in~\autoref{fig:sota_progress}(d), this area highlights the challenge of non-standardized benchmarks. Different works use different datasets and baseline compiler versions, making direct comparisons difficult. Nonetheless, a clear trend of improvement is visible. 
    The initial Meta-LLMCompiler work~\citep{metallmcompiler-arxiv23} achieved a 3.01\% reduction over LLVM -Oz via large-scale pre-training.
    The subsequent Meta-LLMCompiler~\citep{metallmcompiler-cc25} improved this to 5.26\% by adding multi-step post-training to adapt the model for subtasks like compiler behavior emulation and decompilation. 
    Similarly, CompilerDream~\citep{11_deng2024compilerdream_arxiv} build their own compiler world model with reward smoothing technique, achieves 7.3\% over LLVM -Oz with guided search, which leads the CompilerGym~\citep{compilergym} leaderboard.
    A different, agentic approach with LLM-as-Selector philosophy, Compiler-R1~\citep{45_compilerr1-2025_arxiv}, used GRPO (RL) for compiler flag tuning and reported an 8.46\% reduction over LLVM -Oz.

In summary, this section has critically examined the evaluation methodologies and benchmark-driven progress at the \textbf{intersection of LLMs and compilers}. Our analysis, spanning both a broad survey of the landscape (\autoref{benchmark_metrics_scale}) and a deep dive into SOTA evolution (\autoref{benchmark_evolution}), reveals two key insights. 
\textbf{First}, there is clear and rapid technical progress across diverse tasks, from achieving over 99\% correctness in neural compilation to delivering significant speedups in GPU kernel generation. 
\textbf{Second}, this progress is mirrored by a maturation in evaluation, with the community decisively shifting away from superficial text-based metrics (like BLEU) toward rigorous, execution-based metrics (like Computational Accuracy and performance speedup). 
However, this analysis also highlights significant remaining challenges. As seen in the GPU and IR optimization tasks, there is a \textbf{lack of standardized comparison protocols} (e.g., fixed hardware, baselines, and datasets), which hinders fair, direct evaluation. Furthermore, as our analysis in \autoref{benchmark_metrics_scale} and \autoref{tab:benchmarks-summary} indicates, the vast majority of benchmarks remain \textbf{function-level}. Large-scale, \textbf{project-level benchmarks} with unified standards are still rare, making it difficult to properly evaluate scalability—a critical factor for real-world compiler systems.

\section{Discussions}\label{sec:discussion}

Having categorized the existing body of work according to design philosophy (\autoref{dimension1}) and level of code abstraction (\autoref{dimension2}), we now turn to a broader discussion of the field. This chapter synthesizes our findings to address the final two research questions of this survey. We begin by summarizing the primary advancements offered by LLM-based approaches (\textbf{RQ3}). We then delve into the common challenges and corresponding future opportunities that define the research frontier (\textbf{RQ4}). Finally, we explore several additional topics that are critical for the healthy evolution of this domain.

\subsection{Primary Advancements of LLM-based Approaches (\textbf{RQ3})}

The integration of LLMs into the compilation process has catalyzed several fundamental advancements, moving beyond the capabilities of both traditional handcrafted compilers and earlier machine learning techniques. These advancements primarily stem from the models' ability to learn deep semantic and structural patterns directly from vast corpora of source code.

\begin{itemize}
    \item \textbf{Democratizing Compiler Code Development and Optimization}: LLMs significantly lower the barrier to entry for creating sophisticated code transformation and optimization tools. Instead of requiring years of specialized expertise to design and implement complex compiler/transpiler heuristics, developers can now achieve impressive results by fine-tuning pre-trained models or applying large foundation models. This capability stems from the models being pre-trained on vast code corpora, much more than any code expert can read in its entire life, allowing them to internalize a wide array of programming patterns and techniques far beyond the scope of any single human expert in its weights. Consequently, this accelerates the development of new optimizers and makes bespoke, high-performance compilation accessible to a wider audience.
    \item \textbf{Discovering Novel Optimization Strategies}: For decades, code optimization has been guided by human-designed heuristics. LLMs, with their ability to learn from enormous datasets of real-world code, can identify complex patterns and discover novel optimization strategies that may be non-obvious to human experts. By exploring vast optimization spaces and generalizing from successful examples seen during pre-training, LLMs have the potential to surpass the performance ceilings of existing heuristic-based systems.
    \item \textbf{Broadening the Scope and Utility of ``Compilation''}: The application of LLMs has expanded the traditional definition of a compiler's role. Tasks such as large-scale Code Transpilation and Automated Program Repair are now treated as viable compilation problems. This broadened utility positions the compiler not just as a static tool for translation and optimization, but as a versatile platform for ongoing code maintenance, migration, and modernization, providing enormous value across the entire software lifecycle.
\end{itemize}

\begin{tcolorbox}[colback=white,colframe=black]
    \textbf{RQ3}: What are the primary advancements offered by LLM-based approaches?

    \textbf{Answer}: The primary advancements are threefold: (1) They \textbf{democratize} compiler code development by lowering the required expertise and accelerating implementation; (2) they can discover \textbf{novel optimization strategies} beyond human-designed heuristics by learning from vast codebases; and (3) they \textbf{broaden the utility} of compilers, turning them into versatile tools for tasks like code transpilation and repair.
\end{tcolorbox}

\subsection{Common Challenges and Future Opportunities (\textbf{RQ4})}

Despite the rapid progress, the field faces significant challenges that must be addressed to move from academic research to production-grade, reliable tools. These challenges, in turn, highlight promising avenues for future research.

\subsubsection{Common Challenges and Task-Specific Approaches}

Despite the rapid progress, the field faces significant challenges to bridge the gap to production-grade, reliable tools. Addressing them requires moving beyond isolated, function-level studies. However, as we will discuss, the proposed solutions are \textbf{not universal paradigms} but rather \textbf{highly task-specific approaches}. The fundamental gap between current LLM capabilities and the robustness of traditional compilers remains significant.

\textbf{Ensuring Correctness and Verifiability}
This remains the most critical challenge. Traditional compilers must guarantee semantic equivalence, but the probabilistic nature of LLMs can introduce subtle bugs. Approaches to mitigate this are emerging.

One strategy is \textbf{constraining the generation process} itself~\citep{typeconstrainedgeneration2025pldi}. Methods like grammar-guided constrained decoding~\citep{outlines2023arxiv,xgrammar2024arxiv} can force the LLM to produce outputs that adhere to the given grammar constraints, which are therefore \textit{syntactically} valid, guaranteeing compilability. However, this only addresses syntax, not semantic correctness, which remains the harder challenge.

A more dominant strategy is the \textbf{``Translator + Verifier'' hybrid pattern}, where the LLM's generative output is checked by a deterministic component after generation. This pattern primarily manifests in two forms. The most common form is \textbf{dynamic functional validation} via test suites. This is often placed within an iterative feedback loop~\citep{123_ye2024problem, 173_wong2025decllm}, which serves a similar purpose to constrained decoding by iteratively correcting errors, albeit at a higher cost.

The second, more rigorous form employs \textbf{formal verification} in a task-specific manner: \textbf{LLMLift}~\citep{81_bhatia2024_neurips} uses Floyd-Hoare Logic, \textbf{Qimeng-Xpiler}~\citep{10_qimengxpiler2025_osdi} uses SMT-solvers for key transformations, and \textbf{LLMVectorizer}~\citep{47_llmvectorizer2025_cgo} formally verifies generated SIMD intrinsics using the Alive2 validator. While combining these strategies is promising, applying rigorous formal verification at scale for all transformations remains a costly and open problem.

\textbf{Scalability to Large, Real-World Codebases}
This is a critical barrier where the very definition of ``scalability'' is task-dependent. For tasks like \textit{GPU kernel optimization}, e.g., CUDA-L1~\citep{cudal1_2025_arxiv} and KernelBench~\citep{kernelbench_2025_icml}, the scope is naturally constrained to a single function/kernel, making repository-level scalability a non-issue.

However, for \textit{legacy code transpilation}, the challenge becomes managing inter-procedural context across the codebase. Here, divide-and-conquer strategies are employed, such as project partitioning in \textbf{AlphaTrans}~\citep{91_alphatrans2025_fse} or type-checking feature mapping in \textbf{Oxidizer}~\citep{60_zhang2025_pldi}. For project-level \textit{repair}, \textbf{CoCoGen}~\citep{110_bi-etal-2024-iterative} relies on RAG-based context retrieval.

A more fundamental challenge lies in a \textbf{gap in design assumptions}. LLMs are pre-trained on vast public codebases that, for the most part, adhere to ``Clean Code'' principles or the Single Responsibility Principle (SRP). This biases their training data toward short, focused functions. \textbf{Real compilers, however, cannot make this assumption}; they must robustly handle any syntactically valid code, including massive, monolithic functions (``God functions'') that violate these human-centric principles. This creates a critical gap for any LLM aiming to replace or complement core compiler components.

\textbf{LEGO-Compiler}~\citep{5_legocompiler-arxiv25} is a notable work that addresses this \textit{compiler-centric} problem by decomposing functions into finer-grained basic blocks or statements. This semantic-preserving decomposition, as a further supplement to the task-dependent nature of scalability, is itself only effective for non-optimization scenarios or translations with only local, intra-unit optimizations; how to scale this decomposition strategy to global optimization scenarios remains unclear and challenging.

Ultimately, even with these decomposition strategies, the finite context window of LLMs remains the hard bottleneck. Real-world codebases like the LLVM or Linux Kernel projects, with their sheer size and complexity, are orders of magnitude beyond the scalability of current LLM-based approaches.

\textbf{Interpretability and Debuggability}
The ``black box'' nature of LLMs makes their decisions difficult to trust. This challenge is less mature, but specific strategies offer paths forward. One approach is to force the model to ``show its work'' using Chain-of-Thought (CoT) prompting, as seen in \textbf{CodeOptCoT}~\citep{116_xu2024code} and \textbf{SBLLM}~\citep{17_gao2025search}.

Another strategy focuses on debugging and explanation: \textbf{CompilerGPT}~\citep{3_compilergpt2025_arxiv} analyzes compiler reports, while \textbf{DCC}~\citep{80_dcc2024} generates novice-friendly explanations for errors. Perhaps the most robust solution is the \textbf{LLM-as-Generator} philosophy (\autoref{llm_as_generator}), exemplified by \textbf{CodeTransform}~\citep{178_cummins2024donttransformcodecode}, where the generated artifact (a transformation script) is itself human-readable and debuggable.

\textbf{Performance and Cost-Effectiveness}
The inference cost of large models can be substantial. For an LLM-based optimizer to be practical, the performance gains it provides must outweigh the computational cost and latency of its own execution. Striking the right balance between model size, inference speed, and optimization quality is an ongoing challenge.

Strategies to manage this cost-benefit trade-off are emerging. One strategy focuses on \textit{reducing the LLM's own cost} by using smaller, specialized models; \textbf{Perfcodegen}~\citep{52_peng2025perfcodegen} and \textbf{PerfRL}~\citep{134_duan2025perfrlsmalllanguagemodel} both show that small models trained with execution feedback can achieve strong performance, avoiding the expense of large-scale models.

A different strategy seeks to \textit{offset the LLM's cost} by using it as an accelerator for existing, expensive processes. In auto-tuning, for example, \textbf{ReasoningCompiler}~\citep{4_tang2025compiler_arxiv} and \textbf{TLM}~\citep{167_zhai2024_osdi} use an LLM to intelligently guide a search. In this context, the LLM's inference cost is negligible compared to the hours of compilation and benchmarking time it saves.

\subsubsection{Future Opportunities}

\begin{itemize}
    \item \textbf{Hybrid Compiler Systems}: The most promising near-term future lies not in replacing traditional compilers, but in augmenting them. Hybrid systems that combine the creative pattern-matching of LLMs with the rigor and speed of formal, deterministic compiler algorithms could achieve the best of both worlds. This can be realized through the \textbf{Selector} or \textbf{Generator} philosophies, but it also applies powerfully to the \textbf{Translator} model. For instance, a system could delegate the bulk of code compilation to a fast and reliable traditional compiler, while invoking an LLM to handle specific, challenging portions that the compiler cannot. This allows the system to support tasks it was not originally designed for, such as compiling projects with mixed-language codebases or extending support to new and emerging hardware architectures for which a mature backend does not yet exist.
    \item \textbf{Self-Improving and ``Learning'' Compilers}: A significant opportunity lies in creating compilers that learn and evolve over time by transforming the creative, non-deterministic discoveries of LLMs into permanent, deterministic compiler capabilities. This could be realized through a multi-stage process: 
    \begin{itemize}
        \item First, an \textbf{LLM as a Translator} could be used in an exploratory capacity to discover novel, ad-hoc optimizations for specific code snippets that traditional heuristics miss.
        \item Next, these successful and verified transformations would be collected into a specialized dataset of high-quality optimization examples.
        \item Finally, this dataset would be used to task an \textbf{LLM as a Generator} with a more ambitious goal: not just to perform another one-off translation, but to write the source code for a new, deterministic compiler pass or component that systematically implements the discovered optimization strategy.
    \end{itemize}
    This newly generated component can then be validated and integrated into the traditional compiler framework. The result is a compiler that has permanently ``learned'' a new skill, effectively creating a powerful paradigm for compiler evolution.
    \item \textbf{A New Generation of Interactive Developer Tools}: LLMs can transform how developers interact with compilers. We can imagine future IDEs where an LLM-powered compiler agent not only optimizes code but also explains performance bottlenecks in natural language, suggests complex refactorings, and interactively works with the developer to improve their code.
\end{itemize}

Based on the detailed discussion of the key obstacles and the corresponding research avenues, we can now synthesize these findings to provide a concise answer to our fourth research question (\textbf{RQ4}).

\begin{tcolorbox}[colback=white,colframe=black]
    \textbf{RQ4}: What are the common challenges and future opportunities in this emerging field?

    \textbf{Answer}:

    \textbf{Challenges}: Correctness \& Verifiability, Scalability, Interpretability, and Performance Cost.
    
    \textbf{Opportunities}: Hybrid Systems, Self-Improving Compilers, and Interactive Developer Tools.
\end{tcolorbox}

\subsubsection{Further Discussion Points}
\begin{itemize}
    \item \textbf{The Need for Standardized Benchmarks:} The field's progress is hampered by the lack of benchmarks designed for the unique challenges of LLM-based compilers. While several benchmarks have emerged for specific downstream tasks, such as ExeBench~\citep{exebench_2022_maps}, TritonBench~\citep{tritonbench_2025_acl}, and VerilogEval~\citep{verilogeval_2023_iccad}, they often fall short in adequately covering the dimensions of complexity and, most critically, scalability. This gap is a significant obstacle for evaluating LLM-based translators and optimizers on realistic, large-scale applications. Conversely, while traditional suites like SPEC~\citep{spec_2006} possess the required scale and complexity, they are not designed for LLM-based workflows and their end-to-end difficulty can be prohibitive for current models. This suggests a crucial need for a new class of ``LLM-friendly'' benchmarks designed for hybrid evaluation, where external processes handle boilerplate code, allowing the benchmark to focus specifically on evaluating the LLM's core capability in translating or optimizing the most critical sections of a program.
    \item \textbf{Synergy Between PL/Compiler and ML/LLM Communities:} Meaningful progress requires deep, symbiotic collaboration. The ML community can build more powerful and code-aware models, but the Programming Language (PL) and compiler community is essential for defining the right problems, providing domain-specific knowledge (e.g., program semantics, IR structures), curating high-quality datasets, and developing the rigorous verification techniques necessary to ensure the correctness of the final output.
    \item \textbf{The Evolving Role of the Compiler Engineer:} The rise of LLMs is poised to shift the role of the compiler engineer. The focus may move from manually writing complex, handcrafted heuristic algorithms to a new set of responsibilities. These could include curating massive code datasets for model training, designing effective prompting strategies, developing robust verification systems for LLM outputs, and analyzing the novel optimizations discovered by these models to gain new insights into program performance.
\end{itemize}

\section{Conclusion}\label{sec:conclusion}

In this survey, we presented a systematic overview of the emerging application of Large Language Models to the field of compilation, a domain traditionally governed by handcrafted heuristics. We introduced a multi-dimensional taxonomy to structure this diverse landscape, categorizing existing works by their Design Philosophy, LLM Methodology, Level of Code Abstraction, and specific Task Type. Our analysis highlights that LLMs are making significant advancements by democratizing compiler development, discovering novel optimization strategies, and broadening the compiler's utility to include complex tasks like code transpilation and repair. Despite this progress, the field also faces critical challenges in ensuring the correctness and verifiability of generated code, achieving scalability for large-scale software, and improving model interpretability. 
By systematically categorizing the state-of-the-art and synthesizing its primary advancements and challenges, this survey serves as a foundational roadmap for researchers and practitioners navigating this exciting and transformative field.

\section*{Declarations}

\textbf{Conflict of interest} On behalf of all authors, the corresponding author
states that there is no Conflict of interest.

\textbf{Funding} This work was partially supported by National R\&D Program of China (2024YFB4505603), the Jiangsu Province Key R\&D Program (Grant No. BG2024028) and National Natural Science Foundation of China (U23B2020, 62302479, 62232015).

\backmatter


\bibliography{sn-bibliography}

@article{2_HPCTransCompile2025_arxiv,
  title={HPCTransCompile: An AI Compiler Generated Dataset for High-Performance CUDA Transpilation and LLM Preliminary Exploration},
  author={Lv, Jiaqi and He, Xufeng and Liu, Yanchen and Dai, Xu and Shen, Aocheng and Li, Yinghao and Hao, Jiachen and Ding, Jianrong and Hu, Yang and Yin, Shouyi},
  journal={arXiv preprint arXiv:2506.10401},
  year={2025}
}

@article{3_compilergpt2025_arxiv,
  title={CompilerGPT: Leveraging Large Language Models for Analyzing and Acting on Compiler Optimization Reports},
  author={Pirkelbauer, Peter and Liao, Chunhua},
  journal={arXiv preprint arXiv:2506.06227},
  year={2025}
}

@article{4_tang2025compiler_arxiv,
  title={Compiler Optimization via LLM Reasoning for Efficient Model Serving},
  author={Tang, Sujun and Priebe, Christopher and Mahapatra, Rohan and Qin, Lianhui and Esmaeilzadeh, Hadi},
  journal={arXiv preprint arXiv:2506.01374},
  year={2025}
}

@misc{5_legocompiler-arxiv25,
      title={LEGO-Compiler: Enhancing Neural Compilation Through Translation Composability}, 
      author={Shuoming Zhang and Jiacheng Zhao and Chunwei Xia and Zheng Wang and Yunji Chen and Xiaobing Feng and Huimin Cui},
      year={2025},
      eprint={2505.20356},
      archivePrefix={arXiv},
      primaryClass={cs.PL},
      url={https://arxiv.org/abs/2505.20356}, 
}

@article{6_hong2025autocomp,
  title={Autocomp: LLM-Driven Code Optimization for Tensor Accelerators},
  author={Hong, Charles and Bhatia, Sahil and Cheung, Alvin and Shao, Yakun Sophia},
  journal={arXiv preprint arXiv:2505.18574},
  year={2025}
}

@inproceedings{8_hu2025compileagent_acl,
    title = "{C}ompile{A}gent: Automated Real-World Repo-Level Compilation with Tool-Integrated {LLM}-based Agent System",
    author = "Hu, Li  and
      Chen, Guoqiang  and
      Shang, Xiuwei  and
      Cheng, Shaoyin  and
      Wu, Benlong  and
      LiGangyang, LiGangyang  and
      Zhu, Xu  and
      Zhang, Weiming  and
      Yu, Nenghai",
    editor = "Che, Wanxiang  and
      Nabende, Joyce  and
      Shutova, Ekaterina  and
      Pilehvar, Mohammad Taher",
    booktitle = "Proceedings of the 63rd Annual Meeting of the Association for Computational Linguistics (Volume 1: Long Papers)",
    month = jul,
    year = "2025",
    address = "Vienna, Austria",
    publisher = "Association for Computational Linguistics",
    url = "https://aclanthology.org/2025.acl-long.103/",
    pages = "2078--2091",
    ISBN = "979-8-89176-251-0"
}

@article{9_wang2025symrtlo,
  title={SymRTLO: Enhancing RTL Code Optimization with LLMs and Neuron-Inspired Symbolic Reasoning},
  author={Wang, Yiting and Ye, Wanghao and Guo, Ping and He, Yexiao and Wang, Ziyao and Tian, Bowei and He, Shwai and Sun, Guoheng and Shen, Zheyu and Chen, Sihan and others},
  journal={arXiv preprint arXiv:2504.10369},
  year={2025}
}

@inproceedings{10_qimengxpiler2025_osdi,
  title={QiMeng-Xpiler: Transcompiling tensor programs for deep learning systems with a neural-symbolic approach},
  author={Dong, Shouyang and Wen, Yuanbo and Bi, Jun and Huang, Di and Guo, Jiaming and Xu, Jianxing and Xu, Ruibai and Song, Xinkai and Hao, Yifan and Zhou, Xuehai and others},
  booktitle={19th USENIX Symposium on Operating Systems Design and Implementation (OSDI 25)},
  year={2025}
}

@article{11_deng2024compilerdream_arxiv,
  title={CompilerDream: Learning a Compiler World Model for General Code Optimization},
  author={Deng, Chaoyi and Wu, Jialong and Feng, Ningya and Wang, Jianmin and Long, Mingsheng},
  journal={arXiv preprint arXiv:2404.16077},
  year={2024}
}

@article{12_lin2025eco,
  title={ECO: An LLM-driven efficient code optimizer for warehouse scale computers},
  author={Lin, Hannah and Maas, Martin and Roquemore, Maximilian and Hasanzadeh, Arman and Lewis, Fred and Simonson, Yusuf and Yang, Tzu-Wei and Yazdanbakhsh, Amir and Altinb{\"u}ken, Deniz and Papa, Florin and others},
  journal={arXiv preprint arXiv:2503.15669},
  year={2025}
}

@article{13_jiang2025can,
  title={Can Large Language Models Understand Intermediate Representations in Compilers?},
  author={Jiang, Hailong and Zhu, Jianfeng and Wan, Yao and Fang, Bo and Zhang, Hongyu and Jin, Ruoming and Guan, Qiang},
  journal={arXiv preprint arXiv:2502.06854},
  year={2025}
}

@inproceedings{14_italiano2025finding,
  title={Finding missed code size optimizations in compilers using large language models},
  author={Italiano, Davide and Cummins, Chris},
  booktitle={Proceedings of the 34th ACM SIGPLAN International Conference on Compiler Construction},
  pages={81--91},
  year={2025}
}

@article{15_fang2024towards,
  title={Towards LLM-based optimization compilers. Can LLMs learn how to apply a single peephole optimization? Reasoning is all LLMs need!},
  author={Fang, Xiangxin and Mukhanov, Lev},
  journal={arXiv preprint arXiv:2412.12163},
  year={2024}
}

@article{16_chen2024test,
  title={A test-free semantic mistakes localization framework in Neural Code Translation},
  author={Chen, Lei and Zhang, Sai and Xu, Fangzhou and Xing, Zhenchang and Wan, Liang and Zhang, Xiaowang and Feng, Zhiyong},
  journal={arXiv preprint arXiv:2410.22818},
  year={2024}
}

@inproceedings{17_gao2025search,
  title={Search-Based LLMs for Code Optimization},
  author={Gao, Shuzheng and Gao, Cuiyun and Gu, Wenchao and Lyu, Michael R},
  booktitle={2025 IEEE/ACM 47th International Conference on Software Engineering (ICSE)},
  pages={578--590},
  year={2025},
  organization={IEEE}
}

@article{19_albuquerque2024evaluating,
  title={Evaluating the capability of llms in identifying compilation errors in configurable systems},
  author={Albuquerque, Lucas and Gheyi, Rohit and Ribeiro, M{\'a}rcio},
  journal={arXiv preprint arXiv:2407.19087},
  year={2024}
}

@misc{20_saldyt2025algorithmiclanguagemodelsneurally,
      title={Algorithmic Language Models with Neurally Compiled Libraries}, 
      author={Lucas Saldyt and Subbarao Kambhampati},
      year={2025},
      eprint={2407.04899},
      archivePrefix={arXiv},
      primaryClass={cs.AI},
      url={https://arxiv.org/abs/2407.04899}, 
}

@inproceedings{22_romero2025should,
  title={Should AI Optimize Your Code? A Comparative Study of Classical Optimizing Compilers Versus Current Large Language Models},
  author={Romero Rosas, Miguel Angel and Torres Sanchez, Miguel Angel and Eigenmann, Rudolf},
  booktitle={Proceedings of the 2025 Supercomputing Asia Conference},
  pages={22--29},
  year={2025}
}

@article{23_xu2024aios,
  title={Aios compiler: Llm as interpreter for natural language programming and flow programming of ai agents},
  author={Xu, Shuyuan and Li, Zelong and Mei, Kai and Zhang, Yongfeng},
  journal={arXiv preprint arXiv:2405.06907},
  year={2024}
}

@article{25_sibaee2024llms,
  title={LLMs as Compiler for Arabic Programming Language},
  author={Sibaee, Serry and Najar, Omar and Ghouti, Lahouri and Koubaa, Anis},
  journal={arXiv preprint arXiv:2403.16087},
  year={2024}
}

@inproceedings{27_grubisic2024priority_arxiv,
  title={Priority sampling of large language models for compilers},
  author={Grubisic, Dejan and Seeker, Volker and Synnaeve, Gabriel and Leather, Hugh and Mellor-Crummey, John and Cummins, Chris},
  booktitle={Proceedings of the 4th Workshop on Machine Learning and Systems},
  pages={91--97},
  year={2024}
}

@article{29_kabir2025zs4c,
  title={ZS4C: Zero-Shot Synthesis of Compilable Code for Incomplete Code Snippets using LLMs},
  author={Kabir, Azmain and Wang, Shaowei and Tian, Yuan and Chen, Tse-Hsun and Asaduzzaman, Muhammad and Zhang, Wenbin},
  journal={ACM Transactions on Software Engineering and Methodology},
  volume={34},
  number={4},
  pages={1--30},
  year={2025},
  publisher={ACM New York, NY}
}

@inproceedings{
30_ishida2024langprop,
title={LangProp: A code optimization framework using Large Language Models applied to driving},
author={Shu Ishida and Gianluca Corrado and George Fedoseev and Hudson Yeo and Lloyd Russell and Jamie Shotton and Joao F. Henriques and Anthony Hu},
booktitle={ICLR 2024 Workshop on Large Language Model (LLM) Agents},
year={2024},
url={https://openreview.net/forum?id=JQJJ9PkdYC}
}

@article{31_yang2024whitefox,
  title={Whitefox: White-box compiler fuzzing empowered by large language models},
  author={Yang, Chenyuan and Deng, Yinlin and Lu, Runyu and Yao, Jiayi and Liu, Jiawei and Jabbarvand, Reyhaneh and Zhang, Lingming},
  journal={Proceedings of the ACM on Programming Languages},
  volume={8},
  number={OOPSLA2},
  pages={709--735},
  year={2024},
  publisher={ACM New York, NY, USA}
}

@inproceedings{34_jiao2023_ase,
  title={On the Evaluation of Neural Code Translation: Taxonomy and Benchmark},
  author={Jiao, Mingsheng and Yu, Tingrui and Li, Xuan and Qiu, Guanjie and Gu, Xiaodong and Shen, Beijun},
  booktitle={Proceedings of the 38th IEEE/ACM International Conference on Automated Software Engineering},
  pages={1529--1541},
  year={2023}
}

@inproceedings{38_jana2024cotran,
  title={CoTran: An LLM-Based Code Translator Using Reinforcement Learning with Feedback from Compiler and Symbolic Execution},
  author={Jana, Prithwish and Jha, Piyush and Ju, Haoyang and Kishore, Gautham and Mahajan, Aryan and Ganesh, Vijay},
  booktitle={ECAI},
  year={2024}
}

@inproceedings{39_armengol-estape2021learning,
  title={Learning C to x86 Translation: An Experiment in Neural Compilation},
  author={Jordi Armengol-Estape and Michael O'Boyle},
  booktitle={Advances in Programming Languages and Neurosymbolic Systems Workshop},
  year={2021},
}

@inproceedings{40_mammadli2020static_llvmhpc,
  title={Static neural compiler optimization via deep reinforcement learning},
  author={Mammadli, Rahim and Jannesari, Ali and Wolf, Felix},
  booktitle={2020 IEEE/ACM 6th Workshop on the LLVM Compiler Infrastructure in HPC (LLVM-HPC) and Workshop on Hierarchical Parallelism for Exascale Computing (HiPar)},
  pages={1--11},
  year={2020},
  organization={IEEE}
}

@inproceedings{44_vega2025_cgo,
author = {Zhong, Ming and Lv, Fang and Wang, Lulin and Qiu, Lei and Wang, Yingying and Liu, Ying and Cui, Huimin and Feng, Xiaobing and Xue, Jingling},
title = {VEGA: Automatically Generating Compiler Backends using a Pre-trained Transformer Model},
year = {2025},
isbn = {9798400712753},
publisher = {Association for Computing Machinery},
address = {New York, NY, USA},
url = {https://doi.org/10.1145/3696443.3708931},
doi = {10.1145/3696443.3708931},
booktitle = {Proceedings of the 23rd ACM/IEEE International Symposium on Code Generation and Optimization},
pages = {90–106},
numpages = {17},
location = {Las Vegas, NV, USA},
series = {CGO '25}
}

@article{45_compilerr1-2025_arxiv,
  title={Compiler-R1: Towards Agentic Compiler Auto-tuning with Reinforcement Learning},
  author={Pan, Haolin and Lin, Hongyu and Luo, Haoran and Liu, Yang and Yao, Kaichun and Zhang, Libo and Xing, Mingjie and Wu, Yanjun},
  journal={arXiv preprint arXiv:2506.15701},
  year={2025}
}

@article{46_wang2025_pldi,
author = {Wang, Xiangwei and Hui, Xinning and Liao, Chunhua and Shen, Xipeng},
title = {Reductive Analysis with Compiler-Guided Large Language Models for Input-Centric Code Optimizations},
year = {2025},
issue_date = {June 2025},
publisher = {Association for Computing Machinery},
address = {New York, NY, USA},
volume = {9},
number = {PLDI},
url = {https://doi.org/10.1145/3729282},
doi = {10.1145/3729282},
journal = {Proc. ACM Program. Lang.},
month = jun,
articleno = {179},
numpages = {25}
}

@inproceedings{47_llmvectorizer2025_cgo,
author = {Taneja, Jubi and Laird, Avery and Yan, Cong and Musuvathi, Madan and Lahiri, Shuvendu K.},
title = {LLM-Vectorizer: LLM-Based Verified Loop Vectorizer},
year = {2025},
isbn = {9798400712753},
publisher = {Association for Computing Machinery},
address = {New York, NY, USA},
url = {https://doi.org/10.1145/3696443.3708929},
doi = {10.1145/3696443.3708929},
booktitle = {Proceedings of the 23rd ACM/IEEE International Symposium on Code Generation and Optimization},
pages = {137–149},
numpages = {13},
location = {Las Vegas, NV, USA},
series = {CGO '25}
}

@inproceedings{51_decos2025_ics,
  author = {Cui, Tianming and Yew, Pen-Chung and McCamant, Stephen and Zhai, Antonia},
  title = {{DeCOS}: Data-Efficient Reinforcement Learning for Compiler Optimization Selection Ignited by {LLM}},
  year = {2025},
  booktitle = {Proceedings of the 2025 International Conference on Supercomputing},
  series = {ICS '25},
  publisher = {Association for Computing Machinery},
  address = {New York, NY, USA},
  location = {Salt Lake City, UT, USA},
  numpages = {16},
  doi = {10.1145/3721145.3725765},
  url = {https://doi.org/10.1145/3721145.3725765}
}

@inproceedings{52_peng2025perfcodegen,
  title={Perfcodegen: Improving performance of llm generated code with execution feedback},
  author={Peng, Yun and Gotmare, Akhilesh Deepak and Lyu, Michael R and Xiong, Caiming and Savarese, Silvio and Sahoo, Doyen},
  booktitle={2025 IEEE/ACM Second International Conference on AI Foundation Models and Software Engineering (Forge)},
  pages={1--13},
  year={2025},
  organization={IEEE}
}

@inproceedings{53_gao2025clozemaster,
  title={Clozemaster: Fuzzing rust compiler by harnessing llms for infilling masked real programs},
  author={Gao, Hongyan and Yang, Yibiao and Sun, Maolin and Wu, Jiangchang and Zhou, Yuming and Xu, Baowen},
  booktitle={2025 IEEE/ACM 47th International Conference on Software Engineering (ICSE)},
  pages={712--712},
  year={2025},
  organization={IEEE Computer Society}
}

@article{54_rong2025integrating,
  title={Integrating LLM-based code optimization with human-like exclusionary reasoning for computational education},
  author={Rong, Yi and Du, Tianfeng and Li, Roubing and Bao, Wenting},
  journal={Journal of King Saud University Computer and Information Sciences},
  volume={37},
  number={5},
  pages={87},
  year={2025},
  publisher={Springer}
}

@inproceedings{58_purschke2025speedgen,
  title={SpeedGen: Enhancing Code Efficiency through Large Language Model-Based Performance Optimization},
  author={Purschke, Nils and Kirchner, Sven and Knoll, Alois},
  booktitle={2025 IEEE International Conference on Software Analysis, Evolution and Reengineering (SANER)},
  pages={1--12},
  year={2025},
  organization={IEEE}
}

@inproceedings{59_autoiot2025_mobicom,
  title={AutoIOT: LLM-Driven Automated Natural Language Programming for AIoT Applications},
  author={Shen, Leming and Yang, Qiang and Zheng, Yuanqing and Li, Mo},
  booktitle={Mobicom 2025},
  year={2025}
}

@article{60_zhang2025_pldi,
  title={Scalable, validated code translation of entire projects using large language models},
  author={Zhang, Hanliang and David, Cristina and Wang, Meng and Paulsen, Brandon and Kroening, Daniel},
  journal={Proceedings of the ACM on Programming Languages},
  volume={9},
  number={PLDI},
  pages={1616--1641},
  year={2025},
  publisher={ACM New York, NY, USA}
}

@article{63_xu2025efficient,
  title={Efficient program optimization through knowledge-enhanced LoRA fine-tuning of large language models},
  author={Xu, Caixu and Guo, Hui and Cen, Caicun and Chen, Minglang and Tao, Xiongjie and He, Jie},
  journal={The Journal of Supercomputing},
  volume={81},
  number={8},
  pages={1006},
  year={2025},
  publisher={Springer}
}

@article{64_wang2025insights,
  title={Insights from verification: Training a verilog generation LLM with reinforcement learning with testbench feedback},
  author={Wang, Ning and Yao, Bingkun and Zhou, Jie and Hu, Yuchen and Wang, Xi and Guan, Nan and Jiang, Zhe},
  journal={arXiv preprint arXiv:2504.15804},
  year={2025}
}

@inproceedings{70_c-x86-emnlp24,
    title = "Introducing Compiler Semantics into Large Language Models as Programming Language Translators: A Case Study of {C} to x86 Assembly",
    author = "Zhang, Shuoming  and
      Zhao, Jiacheng  and
      Xia, Chunwei  and
      Wang, Zheng  and
      Chen, Yunji  and
      Cui, Huimin",
    editor = "Al-Onaizan, Yaser  and
      Bansal, Mohit  and
      Chen, Yun-Nung",
    booktitle = "Findings of the Association for Computational Linguistics: EMNLP 2024",
    month = nov,
    year = "2024",
    address = "Miami, Florida, USA",
    publisher = "Association for Computational Linguistics",
    url = "https://aclanthology.org/2024.findings-emnlp.55/",
    doi = "10.18653/v1/2024.findings-emnlp.55",
    pages = "996--1011"
}

@inproceedings{71_gao-etal-2024-virtual,
    title = "Virtual Compiler Is All You Need For Assembly Code Search",
    author = "Gao, Zeyu  and
      Wang, Hao  and
      Wang, Yuanda  and
      Zhang, Chao",
    editor = "Ku, Lun-Wei  and
      Martins, Andre  and
      Srikumar, Vivek",
    booktitle = "Proceedings of the 62nd Annual Meeting of the Association for Computational Linguistics (Volume 1: Long Papers)",
    month = aug,
    year = "2024",
    address = "Bangkok, Thailand",
    publisher = "Association for Computational Linguistics",
    url = "https://aclanthology.org/2024.acl-long.167/",
    doi = "10.18653/v1/2024.acl-long.167",
    pages = "3040--3051"
}

@inproceedings{72_armengol2024slade,
  title={Slade: A portable small language model decompiler for optimized assembly},
  author={Armengol-Estape, Jordi and Woodruff, Jackson and Cummins, Chris and O'Boyle, Michael FP},
  booktitle={2024 IEEE/ACM International Symposium on Code Generation and Optimization (CGO)},
  pages={67--80},
  year={2024},
  organization={IEEE}
}

@inproceedings{73_tan-etal-2024-llm4decompile,
    title = "{LLM}4{D}ecompile: Decompiling Binary Code with Large Language Models",
    author = "Tan, Hanzhuo  and
      Luo, Qi  and
      Li, Jing  and
      Zhang, Yuqun",
    editor = "Al-Onaizan, Yaser  and
      Bansal, Mohit  and
      Chen, Yun-Nung",
    booktitle = "Proceedings of the 2024 Conference on Empirical Methods in Natural Language Processing",
    month = nov,
    year = "2024",
    address = "Miami, Florida, USA",
    publisher = "Association for Computational Linguistics",
    url = "https://aclanthology.org/2024.emnlp-main.203/",
    doi = "10.18653/v1/2024.emnlp-main.203",
    pages = "3473--3487"
}

@inproceedings{74_palkowski2024automatic,
  title={Automatic Generation of OpenCL Code through Polyhedral Compilation with LLM},
  author={Palkowski, Marek and Gruzewski, Mateusz},
  booktitle={2024 19th Conference on Computer Science and Intelligence Systems (FedCSIS)},
  pages={671--676},
  year={2024},
  organization={IEEE}
}

@inproceedings{75_hu2024degpt,
  title={Degpt: Optimizing decompiler output with llm},
  author={Hu, Peiwei and Liang, Ruigang and Chen, Kai},
  booktitle={Proceedings 2024 Network and Distributed System Security Symposium},
  volume={267622140},
  year={2024}
}

@inproceedings{76_ou2024mutators_asplos,
  title={The mutators reloaded: Fuzzing compilers with large language model generated mutation operators},
  author={Ou, Xianfei and Li, Cong and Jiang, Yanyan and Xu, Chang},
  booktitle={Proceedings of the 29th ACM International Conference on Architectural Support for Programming Languages and Operating Systems, Volume 4},
  pages={298--312},
  year={2024}
}

@article{79_huang2024effilearner,
  title={Effilearner: Enhancing efficiency of generated code via self-optimization},
  author={Huang, Dong and Dai, Jianbo and Weng, Han and Wu, Puzhen and Qing, Yuhao and Cui, Heming and Guo, Zhijiang and Zhang, Jie},
  journal={Advances in Neural Information Processing Systems},
  volume={37},
  pages={84482--84522},
  year={2024}
}

@inproceedings{80_dcc2024,
author = {Taylor, Andrew and Vassar, Alexandra and Renzella, Jake and Pearce, Hammond},
title = {dcc --help: Transforming the Role of the Compiler by Generating Context-Aware Error Explanations with Large Language Models},
year = {2024},
isbn = {9798400704239},
publisher = {Association for Computing Machinery},
address = {New York, NY, USA},
url = {https://doi.org/10.1145/3626252.3630822},
doi = {10.1145/3626252.3630822},
booktitle = {Proceedings of the 55th ACM Technical Symposium on Computer Science Education V. 1},
pages = {1314–1320},
numpages = {7},
location = {Portland, OR, USA},
series = {SIGCSE 2024}
}

@article{81_bhatia2024_neurips,
  title={Verified code transpilation with LLMs},
  author={Bhatia, Sahil and Qiu, Jie and Hasabnis, Niranjan and Seshia, Sanjit A and Cheung, Alvin},
  journal={Advances in Neural Information Processing Systems},
  volume={37},
  pages={41394--41424},
  year={2024}
}

@article{83_shaw2025alta,
title={{ALTA}: Compiler-Based Analysis of Transformers},
author={Peter Shaw and James Cohan and Jacob Eisenstein and Kenton Lee and Jonathan Berant and Kristina Toutanova},
journal={Transactions on Machine Learning Research},
issn={2835-8856},
year={2025},
url={https://openreview.net/forum?id=h751wl9xiR},
note={}
}

@article{84_yin2024rectifier,
  title={Rectifier: Code translation with corrector via llms},
  author={Yin, Xin and Ni, Chao and Nguyen, Tien N and Wang, Shaohua and Yang, Xiaohu},
  journal={arXiv preprint arXiv:2407.07472},
  year={2024}
}

@article{85_jin2025verilocc,
  title={VeriLocc: End-to-End Cross-Architecture Register Allocation via LLM},
  author={Jin, Lesheng and Ruan, Zhenyuan and Mai, Haohui and Shang, Jingbo},
  journal={arXiv preprint arXiv:2506.17506},
  year={2025}
}

@inproceedings{88_comback2024_neurips,
title={ComBack: A Versatile Dataset for Enhancing Compiler Backend Development Efficiency},
author={Ming Zhong and FANG LYU and Lulin Wang and Hongna Geng and Lei Qiu and Huimin Cui and Xiaobing Feng},
booktitle={The Thirty-eight Conference on Neural Information Processing Systems Datasets and Benchmarks Track},
year={2024},
url={https://openreview.net/forum?id=vfju5hjrJw}
}

@inproceedings{89_quantum2024_dac,
  title={Leveraging machine learning for quantum compilation optimization},
  author={Ren, Xiangyu and Zhang, Tianyu and Xu, Xiong and Zheng, Yi-Cong and Zhang, Shengyu},
  booktitle={Proceedings of the 61st ACM/IEEE Design Automation Conference},
  pages={1--4},
  year={2024}
}

@article{91_alphatrans2025_fse,
  title={AlphaTrans: A Neuro-Symbolic Compositional Approach for Repository-Level Code Translation and Validation},
  author={Ibrahimzada, Ali Reza and Ke, Kaiyao and Pawagi, Mrigank and Abid, Muhammad Salman and Pan, Rangeet and Sinha, Saurabh and Jabbarvand, Reyhaneh},
  journal={Proceedings of the ACM on Software Engineering},
  volume={2},
  number={FSE},
  pages={2454--2476},
  year={2025},
  publisher={ACM New York, NY, USA}
}

@inproceedings{92_sun-etal-2024-unicoder,
    title = "{U}ni{C}oder: Scaling Code Large Language Model via Universal Code",
    author = "Sun, Tao  and
      Chai, Linzheng  and
      Yang, Jian  and
      Yin, Yuwei  and
      Guo, Hongcheng  and
      Liu, Jiaheng  and
      Wang, Bing  and
      Yang, Liqun  and
      Li, Zhoujun",
    editor = "Ku, Lun-Wei  and
      Martins, Andre  and
      Srikumar, Vivek",
    booktitle = "Proceedings of the 62nd Annual Meeting of the Association for Computational Linguistics (Volume 1: Long Papers)",
    month = aug,
    year = "2024",
    address = "Bangkok, Thailand",
    publisher = "Association for Computational Linguistics",
    url = "https://aclanthology.org/2024.acl-long.100/",
    doi = "10.18653/v1/2024.acl-long.100",
    pages = "1812--1824"
}

@misc{93_kadosh2024omparautomaticparallelizationaidriven,
      title={OMPar: Automatic Parallelization with AI-Driven Source-to-Source Compilation}, 
      author={Tal Kadosh and Niranjan Hasabnis and Prema Soundararajan and Vy A. Vo and Mihai Capota and Nesreen Ahmed and Yuval Pinter and Gal Oren},
      year={2024},
      eprint={2409.14771},
      archivePrefix={arXiv},
      primaryClass={cs.CL},
      url={https://arxiv.org/abs/2409.14771}, 
}

@inproceedings{94_compat2024,
  author={Shubin Cai and Honglong Chen and Youyi Huang and Zhong Ming},
  title={ComPAT: A Compiler Principles Course Assistant},
  year={2024},
  cdate={1704067200000},
  pages={74-83},
  booktitle={KSEM (5)},
}

@inproceedings{95_macedo2024exploring,
  title={Exploring the impact of the output format on the evaluation of large language models for code translation},
  author={Macedo, Marcos and Tian, Yuan and Cogo, Filipe and Adams, Bram},
  booktitle={Proceedings of the 2024 IEEE/ACM First International Conference on AI Foundation Models and Software Engineering},
  pages={57--68},
  year={2024}
}

@inproceedings{98_xiong2024hlspilot,
  title={Hlspilot: Llm-based high-level synthesis},
  author={Xiong, Chenwei and Liu, Cheng and Li, Huawei and Li, Xiaowei},
  booktitle={Proceedings of the 43rd IEEE/ACM International Conference on Computer-Aided Design},
  pages={1--9},
  year={2024}
}

@inproceedings{99_pan2024_ICSE,
  title={Lost in translation: A study of bugs introduced by large language models while translating code},
  author={Pan, Rangeet and Ibrahimzada, Ali Reza and Krishna, Rahul and Sankar, Divya and Wassi, Lambert Pouguem and Merler, Michele and Sobolev, Boris and Pavuluri, Raju and Sinha, Saurabh and Jabbarvand, Reyhaneh},
  booktitle={Proceedings of the IEEE/ACM 46th International Conference on Software Engineering},
  pages={1--13},
  year={2024}
}

@article{104_verigen2024,
author = {Thakur, Shailja and Ahmad, Baleegh and Pearce, Hammond and Tan, Benjamin and Dolan-Gavitt, Brendan and Karri, Ramesh and Garg, Siddharth},
title = {VeriGen: A Large Language Model for Verilog Code Generation},
year = {2024},
issue_date = {May 2024},
publisher = {Association for Computing Machinery},
address = {New York, NY, USA},
volume = {29},
number = {3},
issn = {1084-4309},
url = {https://doi.org/10.1145/3643681},
doi = {10.1145/3643681},
journal = {ACM Trans. Des. Autom. Electron. Syst.},
month = apr,
articleno = {46},
numpages = {31}
}

@article{106_knowledgetransfer2024_oopsla,
author = {Cassano, Federico and Gouwar, John and Lucchetti, Francesca and Schlesinger, Claire and Freeman, Anders and Anderson, Carolyn Jane and Feldman, Molly Q and Greenberg, Michael and Jangda, Abhinav and Guha, Arjun},
title = {Knowledge Transfer from High-Resource to Low-Resource Programming Languages for Code LLMs},
year = {2024},
issue_date = {October 2024},
publisher = {Association for Computing Machinery},
address = {New York, NY, USA},
volume = {8},
number = {OOPSLA2},
url = {https://doi.org/10.1145/3689735},
doi = {10.1145/3689735},
journal = {Proc. ACM Program. Lang.},
month = oct,
articleno = {295},
numpages = {32}
}

@inproceedings{107_wang2024enhancing_issre,
  title={Enhancing Black-box Compiler Option Fuzzing with LLM through Command Feedback},
  author={Wang, Taiyan and Wang, Ruipeng and Chen, Yu and Yu, Lu and Pan, Zulie and Zhang, Min and Ma, Huimin and Zheng, Jinghua},
  booktitle={2024 IEEE 35th International Symposium on Software Reliability Engineering (ISSRE)},
  pages={319--330},
  year={2024},
  organization={IEEE}
}

@inproceedings{108_nichols2024can,
  title={Can large language models write parallel code?},
  author={Nichols, Daniel and Davis, Joshua H and Xie, Zhaojun and Rajaram, Arjun and Bhatele, Abhinav},
  booktitle={Proceedings of the 33rd International Symposium on High-Performance Parallel and Distributed Computing},
  pages={281--294},
  year={2024}
}

@inproceedings{110_bi-etal-2024-iterative,
    title = "Iterative Refinement of Project-Level Code Context for Precise Code Generation with Compiler Feedback",
    author = "Bi, Zhangqian  and
      Wan, Yao  and
      Wang, Zheng  and
      Zhang, Hongyu  and
      Guan, Batu  and
      Lu, Fangxin  and
      Zhang, Zili  and
      Sui, Yulei  and
      Jin, Hai  and
      Shi, Xuanhua",
    editor = "Ku, Lun-Wei  and
      Martins, Andre  and
      Srikumar, Vivek",
    booktitle = "Findings of the Association for Computational Linguistics: ACL 2024",
    month = aug,
    year = "2024",
    address = "Bangkok, Thailand",
    publisher = "Association for Computational Linguistics",
    url = "https://aclanthology.org/2024.findings-acl.138/",
    doi = "10.18653/v1/2024.findings-acl.138",
    pages = "2336--2353"
}

@article{111_delorenzo2024make,
  title={Make every move count: Llm-based high-quality rtl code generation using mcts},
  author={DeLorenzo, Matthew and Chowdhury, Animesh Basak and Gohil, Vasudev and Thakur, Shailja and Karri, Ramesh and Garg, Siddharth and Rajendran, Jeyavijayan},
  journal={arXiv preprint arXiv:2402.03289},
  year={2024}
}

@inproceedings{113_deligiannis2024rustassistant,
  title={RustAssistant: Using LLMs to fix compilation errors in Rust code},
  author={Deligiannis, Pantazis and Lal, Akash and Mehrotra, Nikita and Poddar, Rishi and Rastogi, Aseem},
  booktitle={2025 IEEE/ACM 47th International Conference on Software Engineering (ICSE)},
  pages={267--279},
  year={2024},
  organization={IEEE Computer Society}
}

@inproceedings{114_lu2024rtllm,
  title={Rtllm: An open-source benchmark for design rtl generation with large language model},
  author={Lu, Yao and Liu, Shang and Zhang, Qijun and Xie, Zhiyao},
  booktitle={2024 29th Asia and South Pacific Design Automation Conference (ASP-DAC)},
  pages={722--727},
  year={2024},
  organization={IEEE}
}

@inproceedings{116_xu2024code,
  title={Code optimization chain-of-thought: Structured understanding and self-checking},
  author={Xu, Qingyao and Yang, Dingkang and Zhang, Lihua},
  booktitle={Proceedings of the 2024 4th International Conference on Artificial Intelligence, Big Data and Algorithms},
  pages={425--430},
  year={2024}
}

@Article{117_2024,
AUTHOR = {Gao, Yang and Liang, Lunjin and Li, Yifei and Li, Rui and Wang, Yu},
TITLE = {Function-Level Compilation Provenance Identification with Multi-Faceted Neural Feature Distillation and Fusion},
JOURNAL = {Electronics},
VOLUME = {13},
YEAR = {2024},
NUMBER = {9},
ARTICLE-NUMBER = {1692},
URL = {https://www.mdpi.com/2079-9292/13/9/1692},
ISSN = {2079-9292},
DOI = {10.3390/electronics13091692}
}

@article{118_2024,
title = {Compiler-provenance identification in obfuscated binaries using vision transformers},
journal = {Forensic Science International: Digital Investigation},
volume = {49},
pages = {301764},
year = {2024},
note = {DFRWS USA 2024 - Selected Papers from the 24th Annual Digital Forensics Research Conference USA},
issn = {2666-2817},
doi = {https://doi.org/10.1016/j.fsidi.2024.301764},
url = {https://www.sciencedirect.com/science/article/pii/S2666281724000830},
author = {Wasif Khan and Saed Alrabaee and Mousa Al-kfairy and Jie Tang and Kim-Kwang {Raymond Choo}}
}

@inproceedings{120_nakkab2024rome,
  title={Rome was not built in a single step: Hierarchical prompting for llm-based chip design},
  author={Nakkab, Andre and Zhang, Sai Qian and Karri, Ramesh and Garg, Siddharth},
  booktitle={Proceedings of the 2024 ACM/IEEE International Symposium on Machine Learning for CAD},
  pages={1--11},
  year={2024}
}

@inproceedings{122_fair2024_icse,
author = {Niu, Changan and Li, Chuanyi and Ng, Vincent and Lo, David and Luo, Bin},
title = {FAIR: Flow Type-Aware Pre-Training of Compiler Intermediate Representations},
year = {2024},
isbn = {9798400702174},
publisher = {Association for Computing Machinery},
address = {New York, NY, USA},
url = {https://doi.org/10.1145/3597503.3608136},
doi = {10.1145/3597503.3608136},
booktitle = {Proceedings of the IEEE/ACM 46th International Conference on Software Engineering},
articleno = {33},
numpages = {12},
location = {Lisbon, Portugal},
series = {ICSE '24}
}

@article{123_ye2024problem,
  title={A problem-oriented perspective and anchor verification for code optimization},
  author={Ye, Tong and Ma, Tengfei and Zhang, Xuhong and Yu, Hang and Yin, Jianwei and Wang, Wenhai},
  journal={arXiv preprint arXiv:2406.11935},
  year={2024}
}

@inproceedings{124_cui2024origen,
  title={Origen: Enhancing rtl code generation with code-to-code augmentation and self-reflection},
  author={Cui, Fan and Yin, Chenyang and Zhou, Kexing and Xiao, Youwei and Sun, Guangyu and Xu, Qiang and Guo, Qipeng and Liang, Yun and Zhang, Xingcheng and Song, Demin and others},
  booktitle={Proceedings of the 43rd IEEE/ACM International Conference on Computer-Aided Design},
  pages={1--9},
  year={2024}
}

@inproceedings{125_graphbased2024_hpec,
  title={A Graph-Based Algorithm for Optimizing GCC Compiler Flag Settings},
  author={Sajjadinasab, Reza and Arora, Sanjay and Drepper, Ulrich and Sanaullah, Ahmed and Herbordt, Martin},
  booktitle={2024 IEEE High Performance Extreme Computing Conference (HPEC)},
  pages={1--8},
  year={2024},
  organization={IEEE}
}

@inproceedings{127_neuroevolutionary2023_gecco,
  title={Neuroevolutionary Compiler Control for Code Optimization},
  author={Heckel, Kade},
  booktitle={Proceedings of the Companion Conference on Genetic and Evolutionary Computation},
  pages={2362--2365},
  year={2023}
}

@misc{129_binsum2023_arxiv,
      title={Binary Code Summarization: Benchmarking ChatGPT/GPT-4 and Other Large Language Models}, 
      author={Xin Jin and Jonathan Larson and Weiwei Yang and Zhiqiang Lin},
      year={2023},
      eprint={2312.09601},
      archivePrefix={arXiv},
      primaryClass={cs.CR},
      url={https://arxiv.org/abs/2312.09601}, 
}

@article{130_xu2023lmpa,
  title={Lmpa: Improving decompilation by synergy of large language model and program analysis},
  author={Xu, Xiangzhe and Zhang, Zhuo and Feng, Shiwei and Ye, Yapeng and Su, Zian and Jiang, Nan and Cheng, Siyuan and Tan, Lin and Zhang, Xiangyu},
  journal={arXiv preprint arXiv:2306.02546},
  year={2023}
}

@inproceedings{131_wei2023_fse,
author = {Wei, Yuxiang and Xia, Chunqiu Steven and Zhang, Lingming},
title = {Copiloting the Copilots: Fusing Large Language Models with Completion Engines for Automated Program Repair},
year = {2023},
isbn = {9798400703270},
publisher = {Association for Computing Machinery},
address = {New York, NY, USA},
url = {https://doi.org/10.1145/3611643.3616271},
doi = {10.1145/3611643.3616271},
booktitle = {Proceedings of the 31st ACM Joint European Software Engineering Conference and Symposium on the Foundations of Software Engineering},
pages = {172–184},
numpages = {13},
location = {San Francisco, CA, USA},
series = {ESEC/FSE 2023}
}

@article{132_lindner2023tracr,
  title={Tracr: Compiled transformers as a laboratory for interpretability},
  author={Lindner, David and Kram{\'a}r, J{\'a}nos and Farquhar, Sebastian and Rahtz, Matthew and McGrath, Tom and Mikulik, Vladimir},
  journal={Advances in Neural Information Processing Systems},
  volume={36},
  pages={37876--37899},
  year={2023}
}

@INPROCEEDINGS{133_xia2023_icse,
  author={Xia, Chunqiu Steven and Wei, Yuxiang and Zhang, Lingming},
  booktitle={2023 IEEE/ACM 45th International Conference on Software Engineering (ICSE)}, 
  title={Automated Program Repair in the Era of Large Pre-trained Language Models}, 
  year={2023},
  volume={},
  number={},
  pages={1482-1494},
  doi={10.1109/ICSE48619.2023.00129}
}

@misc{134_duan2025perfrlsmalllanguagemodel,
      title={PerfRL: A Small Language Model Framework for Efficient Code Optimization}, 
      author={Shukai Duan and Nikos Kanakaris and Xiongye Xiao and Heng Ping and Chenyu Zhou and Nesreen K. Ahmed and Guixiang Ma and Mihai Capota and Theodore L. Willke and Shahin Nazarian and Paul Bogdan},
      year={2025},
      eprint={2312.05657},
      archivePrefix={arXiv},
      primaryClass={cs.LG},
      url={https://arxiv.org/abs/2312.05657}, 
}

@article{135_thakur2023autochip,
  title={Autochip: Automating hdl generation using llm feedback},
  author={Thakur, Shailja and Blocklove, Jason and Pearce, Hammond and Tan, Benjamin and Garg, Siddharth and Karri, Ramesh},
  journal={arXiv preprint arXiv:2311.04887},
  year={2023}
}

@inproceedings{137_2023_issta,
author = {Xu, Xiangzhe and Feng, Shiwei and Ye, Yapeng and Shen, Guangyu and Su, Zian and Cheng, Siyuan and Tao, Guanhong and Shi, Qingkai and Zhang, Zhuo and Zhang, Xiangyu},
title = {Improving Binary Code Similarity Transformer Models by Semantics-Driven Instruction Deemphasis},
year = {2023},
isbn = {9798400702211},
publisher = {Association for Computing Machinery},
address = {New York, NY, USA},
url = {https://doi.org/10.1145/3597926.3598121},
doi = {10.1145/3597926.3598121},
booktitle = {Proceedings of the 32nd ACM SIGSOFT International Symposium on Software Testing and Analysis},
pages = {1106–1118},
numpages = {13},
location = {Seattle, WA, USA},
series = {ISSTA 2023}
}

@inproceedings{138_ANPL23_neurips,
author = {Huang, Di and Nan, Ziyuan and Hu, Xing and Jin, Pengwei and Peng, Shaohui and Wen, Yuanbo and Zhang, Rui and Du, Zidong and Guo, Qi and Pu, Yewen and Chen, Yunji},
title = {ANPL: towards natural programming with interactive decomposition},
year = {2023},
publisher = {Curran Associates Inc.},
address = {Red Hook, NY, USA},
booktitle = {Proceedings of the 37th International Conference on Neural Information Processing Systems},
articleno = {3040},
numpages = {37},
location = {New Orleans, LA, USA},
series = {NIPS '23}
}

@misc{139_tsimpourlas2023benchdirectdirectedlanguagemodel,
      title={BenchDirect: A Directed Language Model for Compiler Benchmarks}, 
      author={Foivos Tsimpourlas and Pavlos Petoumenos and Min Xu and Chris Cummins and Kim Hazelwood and Ajitha Rajan and Hugh Leather},
      year={2023},
      eprint={2303.01557},
      archivePrefix={arXiv},
      primaryClass={cs.LG},
      url={https://arxiv.org/abs/2303.01557}, 
}

@article{143_friedman2023learning,
  title={Learning transformer programs},
  author={Friedman, Dan and Wettig, Alexander and Chen, Danqi},
  journal={Advances in Neural Information Processing Systems},
  volume={36},
  pages={49044--49067},
  year={2023}
}

@inproceedings{141_optango2023_issre,
  author={Hongna Geng and Ming Zhong and Peihua Zhang and Fang Lv and Xiaobing Feng},
  title={OPTango: Multi-central Representation Learning against Innumerable Compiler Optimization for Binary Diffing},
  year={2023},
  cdate={1672531200000},
  pages={774-785},
  url={https://doi.org/10.1109/ISSRE59848.2023.00013},
  booktitle={ISSRE},
}

@misc{144_zhang2023doesllmgeneratesecurity,
      title={How well does LLM generate security tests?}, 
      author={Ying Zhang and Wenjia Song and Zhengjie Ji and Danfeng and Yao and Na Meng},
      year={2023},
      eprint={2310.00710},
      archivePrefix={arXiv},
      primaryClass={cs.CR},
      url={https://arxiv.org/abs/2310.00710}, 
}

@article{145_luo2023neural,
title={Neural Shape Compiler: A Unified Framework for Transforming between Text, Point Cloud, and Program},
author={Tiange Luo and Honglak Lee and Justin Johnson},
journal={Transactions on Machine Learning Research},
issn={2835-8856},
year={2023},
url={https://openreview.net/forum?id=gR9UVgH8PZ},
note={}
}

@inproceedings{146_sallm_asew,
author = {Siddiq, Mohammed Latif and da Silva Santos, Joanna Cecilia and Devareddy, Sajith and Muller, Anna},
title = {SALLM: Security Assessment of Generated Code},
year = {2024},
isbn = {9798400712494},
publisher = {Association for Computing Machinery},
address = {New York, NY, USA},
url = {https://doi.org/10.1145/3691621.3694934},
doi = {10.1145/3691621.3694934},
booktitle = {Proceedings of the 39th IEEE/ACM International Conference on Automated Software Engineering Workshops},
pages = {54–65},
numpages = {12},
location = {Sacramento, CA, USA},
series = {ASEW '24}
}

@inproceedings{147_brownlee2023enhancing,
  title={Enhancing genetic improvement mutations using large language models},
  author={Brownlee, Alexander EI and Callan, James and Even-Mendoza, Karine and Geiger, Alina and Hanna, Carol and Petke, Justyna and Sarro, Federica and Sobania, Dominik},
  booktitle={International symposium on search based software engineering},
  pages={153--159},
  year={2023},
  organization={Springer}
}

@inproceedings{150_kang2023large,
  title={Large language models are few-shot testers: Exploring llm-based general bug reproduction},
  author={Kang, Sungmin and Yoon, Juyeon and Yoo, Shin},
  booktitle={2023 IEEE/ACM 45th International Conference on Software Engineering (ICSE)},
  pages={2312--2323},
  year={2023},
  organization={IEEE}
}

@inproceedings{149_hpcgpt2023_scw,
author = {Ding, Xianzhong and Chen, Le and Emani, Murali and Liao, Chunhua and Lin, Pei-Hung and Vanderbruggen, Tristan and Xie, Zhen and Cerpa, Alberto and Du, Wan},
title = {HPC-GPT: Integrating Large Language Model for High-Performance Computing},
year = {2023},
isbn = {9798400707858},
publisher = {Association for Computing Machinery},
address = {New York, NY, USA},
url = {https://doi.org/10.1145/3624062.3624172},
doi = {10.1145/3624062.3624172},
booktitle = {Proceedings of the SC '23 Workshops of the International Conference on High Performance Computing, Network, Storage, and Analysis},
pages = {951–960},
numpages = {10},
location = {Denver, CO, USA},
series = {SC-W '23}
}

@article{148_zhang2023critical,
  title={A critical review of large language model on software engineering: An example from chatgpt and automated program repair},
  author={Zhang, Quanjun and Zhang, Tongke and Zhai, Juan and Fang, Chunrong and Yu, Bowen and Sun, Weisong and Chen, Zhenyu},
  journal={arXiv preprint arXiv:2310.08879},
  year={2023}
}

@misc{152_mannarswamy2022learningcombineinstructionsllvm,
      title={Learning to Combine Instructions in LLVM Compiler}, 
      author={Sandya Mannarswamy and Dibyendu Das},
      year={2022},
      eprint={2202.12379},
      archivePrefix={arXiv},
      primaryClass={cs.LG},
      url={https://arxiv.org/abs/2202.12379}, 
}

@inproceedings{153_babeltower2022_icml,
  title={Babeltower: Learning to auto-parallelized program translation},
  author={Wen, Yuanbo and Guo, Qi and Fu, Qiang and Li, Xiaqing and Xu, Jianxing and Tang, Yanlin and Zhao, Yongwei and Hu, Xing and Du, Zidong and Li, Ling and others},
  booktitle={International Conference on Machine Learning},
  pages={23685--23700},
  year={2022},
  organization={PMLR}
}

@inproceedings{154_transcoderir2023_iclr,
title={Code Translation with Compiler Representations},
author={Marc Szafraniec and Baptiste Roziere and Hugh James Leather and Patrick Labatut and Francois Charton and Gabriel Synnaeve},
booktitle={The Eleventh International Conference on Learning Representations },
year={2023},
url={https://openreview.net/forum?id=XomEU3eNeSQ}
}

@inproceedings{155_cao2022boosting,
  title={Boosting neural networks to decompile optimized binaries},
  author={Cao, Ying and Liang, Ruigang and Chen, Kai and Hu, Peiwei},
  booktitle={proceedings of the 38th annual computer security applications conference},
  pages={508--518},
  year={2022}
}

@inproceedings{160_jtrans2022_issta,
author = {Wang, Hao and Qu, Wenjie and Katz, Gilad and Zhu, Wenyu and Gao, Zeyu and Qiu, Han and Zhuge, Jianwei and Zhang, Chao},
title = {jTrans: jump-aware transformer for binary code similarity detection},
year = {2022},
isbn = {9781450393799},
publisher = {Association for Computing Machinery},
address = {New York, NY, USA},
url = {https://doi.org/10.1145/3533767.3534367},
doi = {10.1145/3533767.3534367},
booktitle = {Proceedings of the 31st ACM SIGSOFT International Symposium on Software Testing and Analysis},
pages = {1–13},
numpages = {13},
location = {Virtual, South Korea},
series = {ISSTA 2022}
}

@article{162_tavarageri2021poly_arxiv,
  title={AI Powered Compiler Techniques for DL Code Optimization},
  author={Tavarageri, Sanket and Goyal, Gagandeep and Avancha, Sasikanth and Kaul, Bharat and Upadrasta, Ramakrishna},
  journal={arXiv preprint arXiv:2104.05573},
  year={2021}
}

@inproceedings{165_weiss2021thinking,
  title={Thinking like transformers},
  author={Weiss, Gail and Goldberg, Yoav and Yahav, Eran},
  booktitle={International Conference on Machine Learning},
  pages={11080--11090},
  year={2021},
  organization={PMLR}
}

@inproceedings{166_tfix2021_icml,
  title={Tfix: Learning to fix coding errors with a text-to-text transformer},
  author={Berabi, Berkay and He, Jingxuan and Raychev, Veselin and Vechev, Martin},
  booktitle={International Conference on Machine Learning},
  pages={780--791},
  year={2021},
  organization={PMLR}
}

@inproceedings{167_zhai2024_osdi,
  title={Enabling Tensor Language Model to Assist in Generating $\{$High-Performance$\}$ Tensor Programs for Deep Learning},
  author={Zhai, Yi and Yang, Sijia and Pan, Keyu and Zhang, Renwei and Liu, Shuo and Liu, Chao and Ye, Zichun and Ji, Jianmin and Zhao, Jie and Zhang, Yu and others},
  booktitle={18th USENIX Symposium on Operating Systems Design and Implementation (OSDI 24)},
  pages={289--305},
  year={2024}
}

@inproceedings{168_unsupervised2023_emnlp,
    title = "Unsupervised Binary Code Translation with Application to Code Clone Detection and Vulnerability Discovery",
    author = "Ahmad, Iftakhar  and
      Luo, Lannan",
    editor = "Bouamor, Houda  and
      Pino, Juan  and
      Bali, Kalika",
    booktitle = "Findings of the Association for Computational Linguistics: EMNLP 2023",
    month = dec,
    year = "2023",
    address = "Singapore",
    publisher = "Association for Computational Linguistics",
    url = "https://aclanthology.org/2023.findings-emnlp.971/",
    doi = "10.18653/v1/2023.findings-emnlp.971",
    pages = "14581--14592"
}

@inproceedings{169_armengol-estape2024forklift,
title={Forklift: An Extensible Neural Lifter},
author={Jordi Armengol-Estape and Rodrigo C. O. Rocha and Jackson Woodruff and Pasquale Minervini and Michael O'Boyle},
booktitle={First Conference on Language Modeling},
year={2024},
url={https://openreview.net/forum?id=LWfDcI6txJ}
}

@inproceedings{170_how2021_icml,
  title={How could neural networks understand programs?},
  author={Peng, Dinglan and Zheng, Shuxin and Li, Yatao and Ke, Guolin and He, Di and Liu, Tie-Yan},
  booktitle={International Conference on Machine Learning},
  pages={8476--8486},
  year={2021},
  organization={PMLR}
}

@inproceedings{171_guo2022enabling,
  title={Enabling transformers to understand low-level programs},
  author={Guo, Zifan Carl and Moses, William S},
  booktitle={2022 IEEE High Performance Extreme Computing Conference (HPEC)},
  pages={1--9},
  year={2022},
  organization={IEEE}
}

@inproceedings{172_chakraborty2022natgen,
  title={Natgen: generative pre-training by “naturalizing” source code},
  author={Chakraborty, Saikat and Ahmed, Toufique and Ding, Yangruibo and Devanbu, Premkumar T and Ray, Baishakhi},
  booktitle={Proceedings of the 30th ACM joint european software engineering conference and symposium on the foundations of software engineering},
  pages={18--30},
  year={2022}
}

@article{173_wong2025decllm,
  title={DecLLM: LLM-Augmented Recompilable Decompilation for Enabling Programmatic Use of Decompiled Code},
  author={Wong, Wai Kin and Wu, Daoyuan and Wang, Huaijin and Li, Zongjie and Liu, Zhibo and Wang, Shuai and Tang, Qiyi and Nie, Sen and Wu, Shi},
  journal={Proceedings of the ACM on Software Engineering},
  volume={2},
  number={ISSTA},
  pages={1841--1864},
  year={2025},
  publisher={ACM New York, NY, USA}
}

@inproceedings{174_coderosseta2024_neurips,
title={CodeRosetta: Pushing the Boundaries of Unsupervised Code Translation for Parallel Programming},
author={Ali TehraniJamsaz and Arijit Bhattacharjee and Le Chen and Nesreen K. Ahmed and Amir Yazdanbakhsh and Ali Jannesari},
booktitle={The Thirty-eighth Annual Conference on Neural Information Processing Systems},
year={2024},
url={https://openreview.net/forum?id=V6hrg4O9gg}
}

@inproceedings{
175_transcoderst,
title={Leveraging Automated Unit Tests for Unsupervised Code Translation},
author={Baptiste Roziere and Jie Zhang and Francois Charton and Mark Harman and Gabriel Synnaeve and Guillaume Lample},
booktitle={International Conference on Learning Representations},
year={2022},
url={https://openreview.net/forum?id=cmt-6KtR4c4}
}

@inproceedings{
176_dobf,
title={{DOBF}: A Deobfuscation Pre-Training Objective for Programming Languages},
author={Marie-anne Lachaux and Baptiste Roziere and Marc Szafraniec and Guillaume Lample},
booktitle={Advances in Neural Information Processing Systems},
editor={A. Beygelzimer and Y. Dauphin and P. Liang and J. Wortman Vaughan},
year={2021},
url={https://openreview.net/forum?id=3ez9BSHTNT}
}

@inproceedings{177_transcoder,
author = {Roziere, Baptiste and Lachaux, Marie-Anne and Chanussot, Lowik and Lample, Guillaume},
title = {Unsupervised translation of programming languages},
year = {2020},
isbn = {9781713829546},
publisher = {Curran Associates Inc.},
address = {Red Hook, NY, USA},
booktitle = {Proceedings of the 34th International Conference on Neural Information Processing Systems},
articleno = {1730},
numpages = {11},
location = {Vancouver, BC, Canada},
series = {NIPS '20}
}

@misc{178_cummins2024donttransformcodecode,
      title={Don't Transform the Code, Code the Transforms: Towards Precise Code Rewriting using LLMs}, 
      author={Chris Cummins and Volker Seeker and Jordi Armengol-Estapé and Aram H. Markosyan and Gabriel Synnaeve and Hugh Leather},
      year={2024},
      eprint={2410.08806},
      archivePrefix={arXiv},
      primaryClass={cs.LG},
      url={https://arxiv.org/abs/2410.08806}, 
}

@misc{kevin2025arxiv,
      title={Kevin: Multi-Turn RL for Generating CUDA Kernels}, 
      author={Carlo Baronio and Pietro Marsella and Ben Pan and Simon Guo and Silas Alberti},
      year={2025},
      eprint={2507.11948},
      archivePrefix={arXiv},
      primaryClass={cs.LG},
      url={https://arxiv.org/abs/2507.11948}, 
}

@misc{stark2025arxiv,
      title={STARK: Strategic Team of Agents for Refining Kernels}, 
      author={Juncheng Dong and Yang Yang and Tao Liu and Yang Wang and Feng Qi and Vahid Tarokh and Kaushik Rangadurai and Shuang Yang},
      year={2025},
      eprint={2510.16996},
      archivePrefix={arXiv},
      primaryClass={cs.AI},
      url={https://arxiv.org/abs/2510.16996}, 
}

@inproceedings{tvm_2018_osdi,
author = {Chen, Tianqi and Moreau, Thierry and Jiang, Ziheng and Zheng, Lianmin and Yan, Eddie and Cowan, Meghan and Shen, Haichen and Wang, Leyuan and Hu, Yuwei and Ceze, Luis and Guestrin, Carlos and Krishnamurthy, Arvind},
title = {TVM: an automated end-to-end optimizing compiler for deep learning},
year = {2018},
isbn = {9781931971478},
publisher = {USENIX Association},
address = {USA},
booktitle = {Proceedings of the 13th USENIX Conference on Operating Systems Design and Implementation},
pages = {579–594},
numpages = {16},
location = {Carlsbad, CA, USA},
series = {OSDI'18}
}

@article{cudal1_2025_arxiv,
  title={CUDA-L1: Improving CUDA Optimization via Contrastive Reinforcement Learning},
  author={Li, Xiaoya and Sun, Xiaofei and Wang, Albert and Li, Jiwei and Chris, Shum},
  journal={arXiv preprint arXiv:2507.14111},
  year={2025}
}

@inproceedings{
kernelbench_2025_icml,
title={KernelBench: Can {LLM}s Write Efficient {GPU} Kernels?},
author={Anne Ouyang and Simon Guo and Simran Arora and Alex L Zhang and William Hu and Christopher Re and Azalia Mirhoseini},
booktitle={Forty-second International Conference on Machine Learning},
year={2025},
url={https://openreview.net/forum?id=yeoN1iQT1x}
}

@inproceedings{alive2_2021_pldi,
author = {Lopes, Nuno P. and Lee, Juneyoung and Hur, Chung-Kil and Liu, Zhengyang and Regehr, John},
title = {Alive2: bounded translation validation for LLVM},
year = {2021},
isbn = {9781450383912},
publisher = {Association for Computing Machinery},
address = {New York, NY, USA},
url = {https://doi.org/10.1145/3453483.3454030},
doi = {10.1145/3453483.3454030},
booktitle = {Proceedings of the 42nd ACM SIGPLAN International Conference on Programming Language Design and Implementation},
pages = {65–79},
numpages = {15},
location = {Virtual, Canada},
series = {PLDI 2021}
}

@article{typeconstrainedgeneration2025pldi,
  title={Type-Constrained Code Generation with Language Models},
  author={M{\"u}ndler, Niels and He, Jingxuan and Wang, Hao and Sen, Koushik and Song, Dawn and Vechev, Martin},
  journal={Proceedings of the ACM on Programming Languages},
  volume={9},
  number={PLDI},
  pages={601--626},
  year={2025},
  publisher={ACM New York, NY, USA}
}

@misc{xgrammar2024arxiv,
      title={XGrammar: Flexible and Efficient Structured Generation Engine for Large Language Models}, 
      author={Yixin Dong and Charlie F. Ruan and Yaxing Cai and Ruihang Lai and Ziyi Xu and Yilong Zhao and Tianqi Chen},
      year={2024},
      eprint={2411.15100},
      archivePrefix={arXiv},
      primaryClass={cs.CL},
      url={https://arxiv.org/abs/2411.15100}, 
}

@misc{outlines2023arxiv,
      title={Efficient Guided Generation for Large Language Models}, 
      author={Brandon T. Willard and Rémi Louf},
      year={2023},
      eprint={2307.09702},
      archivePrefix={arXiv},
      primaryClass={cs.CL},
      url={https://arxiv.org/abs/2307.09702}, 
}

@misc{grpo_2024_arxiv,
      title={DeepSeekMath: Pushing the Limits of Mathematical Reasoning in Open Language Models}, 
      author={Zhihong Shao and Peiyi Wang and Qihao Zhu and Runxin Xu and Junxiao Song and Xiao Bi and Haowei Zhang and Mingchuan Zhang and Y. K. Li and Y. Wu and Daya Guo},
      year={2024},
      eprint={2402.03300},
      archivePrefix={arXiv},
      primaryClass={cs.CL},
      url={https://arxiv.org/abs/2402.03300}, 
}

@inproceedings{cot_2022_neurips,
author = {Wei, Jason and Wang, Xuezhi and Schuurmans, Dale and Bosma, Maarten and Ichter, Brian and Xia, Fei and Chi, Ed H. and Le, Quoc V. and Zhou, Denny},
title = {Chain-of-thought prompting elicits reasoning in large language models},
year = {2022},
isbn = {9781713871088},
publisher = {Curran Associates Inc.},
address = {Red Hook, NY, USA},
booktitle = {Proceedings of the 36th International Conference on Neural Information Processing Systems},
articleno = {1800},
numpages = {14},
location = {New Orleans, LA, USA},
series = {NIPS '22}
}

@inproceedings{rag_2020_neurips,
author = {Lewis, Patrick and Perez, Ethan and Piktus, Aleksandra and Petroni, Fabio and Karpukhin, Vladimir and Goyal, Naman and K\"{u}ttler, Heinrich and Lewis, Mike and Yih, Wen-tau and Rockt\"{a}schel, Tim and Riedel, Sebastian and Kiela, Douwe},
title = {Retrieval-augmented generation for knowledge-intensive NLP tasks},
year = {2020},
isbn = {9781713829546},
publisher = {Curran Associates Inc.},
address = {Red Hook, NY, USA},
booktitle = {Proceedings of the 34th International Conference on Neural Information Processing Systems},
articleno = {793},
numpages = {16},
location = {Vancouver, BC, Canada},
series = {NIPS '20}
}

@inproceedings{z3smt_2008,
author = {De Moura, Leonardo and Bj\o{}rner, Nikolaj},
title = {Z3: an efficient SMT solver},
year = {2008},
isbn = {3540787992},
publisher = {Springer-Verlag},
address = {Berlin, Heidelberg},
booktitle = {Proceedings of the Theory and Practice of Software, 14th International Conference on Tools and Algorithms for the Construction and Analysis of Systems},
pages = {337–340},
numpages = {4},
location = {Budapest, Hungary},
series = {TACAS'08/ETAPS'08}
}

@misc{metallmcompiler-arxiv23,
      title={Large Language Models for Compiler Optimization}, 
      author={Chris Cummins and Volker Seeker and Dejan Grubisic and Mostafa Elhoushi and Youwei Liang and Baptiste Roziere and Jonas Gehring and Fabian Gloeckle and Kim Hazelwood and Gabriel Synnaeve and Hugh Leather},
      year={2023},
      eprint={2309.07062},
      archivePrefix={arXiv},
      primaryClass={cs.PL},
      url={https://arxiv.org/abs/2309.07062}, 
}

@inproceedings{metallmcompiler-cc25,
author = {Cummins, Chris and Seeker, Volker and Grubisic, Dejan and Roziere, Baptiste and Gehring, Jonas and Synnaeve, Gabriel and Leather, Hugh},
title = {LLM Compiler: Foundation Language Models for Compiler Optimization},
year = {2025},
isbn = {9798400714078},
publisher = {Association for Computing Machinery},
address = {New York, NY, USA},
url = {https://doi.org/10.1145/3708493.3712691},
doi = {10.1145/3708493.3712691},
booktitle = {Proceedings of the 34th ACM SIGPLAN International Conference on Compiler Construction},
pages = {141–153},
numpages = {13},
location = {Las Vegas, NV, USA},
series = {CC '25}
}

@article{copilot,
  title={Evaluating Large Language Models Trained on Code},
  author={Chen, Mark and Tworek, Jerry and Jun, Heewoo and Yuan, Qiming and Pinto, Henrique Ponde de Oliveira and Kaplan, Jared and Edwards, Harri and Burda, Yuri and Joseph, Nicholas and Brockman, Greg and others},
  journal={arXiv preprint arXiv:2107.03374},
  year={2021}
}

@misc{tabnine,
  author = {{Tabnine}},
  title = {{Tabnine: AI Code Completion Tool}},
  year = {2022},
  howpublished = {\url{https://www.tabnine.com/}},
  note = {Accessed: August 11, 2025}
}

@misc{cursor_ide,
  author = {{Anysphere, Inc.}},
  title = {Cursor: The AI-first Code Editor},
  year = {2023},
  howpublished = {\url{https://cursor.sh/}},
  note = {Accessed: August 11, 2025}
}

@misc{claude_code,
  author = {{anthropics}},
  title = {{claude-code: A command line interface for Anthropic's Claude AI}},
  year = {2025},
  publisher = {GitHub},
  journal = {GitHub repository},
  howpublished = {\url{https://github.com/anthropics/claude-code}},
  note = {Accessed: August 11, 2025}
}

@misc{gemini_cli,
  author = {{google-gemini}},
  title = {{gemini-cli: A Google Gemini CLI and Python API}},
  year = {2025},
  publisher = {GitHub},
  journal = {GitHub repository},
  howpublished = {\url{https://github.com/google-gemini/gemini-cli}},
  note = {Accessed: August 11, 2025}
}

@inproceedings{triton2019mapl,
author = {Tillet, Philippe and Kung, H. T. and Cox, David},
title = {Triton: an intermediate language and compiler for tiled neural network computations},
year = {2019},
isbn = {9781450367196},
publisher = {Association for Computing Machinery},
address = {New York, NY, USA},
url = {https://doi.org/10.1145/3315508.3329973},
doi = {10.1145/3315508.3329973},
booktitle = {Proceedings of the 3rd ACM SIGPLAN International Workshop on Machine Learning and Programming Languages},
pages = {10–19},
numpages = {10},
location = {Phoenix, AZ, USA},
series = {MAPL 2019}
}

@misc{cute2025web,
  author = {{NVIDIA Corporation}},
  title = {{CUTLASS Python Interface Overview}},
  year = {2025},
  howpublished = {\url{https://docs.nvidia.com/cutlass/media/docs/pythonDSL/overview.html}},
  note = {Accessed: August 11, 2025}
}

@misc{gpt4,
      title={GPT-4 Technical Report}, 
      author={OpenAI},
      year={2024},
      eprint={2303.08774},
      archivePrefix={arXiv},
      primaryClass={cs.CL},
      url={https://arxiv.org/abs/2303.08774}, 
}

@article{gemini2023,
  title={Gemini: A Family of Highly Capable Multimodal Models},
  author={{Gemini Team} and {Google}},
  journal={arXiv preprint arXiv:2312.11805},
  year={2023}
}

@techreport{claude3,
  author = {Anthropic},
  title = {The Claude 3 Model Family: Opus, Sonnet, Haiku},
  institution = {Anthropic},
  year = {2024},
  month = {March},
  note = {Available at \url{https://www.anthropic.com/news/claude-3-family}}
}

@ARTICLE{mlincompilersurvey2018,
  author={Wang, Zheng and O’Boyle, Michael},
  journal={Proceedings of the IEEE}, 
  title={Machine Learning in Compiler Optimization}, 
  year={2018},
  volume={106},
  number={11},
  pages={1879-1901},
  doi={10.1109/JPROC.2018.2817118}}

@article{alexnet2012,
  title={Imagenet classification with deep convolutional neural networks},
  author={Krizhevsky, Alex and Sutskever, Ilya and Hinton, Geoffrey E},
  journal={Advances in neural information processing systems},
  volume={25},
  year={2012}
}

@inproceedings{resnet2016,
  title={Deep residual learning for image recognition},
  author={He, Kaiming and Zhang, Xiangyu and Ren, Shaoqing and Sun, Jian},
  booktitle={Proceedings of the IEEE conference on computer vision and pattern recognition},
  pages={770--778},
  year={2016}
}

@article{literaturereview2007,
  title={Guidelines for performing systematic literature reviews in software engineering},
  author={Kitchenham, Barbara and Charters, Stuart and others},
  year={2007},
  publisher={Keele, UK}
}

@inproceedings{exebench_2022_maps,
author = {Armengol-Estap\'{e}, Jordi and Woodruff, Jackson and Brauckmann, Alexander and Magalh\~{a}es, Jos\'{e} Wesley de Souza and O'Boyle, Michael F. P.},
title = {ExeBench: an ML-scale dataset of executable C functions},
year = {2022},
isbn = {9781450392730},
publisher = {Association for Computing Machinery},
address = {New York, NY, USA},
url = {https://doi.org/10.1145/3520312.3534867},
doi = {10.1145/3520312.3534867},
booktitle = {Proceedings of the 6th ACM SIGPLAN International Symposium on Machine Programming},
pages = {50–59},
numpages = {10},
location = {San Diego, CA, USA},
series = {MAPS 2022}
}

@inproceedings{tritonbench_2025_acl,
    title = "{T}riton{B}ench: Benchmarking Large Language Model Capabilities for Generating Triton Operators",
    author = "Li, Jianling  and
      Li, ShangZhan  and
      Gao, Zhenye  and
      Shi, Qi  and
      Li, Yuxuan  and
      Wang, Zefan  and
      Huang, Jiacheng  and
      WangHaojie, WangHaojie  and
      Wang, Jianrong  and
      Han, Xu  and
      Liu, Zhiyuan  and
      Sun, Maosong",
    editor = "Che, Wanxiang  and
      Nabende, Joyce  and
      Shutova, Ekaterina  and
      Pilehvar, Mohammad Taher",
    booktitle = "Findings of the Association for Computational Linguistics: ACL 2025",
    month = jul,
    year = "2025",
    address = "Vienna, Austria",
    publisher = "Association for Computational Linguistics",
    url = "https://aclanthology.org/2025.findings-acl.1183/",
    doi = "10.18653/v1/2025.findings-acl.1183",
    pages = "23053--23066",
    ISBN = "979-8-89176-256-5"
}

@inproceedings{verilogeval_2023_iccad,
  title={Verilogeval: Evaluating large language models for verilog code generation},
  author={Liu, Mingjie and Pinckney, Nathaniel and Khailany, Brucek and Ren, Haoxing},
  booktitle={2023 IEEE/ACM International Conference on Computer Aided Design (ICCAD)},
  pages={1--8},
  year={2023},
  organization={IEEE}
}

@article{spec_2006,
  title={SPEC CPU2006 benchmark descriptions},
  author={Henning, John L},
  journal={ACM SIGARCH Computer Architecture News},
  volume={34},
  number={4},
  pages={1--17},
  year={2006},
  publisher={ACM New York, NY, USA}
}

@inproceedings{avatar,
    title = "{AVATAR}: A Parallel Corpus for {J}ava-Python Program Translation",
    author = "Ahmad, Wasi Uddin  and
      Tushar, Md Golam Rahman  and
      Chakraborty, Saikat  and
      Chang, Kai-Wei",
    editor = "Rogers, Anna  and
      Boyd-Graber, Jordan  and
      Okazaki, Naoaki",
    booktitle = "Findings of the Association for Computational Linguistics: ACL 2023",
    month = jul,
    year = "2023",
    address = "Toronto, Canada",
    publisher = "Association for Computational Linguistics",
    url = "https://aclanthology.org/2023.findings-acl.143/",
    doi = "10.18653/v1/2023.findings-acl.143",
    pages = "2268--2281"
}

@misc{codenet,
      title={CodeNet: A Large-Scale AI for Code Dataset for Learning a Diversity of Coding Tasks}, 
      author={Ruchir Puri and David S. Kung and Geert Janssen and Wei Zhang and Giacomo Domeniconi and Vladimir Zolotov and Julian Dolby and Jie Chen and Mihir Choudhury and Lindsey Decker and Veronika Thost and Luca Buratti and Saurabh Pujar and Shyam Ramji and Ulrich Finkler and Susan Malaika and Frederick Reiss},
      year={2021},
      eprint={2105.12655},
      archivePrefix={arXiv},
      primaryClass={cs.SE},
      url={https://arxiv.org/abs/2105.12655}, 
}

@article{pie,
    title={Learning Performance-Improving Code Edits},
    author={Madaan, Aman and Shypula, Alexander and Alon, Uri and Hashemi, Milad and Ranganathan, Parthasarathy and Yang, Yiming and Neubig, Graham and Yazdanbakhsh, Amir},
    journal={arXiv preprint arXiv:2302.07867},
    year={2023}
}

@inproceedings{mibench,
  title={MiBench: A free, commercially representative embedded benchmark suite},
  author={Guthaus, Matthew R and Ringenberg, Jeffrey S and Ernst, Dan and Austin, Todd M and Mudge, Trevor and Brown, Richard B},
  booktitle={Proceedings of the fourth annual IEEE international workshop on workload characterization. WWC-4 (Cat. No. 01EX538)},
  pages={3--14},
  year={2001},
  organization={IEEE}
}

@inproceedings{anghabench,
  title={Anghabench: A suite with one million compilable c benchmarks for code-size reduction},
  author={Da Silva, Anderson Faustino and Kind, Bruno Conde and de Souza Magalh{\~a}es, Jos{\'e} Wesley and Rocha, Jer{\^o}nimo Nunes and Guimaraes, Breno Campos Ferreira and Pereira, Fernando Magno Quinao},
  booktitle={2021 IEEE/ACM International Symposium on Code Generation and Optimization (CGO)},
  pages={378--390},
  year={2021},
  organization={IEEE}
}

@misc{humaneval,
      title={Evaluating Large Language Models Trained on Code}, 
      author={Mark Chen and Jerry Tworek and Heewoo Jun and Qiming Yuan and Henrique Ponde de Oliveira Pinto and Jared Kaplan and Harri Edwards and Yuri Burda and Nicholas Joseph and Greg Brockman and Alex Ray and Raul Puri and Gretchen Krueger and Michael Petrov and Heidy Khlaaf and Girish Sastry and Pamela Mishkin and Brooke Chan and Scott Gray and Nick Ryder and Mikhail Pavlov and Alethea Power and Lukasz Kaiser and Mohammad Bavarian and Clemens Winter and Philippe Tillet and Felipe Petroski Such and Dave Cummings and Matthias Plappert and Fotios Chantzis and Elizabeth Barnes and Ariel Herbert-Voss and William Hebgen Guss and Alex Nichol and Alex Paino and Nikolas Tezak and Jie Tang and Igor Babuschkin and Suchir Balaji and Shantanu Jain and William Saunders and Christopher Hesse and Andrew N. Carr and Jan Leike and Josh Achiam and Vedant Misra and Evan Morikawa and Alec Radford and Matthew Knight and Miles Brundage and Mira Murati and Katie Mayer and Peter Welinder and Bob McGrew and Dario Amodei and Sam McCandlish and Ilya Sutskever and Wojciech Zaremba},
      year={2021},
      eprint={2107.03374},
      archivePrefix={arXiv},
      primaryClass={cs.LG},
      url={https://arxiv.org/abs/2107.03374}, 
}

@misc{mbpp,
      title={Program Synthesis with Large Language Models}, 
      author={Jacob Austin and Augustus Odena and Maxwell Nye and Maarten Bosma and Henryk Michalewski and David Dohan and Ellen Jiang and Carrie Cai and Michael Terry and Quoc Le and Charles Sutton},
      year={2021},
      eprint={2108.07732},
      archivePrefix={arXiv},
      primaryClass={cs.PL},
      url={https://arxiv.org/abs/2108.07732}, 
}

@inproceedings{bleu,
  title={Bleu: a method for automatic evaluation of machine translation},
  author={Papineni, Kishore and Roukos, Salim and Ward, Todd and Zhu, Wei-Jing},
  booktitle={Proceedings of the 40th annual meeting of the Association for Computational Linguistics},
  pages={311--318},
  year={2002}
}

@article{codebleu,
  title={Codebleu: a method for automatic evaluation of code synthesis},
  author={Ren, Shuo and Guo, Daya and Lu, Shuai and Zhou, Long and Liu, Shujie and Tang, Duyu and Sundaresan, Neel and Zhou, Ming and Blanco, Ambrosio and Ma, Shuai},
  journal={arXiv preprint arXiv:2009.10297},
  year={2020}
}

@inproceedings{compilergym,
  title={Compilergym: Robust, performant compiler optimization environments for ai research},
  author={Cummins, Chris and Wasti, Bram and Guo, Jiadong and Cui, Brandon and Ansel, Jason and Gomez, Sahir and Jain, Somya and Liu, Jia and Teytaud, Olivier and Steiner, Benoit and others},
  booktitle={2022 IEEE/ACM International Symposium on Code Generation and Optimization (CGO)},
  pages={92--105},
  year={2022},
  organization={IEEE}
}

@inproceedings{defects4j,
  title={Defects4J: A database of existing faults to enable controlled testing studies for Java programs},
  author={Just, Ren{\'e} and Jalali, Darioush and Ernst, Michael D},
  booktitle={Proceedings of the 2014 international symposium on software testing and analysis},
  pages={437--440},
  year={2014}
}

\newpage
\begin{appendices}

\end{appendices}

\end{document}